\documentclass[a4paper,11pt]{article}
\pdfoutput=1

\usepackage[english]{babel}

\usepackage{amsmath, amsfonts, amssymb, amsthm}

\usepackage[mathscr]{euscript}
\usepackage{enumitem}

\usepackage{tikz}
\usetikzlibrary{arrows}
\usetikzlibrary{calc}
\usetikzlibrary{decorations.pathreplacing}

\usepackage{dsfont}

\usepackage{bbm}
\DeclareMathAlphabet{\mathdsl}{U}{bbm}{m}{sl}

\usepackage{cite}

\usepackage{changes}

\usepackage{accents}

\usepackage{amsmath, amsfonts, amssymb}
\usepackage{slashed}
\usepackage{booktabs}
\usepackage[format=hang,labelfont={bf}]{caption}
\usepackage{bm}
\usepackage{jheppub}

\newcommand{\Adj}[1]{{\rm Ad}\, #1}

\newcommand{\Lie}{{\cL}}
\newcommand{\dLie}{\mathbb L}
\newcommand{\eol}{\notag \\}

\newcommand{\ol}{\overline}
\newcommand{\pa}{\partial}
\newcommand{\veps}{\varepsilon}
\newcommand{\ul}{\underline}
\newcommand{\Pair}[2]{\langle \langle #1, #2 \rangle \rangle}
\newcommand{\pair}[2]{\langle #1, #2 \rangle}
\newcommand{\EE}{{\mathbb E}}
\newcommand{\WZW}{\mathrm{WZW}}

\newcommand{\pl}{\langle}

\newcommand{\pr}{\rangle}

\newcommand{\dd}{\mathrm{d}}

\newcommand{\ket}[1]{|#1\rangle}

\DeclareMathOperator{\sdet}{sdet}
\DeclareMathOperator{\STr}{STr}

\newcommand{\cA}{{\mathcal A}}
\newcommand{\cB}{{\mathcal B}}
\newcommand{\cC}{{\mathcal C}}
\newcommand{\cD}{{\mathcal D}}
\newcommand{\cE}{{\mathcal E}}
\newcommand{\cF}{{\mathcal F}}
\newcommand{\cH}{{\mathcal H}}
\newcommand{\cL}{{\mathcal L}}
\newcommand{\cM}{{\mathcal M}}
\newcommand{\cN}{{\mathcal N}}
\newcommand{\cO}{{\mathcal O}}
\newcommand{\cP}{{\mathcal P}}
\newcommand{\cQ}{{\mathcal Q}}
\newcommand{\cR}{{\mathcal R}}
\newcommand{\cT}{{\mathcal T}}
\newcommand{\cU}{{\mathcal U}}
\newcommand{\cV}{{\mathcal V}}
\newcommand{\cW}{{\mathcal W}}
\newcommand{\cX}{{\mathcal X}}
\newcommand{\cZ}{{\mathcal Z}}

\newcommand{\wtM}{\widetilde M}

\newcommand{\wtcM}{{\widetilde\cM}}
\newcommand{\wtcN}{{\widetilde\cN}}
\newcommand{\wtcP}{{\widetilde\cP}}
\newcommand{\wtcV}{{\widetilde\cV}}

\newcommand{\ra}{{\mathrm a}}
\newcommand{\rb}{{\mathrm b}}
\newcommand{\rc}{{\mathrm c}}
\newcommand{\rd}{{\mathrm d}}

\newcommand{\rba}{{\ol{\ra}}}
\newcommand{\rbb}{{\ol{\rb}}}
\newcommand{\rbc}{{\ol{\rc}}}

\newcommand{\halpha}{{\hat{\alpha}}}
\newcommand{\hbeta}{{\hat{\beta}}}
\newcommand{\hgamma}{{\hat{\gamma}}}

\newcommand{\balpha}{{\bar\alpha}}
\newcommand{\bbeta}{{\bar\beta}}
\newcommand{\bgamma}{{\bar\gamma}}
\newcommand{\bdelta}{{\bar\delta}}

\newcommand{\talpha}{{\tilde\alpha}}
\newcommand{\tbeta}{{\tilde\beta}}
\newcommand{\tgamma}{{\tilde\gamma}}

\newcommand{\hbbeta}{{\hat\bbeta}}
\newcommand{\hbgamma}{{\hat\bgamma}}

\ifdefined\hm
\renewcommand{\hm}{{\hat{m}}}
\else
\newcommand{\hm}{{\hat{m}}}
\fi
\newcommand{\hn}{{\hat{n}}}

\newcommand{\ha}{{\hat{a}}}
\newcommand{\hb}{{\hat{b}}}

\newcommand{\hA}{{\widehat{A}}}
\newcommand{\hB}{{\widehat{B}}}
\newcommand{\hC}{{\widehat{C}}}
\newcommand{\hD}{{\widehat{D}}}
\newcommand{\hM}{{\widehat{M}}}
\newcommand{\hN}{{\widehat{N}}}

\newcommand{\hcA}{{\widehat\cA}}
\newcommand{\hcB}{{\widehat\cB}}
\newcommand{\hcC}{{\widehat\cC}}
\newcommand{\hcM}{{\widehat\cM}}
\newcommand{\hcN}{{\widehat\cN}}

\newcommand{\uM}{{\ul{M}}}
\newcommand{\uN}{{\ul{N}}}
\newcommand{\uP}{{\ul{P}}}

\newcommand{\um}{{\ul{m}}}
\newcommand{\un}{{\ul{n}}}

\newcommand{\g}[1]{\mathsf{#1}}

\DeclareRobustCommand{\Iso}[1]{
    \IfEqCase{#1}{%
        {1}{{\scriptscriptstyle{\mathbf R}}}
        {2}{{\scriptscriptstyle{\mathbf S}}}
        {3}{{\scriptscriptstyle{\mathbf T}}}
        {4}{{\scriptscriptstyle{\mathbf U}}}
        {5}{{\scriptscriptstyle{\mathbf V}}}
    }[\PackageError{Iso}{Undefined option to Iso: #1}{}]%
}

\DeclareRobustCommand{\bigIso}[1]{
    \IfEqCase{#1}{%
        {1}{{{\mathbf R}}}
        {2}{{{\mathbf S}}}
        {3}{{{\mathbf T}}}
        {4}{{{\mathbf U}}}
    }[\PackageError{Iso}{Undefined option to Iso: #1}{}]%
}

\DeclareRobustCommand{\iso}[1]{
    \IfEqCase{#1}{%
        {1}{{\mathbf r}}
        {2}{{\mathbf s}}
        {3}{{\mathbf t}}
        {4}{{\mathbf u}}
    }[\PackageError{iso}{Undefined option to ciso: #1}{}]%
}

\DeclareRobustCommand{\fIso}[1]{
    \IfEqCase{#1}{%
        {1}{{\scriptscriptstyle{\bm \rho}}}
        {2}{{\scriptscriptstyle{\bm \sigma}}}
        {3}{{\scriptscriptstyle{\bm \tau}}}
        {4}{{\scriptscriptstyle{\bm \upsilon}}}
    }[\PackageError{fIso}{Undefined option to sIso: #1}{}]%
}

\DeclareRobustCommand{\ciso}[1]{
    \IfEqCase{#1}{%
        {1}{{\dot m}}
        {2}{{\dot n}}
        {3}{{\dot p}}
        {4}{{\dot q}}
    }[\PackageError{ciso}{Undefined option to ciso: #1}{}]%
}

\DeclareRobustCommand{\cIso}[1]{
    \IfEqCase{#1}{%
        {1}{{\dot M}}
        {2}{{\dot N}}
        {3}{{\dot P}}
        {4}{{\dot Q}}
    }[\PackageError{cIso}{Undefined option to cIso: #1}{}]%
}

\newcommand{\ca}{A}
\newcommand{\cb}{B}
\newcommand{\cc}{C}
\newcommand{\cd}{D}

\newcommand{\sa}{{\ul A}}
\renewcommand{\sb}{{\ul B}}
\renewcommand{\sc}{{\ul C}}
\newcommand{\sd}{{\ul D}}
\newcommand{\se}{{\ul E}}
\renewcommand{\sf}{{\ul F}}

\newcommand{\si}{I}
\newcommand{\sj}{J}

\newcommand{\ci}{M}
\newcommand{\cj}{N}
\newcommand{\ck}{P}

\def\Polacek{Pol\'a\v{c}ek}

\newcommand{\zB}{{\mathbb B}}
\newcommand{\zH}{{\mathbb H}}

\newboolean{default}
\newcommand{\case}{}
\newcommand{\default}{}

\newenvironment{switch}[1]{%
    \setboolean{default}{true}
    \renewcommand{\case}[2]{\ifthenelse{\equal{#1}{##1}}{%
        \setboolean{default}{false}##2}{}}%
    \renewcommand{\default}[1]{\ifthenelse{\boolean{default}}{##1}{}}
}{}

\newcommand{\grad}[1]{\epsilon{(#1)}}

\renewcommand{\grad}[1]{%
    \begin{switch}{#1}
        \case{\Iso1}{r}
        \case{\Iso2}{s}
        \case{\Iso3}{t}
        \case{\Iso4}{u}
        \case{\Iso1'}{r'}
        \case{\Iso2'}{s'}
        \case{\Iso3'}{t'}
        \case{M}{m}
        \case{N}{n}
        \case{P}{p}
        \case{\uM}{m}
        \case{\uN}{n}
        \case{\uP}{p}
        \case{\cIso1}{m}
        \case{\cIso2}{n}
        \case{\cIso3}{p}
        \case{\cM}{m}
        \case{\cN}{n}
        \case{\cP}{p}
        \case{A}{a}
        \case{B}{b}
        \case{C}{c}
        \case{D}{d}
        \case{\hA}{a}
        \case{\hB}{b}
        \case{\hC}{c}
        \case{\hD}{d}
        \case{\sa}{a}
        \case{\sb}{b}
        \case{\sc}{c}
        \case{\sd}{d}
        \case{\cA}{a}
        \case{\cB}{b}
        \case{\cC}{c}
        \case{\cD}{D}
        \default{\epsilon{(#1)}}
    \end{switch}
}

\title{\boldmath Generalized Dualities and Supergroups}
\preprint{MI-HET-807}

\author[a]{Daniel Butter,}
\author[a,b]{Falk Hassler,}
\author[a,c]{Christopher N. Pope,}
\author[a]{and Haoyu Zhang}

\emailAdd{dbutter@gmail.com}
\emailAdd{falk.hassler@uwr.edu.pl}
\emailAdd{pope@physics.tamu.edu}
\emailAdd{zhanghaoyu@tamu.edu}

\affiliation[a]{George P. \& Cynthia Woods Mitchell Institute for Fundamental Physics and Astronomy,\\ Texas A\&M University, College Station, TX 77843, USA}
\affiliation[b]{University of Wroc\l{}aw, Faculty of Physics and Astronomy, Maksa Borna 9, 50-204 Wroc\l{}aw, Poland}
\affiliation[c]{DAMTP, Centre for Mathematical Sciences, Cambridge University, Wilberforce Road,\\ Cambridge CB3 OWA, UK}

\abstract{
Using a recently developed formulation of double field theory in superspace, the  graviton, $B$-field, gravitini, dilatini, and Ramond-Ramond bispinor are encoded in a single generalized supervielbein. Duality transformations are encoded as orthosymplectic transformations, extending the bosonic $\g{O}(D,D)$ duality group, and these act on all constituents of the supervielbein in an easily computable way. We first review conventional non-abelian T-duality in the Green-Schwarz superstring and describe the dual geometries in the language of double superspace. Since dualities are related to super-Killing vectors, this includes as special cases both abelian and non-abelian fermionic T-duality.

We then extend this approach to include Poisson-Lie T-duality and its generalizations, including the generalized coset construction recently discussed in
\href{https://arxiv.org/abs/1912.11036}{\texttt{[arXiv:1912.11036]}}.
As an application, we construct the supergeometries associated with the integrable $\lambda$ and $\eta$ deformations of the $\g{AdS}_5 \times \g{S}^5$ superstring. The deformation parameters $\lambda$ and $\eta$ are identified with the possible one-parameter embeddings of the supergravity frame within the doubled supergeometry. In this framework, the Ramond-Ramond bispinors are directly computable purely from the algebraic data of the supergroup.}

\begin{document}

\maketitle

\section{Introduction}

Abelian T-duality is an exact symmetry of perturbative string theory. Its initial formulation on an $\g{S}^1$ with associated isometries of metric and $B$-field can be straightforwardly extended to a $d$-dimensional torus, where the T-duality group expands to $\g{O}(d,d; \mathbb Z)$. Its modern description was given by Buscher \cite{Buscher:1987sk,Buscher:1987qj}, who couched it in the language of the effective worldsheet $\sigma$-models with commuting isometries; here one can derive the transformation rules of the metric and $B$-field by integrating out the worldsheet one-forms that gauge the isometries. Later work extended this approach to include the fermionic fields and the Ramond-Ramond sector from the target space perspective \cite{Bergshoeff:1995as,Hassan:1999bv,Hassan:1999mm} and from the worldsheet using both the Green-Schwarz superstring \cite{Cvetic:1999zs,Kulik:2000nr} and the pure spinor superstring \cite{Benichou:2008it}.

When these isometries no longer commute, it is no longer clear that the corresponding classical $\sigma$-model duality, known as non-abelian T-duality (NATD), is a full-fledged symmetry of string theory \cite{delaOssa:1992vci,Giveon:1993ai,Alvarez:1993qi,Alvarez:1994np}. A symptom of this is that the dual space typically lacks local isometries that would permit one to invert the duality and recover the original space --  the duality appears to be effectively one-way. Nevertheless, this procedure can still provide a means to systematically generate new supergravity solutions from existing ones.

Klimčík and Ševera showed that one can generalize the notion of duality, so that two or more $\sigma$-models related by NATD are indeed properly dual, in the sense that they can be derived from the same universal $\cE$-model \cite{Klimcik:1995dy, Klimcik:1996nq, Klimcik:2015gba}. In this framework, NATD is just the simplest example of Poisson-Lie T-duality (PLTD) \cite{Klimcik:1995ux, Klimcik:1995jn}, which can be further generalized to include a Wess-Zumino-Witten term \cite{Klimcik:2001vg} and a so-called dressing action \cite{Klimcik:1996np}, where one factors out local symmetries in very close analogy to the construction of $\sigma$-models on coset spaces. In this paper, we will be concerned with an even more general framework, known as a generalized coset \cite{Demulder:2019vvh,Butter:2022iza}. The relations between these various concepts can be summarized as follows:
\begin{equation*}
  \begin{array}{ccccccc}
    \text{ abelian } & \subset & \text{ non-abelian } & \subset & \text{ Poisson-Lie } & \subset & \text{ WZW-Poisson }\\
    & & & & 
    \raisebox{0.6em}{\rotatebox{270}{$\subset$}}
    & &
    \raisebox{0.6em}{\rotatebox{270}{$\subset$}}\\
    & & & & \text{ dressing coset } & \subset & \text{ \underline{generalized coset}\,. }
  \end{array}
\end{equation*}
One specifies a Lie group $\mathdsl{D}$ with a split signature Killing metric $\eta$ and a maximally isotropic subgroup $H$ of half the dimension. In the absence of a dressing action, the physical space lies on the coset $H \backslash \mathdsl{D}$, and in this context $\mathdsl{D}$ is usually called a double Lie group. For the case of a generalized coset, there is an additional ``dressing action'' by another isotropic subgroup $F$, and the physical space is the double coset $H \backslash \mathdsl{D} / F$. Different $\sigma$-models arise when there exist different choices for $H$, and these are related by this more general notion of duality.

In recent years, a modern perspective on these developments has been provided in the language of double field theory (DFT) \cite{Siegel:1993xq,Siegel:1993th,Hull:2009mi,Hull:2009zb,Hohm:2010jy,Hohm:2010pp}.\footnote{The early work of Siegel \cite{Siegel:1993xq,Siegel:1993th} is essentially equivalent to the frame formulation of DFT. This already included a superspace formulation \cite{Siegel:1993th}, although limited to the type I and heterotic cases.} This is a generalization of supergravity incorporating T-duality manifestly in the target space geometry and low energy action of string theory. The coordinates $x^m$ of spacetime are ``doubled'' to include dual coordinates $\tilde x_m$ corresponding to winding modes of the string. The metric and $B$-field are combined into a generalized metric $\cH$. We decompose the coordinates and generalized metric as
\begin{align}
x^\hm = (x^m, \tilde x_m)~, \qquad
\cH_{\hm \hn} =
\begin{pmatrix}
g_{mn} - b_{m k} g^{k l} b_{l n} & b_{m k} g^{k n} \\
- g^{m k} b_{k n} & g^{m n}
\end{pmatrix}~.
\end{align}
In order to ensure that at most half the coordinates are physical, a section condition is imposed
\begin{align}
\eta^{\hm \hn} \pa_\hm \otimes \pa_\hn = 0~, \qquad
\eta^{\hm \hn} =
\begin{pmatrix}
0 & \delta^m{}_n \\
\delta_m{}^n & 0
\end{pmatrix}~, 
\end{align}
where the derivatives act either on the same field or two different fields. The constant metric $\eta$ is the natural split-signature $\g{O}(D,D)$ invariant, and we have decomposed indices with respect to the $\g{GL}(D) \subset \g{O}(D,D)$ subgroup. Typically the section condition is solved by dispensing with all dependence on the winding coordinates $\tilde \pa^m = 0$. Different T-dual geometries are related by choosing different solutions of the section condition; these solutions are related by global $\g{O}(D,D)$ rotations, which act on the generalized metric $\cH$ in the same manner as the Buscher rules \cite{Buscher:1987sk,Buscher:1987qj}. In this sense, double field theory geometrizes T-duality.

This bears a striking similarity to PLTD and indeed the two are intimately related \cite{Hassler:2017yza}, with PLTD and its generalizations corresponding to double field theory on group manifolds \cite{Demulder:2018lmj} or coset spaces \cite{Demulder:2019vvh}. This has been an active area of research in recent years (see e.g. \cite{Hassler:2017yza,Demulder:2018lmj,Demulder:2019bha,Osten:2019ayq,Sakatani:2019jgu,Borsato:2021gma,Sakatani:2021skx} and references therein). As formulated in \cite{Hull:2009mi,Hull:2009zb,Hohm:2010jy,Hohm:2010pp}, DFT encompassed only the NS-NS sector (graviton, $B$-field, and dilaton). It has since been extended \cite{Hohm:2011zr, Hohm:2011dv,Jeon:2012kd, Jeon:2012hp} to include the NS-R fermions (gravitini and dilatini) and the R-R sector (the even/odd $p$-form complexes) of type II string theory, but this extension did not fully unify the fields. The three sectors, NS-NS, NS-R, and R-R are encoded separately in the low energy type II action, and this complicates the construction of the dual supergravity backgrounds since one cannot address all sectors simultaneously using the same methods. Typically, one uses geometric or $\sigma$-model methods to fix some of the fields and then exploits the structure of $\kappa$-symmetry and supersymmetry to uncover the rest. The Ramond-Ramond sector is particularly onerous, since unlike the other bosonic fields, it does not appear \emph{explicitly} as a separate term in the Green-Schwarz $\sigma$-model action.\footnote{After fixing a certain supersymmetric gauge (i.e. normal coordinates) in the Green-Schwarz action, one can perform an order-by-order $\theta$ expansion of the superfields $E_M{}^a$ and $B_{MN}$ and uncover the Ramond-Ramond fields,
see for example \cite{Wulff:2013kga}. In contrast, the pure spinor action contains the Ramond-Ramond sector explicitly witout any gauge fixing, which has been used to cleanly derive its transformation rules under NATD \cite{Benichou:2008it}.}

Naturally, one could consider broader U-duality covariant formulations, which are based on exceptional groups. These include double field theory as subcases: for example, the maximal case of $E_{11(11)}$, when decomposed under its $\g{O}(10,10)$ subgroup, possesses at leading order in the level decomposition the NS-NS and R-R sectors of DFT \cite{West:2003fc, West:2010ev, Rocen:2010bk}. However, the situation with exceptional groups and generalized dualities is not nearly as well developed as their DFT analogues. We will return to this point in the discussion section.

The goal of this paper is to address some of the topics discussed above from the perspective of a manifestly supersymmetric \emph{and} duality covariant formulation.\footnote{We are not the first to discuss Poisson-Lie T-duality on supermanifolds. To our knowledge, this was first addressed in the work of Eghbali and Rezaei-Aghdam (see \cite{Eghbali:2009cp,Eghbali:2011su} and subsequent works by these authors addressing specific examples). A small but important difference in our scheme is that we do not require an invertible supermetric, which is important for applications to the Green-Schwarz superstring.} Such a formulation has recently been constructed by one of us in the language of double superspace \cite{Butter:2022gbc}, building off earlier work on the subject \cite{Siegel:1993th,Hatsuda:2014qqa, Hatsuda:2014aza, Cederwall:2016ukd,Butter:2021dtu}. Double superspace can be understood in a nutshell as simultaneously geometrizing supersymmetry and T-duality. In conventional superspace, the graviton (vielbein) and gravitino are unified into a single supervielbein, which in a certain gauge reads
\begin{align}
E_M{}^A(x,\theta) =
\begin{pmatrix}
e_m{}^a(x) & \psi_m{}^\alpha(x) \\
0 & \delta_\mu{}^\alpha
\end{pmatrix} + \cO(\theta)~.
\end{align}
Diffeomorphisms and supersymmetry are unified into superdiffeomorphisms. In double superspace one is led to consider a generalized (double) supervielbein, which can be written in a certain gauge and duality frame as a product of three factors,
\begin{align}
\cV_\cM{}^\cA(x,\theta, \tilde x, \tilde \theta) =
		\begin{pmatrix}
		\delta_M{}^N & B_{MN} (-)^n \\
		0 & \delta^M{}_N
		\end{pmatrix} \times
		\begin{pmatrix}
		E_N{}^B & 0\\
		0 & E_B{}^N (-)^{b+bn}
		\end{pmatrix} \times
		\begin{pmatrix}
		\delta_B{}^A & 0\\
		S^{B A} & \delta^B{}_A
		\end{pmatrix}
\end{align}
The field $E_M{}^A$ is the supervielbein, $B_{MN}$ is the super two-form (which appears in the Green-Schwarz action), and $S^{BA}$ includes ``matter'' fields, the dilatini and Ramond-Ramond bispinor. The duality group $\g{O}(D,D)$, which governs the geometric structure of double field theory, is replaced by its natural supergroup analogue, the orthosymplectic group $\g{OSp}(D,D|2s)$ with $D$ bosonic coordinates, $s$ fermionic coordinates, and their duals.\footnote{The role of the orthosymplectic group has been explored for dualities of general sigma models with both bosonic and fermionic degrees of freedom in \cite{Fre:2009ki}. It has been discussed in the sigma model context e.g. in \cite{Osten:2016dvf} and in the double field theory context in \cite{Siegel:1993th, Hatsuda:2014qqa, Cederwall:2016ukd}.} Diffeomorphisms, $B$-field gauge transformations, and supersymmetry are all encoded in a single generalized superdiffeomorphism. Because all of the fields of supersymmetric double field theory are described in a single geometric object, one can apply the same techniques to derive how all of them transform under dualities, including abelian, non-abelian, and their generalized cousins.

A crucial point about conventional superspace is that it is \emph{not} simply described by a super-Riemannian geometry with an unconstrained supermetric. Rather, one must employ the supervielbein and impose constraints on its torsion tensor in order to recover the physical field content of supergravity. These constraints involve $\theta$-derivatives, but typically constrain the $x$-dependence as well, placing the geometry on-shell. In the Green-Schwarz superstring, these constraints arise from requiring $\kappa$-symmetry. Analogous statements hold for double superspace --  we need to impose constraints on the generalized flux tensor $\cF_{\cA \cB \cC}$ in order for a supergravity interpretation to be possible, and these will coincide with the $\kappa$-symmetry constraints.

We begin in section \ref{S:SDFT} with a discussion of superspace double field theory, highlighting how the duality group $\g{OSp}(D,D|2s)$ acts on the various constituents of $\cV_\cM{}^\cA$. These transformations provide the generic scaffolding in which all T-dualities must act. In section \ref{S:SNATD}, as the simplest non-trivial example of such a transformation, we review the case of super non-abelian T-duality (NATD) in the Green-Schwarz superstring, where a supergroup $G$ of isometries is dualized \cite{Borsato:2018idb} (see \cite{Cvetic:1999zs, Kulik:2000nr,Bandos:2003bz} for earlier work on abelian T-duality of a single bosonic isometry, \cite{Borsato:2017qsx} for the non-abelian T-dual of supercoset $\sigma$-models, and \cite{Abbott:2015mla} for a discussion of the self-duality of the Green-Schwarz $\sigma$-model on $\g{AdS}_d \times \g{S}^d$ backgrounds). By comparing the dual Green-Schwarz models, one can deduce the form of the orthosymplectic transformation, which immediately yields the transformation rules of the supergravity fields, including the transformations of the Ramond-Ramond fields \cite{Sfetsos:2010uq,Lozano:2011kb}.

As a side benefit of this analysis, we are able to specialize to a fermionic isometry and recover results for fermionic T-dualities, both in the abelian \cite{Berkovits:2008ic, Beisert:2008iq, Bakhmatov:2009be,Grassi:2011zf,Sfetsos:2010xa,OColgain:2012si} and non-abelian \cite{Astrakhantsev:2021rhj,Astrakhantsev:2022mfs} cases. The case of non-abelian fermionic T-duality has been of particular interest recently, and we highlight the origin of the conditions given in \cite{Astrakhantsev:2021rhj,Astrakhantsev:2022mfs} for the Killing spinor from the $\sigma$-model.\footnote{Fermionic T-duality has also been discussed in the context of a doubled $\sigma$-model with T-dual fermionic coordinates \cite{Nikolic:2016nxt}. We will not address doubled $\sigma$-models here, but it is likely super DFT can be formulated there, in analogy to the work of
\cite{Park:2016sbw,Bandos:2015cha,Sakamoto:2018krs}.}

Non-abelian T-duality of the GS superstring provides a concrete example, exhibiting a number of important features that continue to hold for more general cases. In section \ref{S:GPS}, we introduce, following \cite{Demulder:2018lmj,Hassler:2019wvn,Demulder:2019vvh}, the notion of a generalized parallelizable superspace, which is the natural analogue of a group manifold in the doubled setting, requiring only a double Lie group $\mathdsl{D}$ and its maximally isotropic subgroup $H$. In section \ref{S:GCoset}, we extend this framework to generalized supercosets, where an additional isotropic subgroup $F$ is factored out, in direct analogy to the bosonic case \cite{Demulder:2019vvh}.  In both of these discussions, we address two particular examples, $\mathdsl{D} = G \times G$ and $\mathdsl{D} = G^{\mathbb C}$, where $G$ is a real super Lie group admitting an invertible Killing form. Both examples admit maximally isotropic subgroups $H$, the diagonal subgroup $G_{\rm diag}$ for $G\times G$ and the real subgroup $G$ for $G^{\mathbb C}$. The two groups $G \times G$ and $G^{\mathbb C}$ can be analytically continued into each other, and the same holds true for their respective generalized geometries. For $G^{\mathbb C}$, another isotropic subgroup $H$ is sometimes possible: it requires an $R$-matrix satisfying the modified classical Yang-Baxter equation. The two solutions for $G^{\mathbb C}$ lead to backgrounds related by Poisson-Lie T-duality.

The discussion of generalized parallelizable superspaces and generalized supercosets in sections \ref{S:GPS} and \ref{S:GCoset} is not really any different from their bosonic analogues: in effect, we simply insert a grading. In order to apply these results to supergravity, we must further impose additional $\kappa$-symmetry constraints on the generalized flux tensors. We review these in section \ref{S:SUGRA} and discuss how they can be imposed in two specific cases: these are the so-called $\lambda$ and $\eta$ deformations of the $\g{AdS}_5 \times \g{S}^5$ superstring. The $\lambda$ deformation \cite{Hollowood:2014qma} (building off earlier work \cite{Sfetsos:2013wia,Hollowood:2014rla}) arises from a deformation of the non-abelian T-dual of the $\g{AdS}_5 \times \g{S}^5$ superstring. The $\eta$ deformation \cite{Delduc:2013qra, Delduc:2014kha} is a type of Yang-Baxter $\sigma$-model \cite{Klimcik:2002zj, Klimcik:2008eq} (see also \cite{Delduc:2013fga}). Remarkably, these two different deformations preserve the integrable structure of the superstring, and this property has driven interest in them. From our perspective, these models are interesting because they can be very simply understood in the context of Poisson-Lie T-duality for the double Lie groups $G \times G$ and $G^{\mathbb C}$, respectively, where $G$ is the superisometry group $\g{PSU}(2,2|4)$ of the $\g{AdS}_5 \times \g{S}^5$ superstring.\footnote{This particular duality between $\sigma$-models on $G\times G$ and $G^{\mathbb C}$ has been known for some time \cite{Ivanov:1987yv}. We thank Evgeny Ivanov for pointing out this reference.} This interpretation was given in the language of $\cE$-models for the bosonic sector in \cite{Klimcik:2015gba}.
Our main task in section \ref{S:SUGRA} is to extend this to the fully supersymmetric case.

In addressing the $\lambda$ and $\eta$ models, we proceed ahistorically, and in fact, anti-chronologically. Beginning with the underlying double Lie structure of $G \times G$ and $G^{\mathbb C}$, we will seek to build a generalized supervielbein $\cV_\cM{}^\cA$ whose flux tensor $\cF_{\cA \cB \cC}$ obeys the $\kappa$-symmetry constraints \eqref{E:FluxConstraint1} and \eqref{E:FluxConstraint2}.
For each case, there turns out to be a single one-parameter family, and this leads inexorably to the $\lambda$ and $\eta$ models upon identifying the underlying constituents of the Green-Schwarz action. All supergravity fields, including the Ramond-Ramond field strengths, are read directly off from the supervielbein and match the results derived by analyzing the respective Green-Schwarz $\sigma$-models \cite{Borsato:2016ose}. 

In line with our ahistorical approach, we will not directly address issues of integrability or the connection between generalized duality and integrability. For a discussion of integrability, the reader is referred to the recent work \cite{Bielli:2021hzg}, which explored some of these very issues for supersymmetric $\sigma$-models; specifically, it was shown that the Lax connection is preserved (a sufficient condition for integrability) after performing non-abelian T-duality in superspace analogues of the the principal chiral, symmetric space, and semi-symmetric space $\sigma$-models. On the connection between $\cE$-models and integrability, the reader is referred to
\cite{Lacroix:2020flf,Lacroix:2021iit}.

We include several appendices. Our conventions for supergroups, including the orthosymplectic group, can be found in appendix \ref{A:Supergroups}. We sketch some relevant results for type II supergravity in superspace in appendix \ref{A:DemoTypeII}. A concise discussion of gauged superspace $\sigma$-models (whose results we employ in section \ref{S:SNATD}) is given in appendix \ref{A:GaugedSigmaModels}. Finally in appendix \ref{A:FluxTensors} we give the generalized flux tensors for the $\eta$ and $\lambda$ models that are compatible with $\kappa$-symmetry.

\section{Supersymmetric double field theory and the framework of T-duality}
\label{S:SDFT}
We will be employing the supersymmetric formulation of type II double field theory
in superspace recently discussed in \cite{Butter:2022gbc} (see also \cite{Hatsuda:2014qqa, Hatsuda:2014aza} and \cite{Cederwall:2016ukd} for related earlier discussions).
In this section, we will review some basic elements of this approach and explain how T-duality is manifested on the generalized supervielbein. As a first step, we will review some key features of
bosonic double field theory, before showing how these generalize to the supersymmetric setting.

\subsection{Bosonic double field theory and \texorpdfstring{$\g{O}(D,D)$}{O(D,D)} T-duality}
Double field theory \cite{Siegel:1993th, Hull:2009mi, Hohm:2010pp} is formulated on a space with local coordinates $x^\hm$ where fields are subject to a modified notion of generalized diffeomorphism governed by a Lie derivative $\dLie$ which preserves an $\g{O}(D,D)$ structure. For vector fields $V^\hm$, 
\begin{align}\label{E:genLieBos}
\dLie_\xi V^\hm = \xi^\hn \pa_\hn V^\hm - V^\hn (\pa_\hn \xi^\hm - \pa^\hm \xi_\hn)~.
\end{align}
where indices are raised and lowered with the constant $\g{O}(D,D)$ metric $\eta_{\hm \hn}$. The space comes equipped with a generalized metric $\cH_{\hm \hn}$, which is an element of $\g{O}(D,D)$ so that its inverse is $(\cH^{-1})^{\hm \hn} = \cH^{\hm \hn}$. Closure of generalized diffeomorphisms is guaranteed if we universally impose a section condition on all fields and parameters,
$\eta^{\hm \hn} \pa_\hm \otimes \pa_\hn = 0$,
where the derivatives may act either on the same or different fields. The metric and coordinates can be decomposed in terms of the $\g{GL}(D)\subset\g{O}(D,D)$ subgroup as
\begin{align}
\eta_{\hm \hn} =
\begin{pmatrix}
0 & \delta_m{}^n \\
\delta^m{}_n & 0
\end{pmatrix}~, 
\qquad
x^\hm = (x^m, \tilde x_m)~, \qquad
\pa_\hm = (\pa_m, \tilde \pa^m)~.
\end{align}
The section condition can then be solved by choosing $\tilde \pa^m = 0$ universally. Then, the generalized metric $\cH$ is described in terms of a metric $g_{mn}$ and a Kalb-Ramond two-form $b_{mn}$ as
\begin{align}
\cH_{\hm \hn} =
\begin{pmatrix}
g_{mn} - b_{m k} g^{k l} b_{l n} & b_{m k} g^{k n} \\
- g^{m k} b_{k n} & g^{m n}
\end{pmatrix}~,
\end{align}
and the generalized Lie derivative decomposes into the conventional $\g{GL}(D)$ Lie derivative and $B$-field transformations.

The description in terms of a generalized metric turns out to not be particularly useful when passing to superspace. Just as supergravity requires that we exchange a metric $g_{mn}$ for a vielbein $e_m{}^a$, supersymmetric double field theory requires we replace the generalized metric $\cH_{\hm \hn}$ with a generalized vielbein $V_\hm{}^\ha$. These are related by
\begin{align}
\cH_{\hm \hn} = V_\hm{}^\ha V_\hn{}^\hb \cH_{\ha \hb}
\end{align}
where $\cH_{\ha \hb}$ is a constant matrix invariant only under the double Lorentz subgroup $\g{O}(D-1,1) \times \g{O}(1,D-1)$ of $\g{O}(D,D)$. These objects are naturally written in the chiral basis of $\g{O}(D,D)$, where a flat vector $V^\ha = (V^\ra, V^\rba)$ is decomposed into a left-handed vector $V^\ra$ of $\g{O}(D-1,1)$ and a right-handed vector $V^\rba$ of $\g{O}(1,D-1)$. In this chiral basis,
\begin{align}
\eta_{\ha \hb} =
\begin{pmatrix}
\eta_{\ra \rb} & 0 \\
0 & \eta_{\ol{\ra\rb}}
\end{pmatrix}~, \qquad
\cH_{\ha \hb} =
\begin{pmatrix}
\eta_{\ra \rb} & 0 \\
0 & -\eta_{\ol{\ra\rb}}
\end{pmatrix}~,  \qquad
\eta_{\ol{\ra\rb}} = - \eta_{\ra\rb}~.
\end{align}
The generalized vielbein can be decomposed as \cite{Siegel:1993xq, Jeon:2011cn}
\begin{subequations}\label{E:chiralBosonicV}
\begin{alignat}{2}
V_\ra{}^m &= \frac{1}{\sqrt 2} e_\ra{}^m~, &\qquad
V_\ra{}_m &= \frac{1}{\sqrt 2} (e_m{}_\ra - e_\ra{}^n b_{nm})
    = \frac{1}{\sqrt 2} e_\ra{}^n (g_{nm} - b_{nm})~, \\
V_\rba{}^m &= \frac{1}{\sqrt 2} \bar e_\rba{}^m~, &\qquad
V_\rba{}_m &= \frac{1}{\sqrt 2} (\bar e_m{}_\rba - \bar e_\rba{}^n b_{nm})
    = -\frac{1}{\sqrt 2} \bar e_\rba{}^n (g_{nm} + b_{nm})~,
\end{alignat}
\end{subequations}
which is the generic form if one supposes $V_\ra{}^m$ and $V_\rba{}^m$ to both be invertible matrices. This can be expressed as a product of two $\g{O}(D,D)$ factors:
\begin{align}\label{E:bosonicVDecomp}
V_\ha{}^\hm &=
\frac{1}{\sqrt 2}
\begin{pmatrix}
e_\ra{}^n & \,\eta_{\ra \rb} e_{n}{}^\rb \\
\bar e_\rba{}^n & \,\eta_{\ol{\ra\rb}} \bar e_{n}{}^{\rbb}
\end{pmatrix} \times
\begin{pmatrix}
\delta_n{}^m & -b_{nm} \\
0 & \delta^n{}_m
\end{pmatrix}~.
\end{align}
The two vielbeins $e_m{}^\ra$ and $e_m{}^\rba$ describe the same metric,
$g_{mn} = e_m{}^\ra e_n{}^\rb \eta_{\ra\rb} = - \bar e_m{}^\rba \bar e_n{}^\rbb \eta_{\ol{\ra\rb}}$
implying that they are connected by a Lorentz transformation
\begin{align}\label{E:BosonicLambda}
\Lambda_\ra{}^\rbb = e_\ra{}^m \bar e_m{}^\rbb~.
\end{align}
The double Lorentz symmetry can be fixed to a single Lorentz group by adopting the gauge $\Lambda=1$. However, in supergravity this is more subtle because chiral fermions are present, breaking each Lorentz group to its connected (proper  orthochronous) component. This means that $\Lambda$ falls into one of four classes, depending on whether it preserves or reverses the temporal and spatial orientations: this distinguishes the type IIA/IIB/IIA$^*$/IIB$^*$ duality frames \cite{Jeon:2012kd, Jeon:2012hp,Butter:2022sfh}.

Double field theory conveniently packages the $\g{O}(D,D)$ structure of T-duality transformations. To see how, we define $\EE_{nm} := g_{nm} - b_{nm}$ and $\bar \EE_{nm} := g_{nm} + b_{nm} = (\EE^T)_{nm} = \EE_{mn}$.
An $\g{O}(D,D)$ transformation $U_\hn{}^\hm$ acting on the right of $V_\ha{}^\hm$ can be written
\begin{align}
V'_\ha{}^{\hm} = V_\ha{}^\hn U_\hn{}^\hm~, \qquad
U_{\hm}{}^\hn =
\begin{pmatrix}
U_m{}^n & U_{mn} \\
U^{mn} & U^m{}_n
\end{pmatrix}~.
\end{align}
Defining
\begin{alignat}{2}
X_m{}^n &:= U_m{}^n + \EE_{m p} U^{p n} ~, &\qquad
\bar X_m{}^n &:= U_m{}^n - \bar \EE_{m p} U^{p n} ~, \\
Y_{mn} &:= U_{mn} + \EE_{m p} U^p{}_n & \quad
\bar Y_{mn} &:= U_{mn} - \bar \EE_{m p} U^p{}_n~,
\end{alignat}
one can show that
\begin{subequations}
\begin{alignat}{2}
e'_\ra{}^m &= e_\ra{}^n X_n{}^m ~, &\qquad
\bar e'_\rba{}^{m} &= \bar e_\rba{}^n \bar X_n{}^m, \\
\EE'_{mn} &= (X^{-1})_m{}^p Y_{p n}~, &\qquad
\bar \EE'_{mn} &= (\bar X^{-1})_m{}^p \bar Y_{p n}~.
\end{alignat}
\end{subequations}
This recovers the Buscher rules for the metric and $B$-field and has
the form of a fractional linear transformation on $\EE_{nm}$.
The fact that $\bar \EE'_{mn} = \EE'_{nm}$ follows from the $\g{O}(D,D)$ structure. Also encoded above is how the Lorentz transformation $\Lambda'_\ra{}^\rbb$ that defines the type II duality frame is related to the original $\Lambda_\ra{}^\rbb$. 
This can be written alternatively as a left or right Lorentz transformation $\bm\Lambda(U)$,
\begin{align}
\Lambda'_\ra{}^\rbb 
    &= \underbrace{e_\ra{}^m (X \bar X^{-1})_m{}^n e_n{}^\rb}_{{\bm\Lambda}(U)_\ra{}^\rb}
    \times \Lambda_\rb{}^\rbb
    = \Lambda_\ra{}^\rba \times 
    \underbrace{\bar e_\rba{}^m (X \bar X^{-1})_m{}^n \bar e_n{}^\rbb}_{{\bm\Lambda}(U)_\rba{}^\rbb}~.
\end{align}
Again, the fact that this is a Lorentz transformation follows from the $\g{O}(D,D)$ structure.

In addition to the generalized vielbein, double field theory also involves a generalized dilaton $e^{-2d}$. This is a density under $\g{O}(D,D)$ transformations, transforming as
\begin{align}
\dLie_\xi e^{-2d} = \xi^\hm \pa_\hm e^{-2d} + \pa_\hm \xi^\hm e^{-2d}
    = \pa_\hm (\xi^\hm e^{-2d})~.
\end{align}
Upon solving the section condition, the physical dilaton $\varphi$ is
identified by removing a density factor from the generalized dilaton,
$e^{-2d}  = e^{-2\varphi} \times \det e_m{}^\ra$. A generic transformation of the generalized dilaton is simply a scalar factor
\begin{align}
e^{-2d'} = e^{-2d} \,U_\Delta~,
\end{align}
which is \emph{a priori} independent of $U_\hm{}^\hn$. Together $U_\hm{}^\hn$ and $U_\Delta$ encode an $\g{O}(D,D) \times \mathbb R_+$ transformation. It follows that the physical
dilaton transforms as
\begin{align}
e^{-2\varphi'} = e^{-2 \varphi} \times \det X_m{}^n \times U_\Delta~.
\end{align}
Note that $\det \bar X_m{}^n = \det X_m{}^n$ since $X$ and $\bar X$ are related  by a Lorentz transformation.

\subsection{Supersymmetric type II double field theory}
\label{S:SDFT2Review}
We turn now to supersymmetric type II double field theory \cite{Jeon:2012hp}.
At the component level, supersymmetric double field theory consists of the following fields:
\begin{itemize}
\item the generalized vielbein $V_\hm{}^\ha$ and the generalized dilaton $e^{-2d}$;
\item the gravitini $\Psi_\ra{}^\bbeta$ and $\Psi_\rba{}^{\beta}$, which are vectors and Weyl spinors under alternating Lorentz groups, and the dilatini $\rho_\alpha$ and $\rho_\balpha$, which are Weyl spinors of opposite chirality to the gravitini;
\item and the Ramond/Ramond field strengths, which can be described equivalently as an $\g{O}(D,D)$ spinor $\ket{F}$ \cite{Hohm:2011zr, Hohm:2011dv} or a Weyl bispinor $F^{\alpha \bbeta}$ of $\g{O}(D-1,1) \times \g{O}(1,D-1)$ \cite{Jeon:2012kd, Jeon:2012hp}.
\end{itemize}

In order to make contact with conventional superspace (and the Green-Schwarz superstring), a parametrization is needed that naturally leads to a supervielbein $E_M{}^A$ and a Kalb-Ramond super-two-form $B_{MN}$ where $z^M = (x^m, \theta^\mu)$ are the $D$ bosonic and $s$ fermionic coordinates of superspace. This can simply be done by mimicking the structure of bosonic double field theory, but replacing $\g{O}(D,D)$ with its natural graded extension $\g{OSp}(D,D|2s)$, the orthosymplectic supergroup involving $2D$ bosonic and $2s$ fermionic directions \cite{Cederwall:2016ukd}. For type II superspace, we will need $D=10$ and $s=32$. For the details about this supergroup, we refer to appendix \ref{A:Supergroups.OSp}.

One begins by formulating supersymmetric double field theory on a superspace with local coordinates $z^\cM$, with $\cM$ a curved vector index of $\g{OSp}(D,D|2s)$. Generalized diffeomorphisms act on a vector $V^\cM$ as
\begin{align}
\dLie_\xi V^\cM = \xi^\cN \pa_\cN V^\cM - V^\cN 
    \Big(\pa_\cN \xi^\cM - \pa^\cM \xi_\cN (-)^{\grad{\cM} \grad{\cN}}\Big)~.
\end{align}

Here and in the rest of the paper, we use the notation that $(-)^{\grad{\cM} \grad{\cN}}$ is $-1$ if both $\cM$ and $\cN$ are fermionic and $+1$ otherwise. Similarly, we use $(-)^{\grad{\cM}}$ for $-1$ if $\cM$ is fermionic and $+1$ otherwise. This notation follows the classic text \cite{Wess:1992cp} and is a shorthand for the mathematically cleaner but bulkier notations
$(-1)^{\epsilon(\cM) \epsilon(\cN)}$ and $(-1)^{\epsilon(\cM)}$ respectively,
where $\epsilon(\cM)$ is $0$ for bosonic $\cM$ and $1$ for fermionic $\cM$,
see for example \cite{Cederwall:2016ukd}.

Indices are raised and lowered with the graded symmetric orthosymplectic invariant $\eta_{\cM \cN}$ subject to NW-SE rules,
$V_\cM = V^\cN \eta_{\cN \cM}$,
$V^\cM = \eta^{\cM \cN} V_{\cN}$,
and $\eta^{\cM \cP} \eta_{\cP \cN} = \delta_\cN{}^\cM (-)^{\grad{\cM} \grad{\cN}}$.
Closure of the gauge algebra is guaranteed by imposing the section condition
$\eta^{\cM \cN} \pa_\cN \otimes \pa_\cM = 0$, exactly as in bosonic double field theory.

To recover conventional superspace, we decompose all objects carrying curved indices $\cM$ under the $\g{GL}(D|s) \subset \g{OSp}(D,D|2s)$ subgroup. The $\g{OSp}(D,D|2s)$ metric in this basis is
\begin{align}\label{E:eta.GLbasis}
\eta^{\cM \cN} =
\begin{pmatrix}
0 & \delta^M{}_N  \\
\delta_M{}^N (-)^{\grad{M} \grad{N} } & 0
\end{pmatrix}~, \qquad
\eta_{\cM \cN} =
\begin{pmatrix}
0 & \delta_M{}^N  \\
\delta^M{}_N (-)^{\grad{M} \grad{N}} & 0
\end{pmatrix}~.
\end{align}
The coordinates and their derivatives decompose as
\begin{gather}
\pa_\cM = \Big(\pa_M, \tilde \pa^M\Big)~, \qquad
z_\cM = (\tilde z_M, z^M)~, \qquad
\pa_\cM z^\cN = \delta_\cM{}^\cN \quad \implies \eol
\quad \pa_M z^N = \delta_M{}^N~, \qquad
\tilde\pa^M \tilde z_N = \delta_N{}^M (-)^{\grad{N} \grad{M}}
\end{gather}
where $z^M$ is the physical coordinate and $\tilde z_M$ is the winding coordinate. We normally solve the section condition by discarding any dependence on the winding coordinate.

As in bosonic double field theory, we introduce a generalized supervielbein $\cV_\cM{}^\cA$ with which to flatten generalized vectors. We choose it to be an $\g{OSp}(D,D|2s)$ element, so that it is related to its inverse $(\cV^{-1})_\cA{}^\cM \equiv \cV_\cA{}^\cM$ by
$\cV_\cA{}^\cM = \eta^{\cM \cN} \cV_\cN{}^\cB \eta_{\cB \cA} \,(-)^{\grad{\cA} \grad{\cM}}$. For type II superspace, the flat index $\cA$ decomposes in the chiral basis into two vector indices, one for each factor of the double Lorentz group, and four Weyl spinor indices, one of each chirality for each factor. We denote this for a vector $V_\cA$ as
\begin{equation}
\begin{array}{rccccccccc}
    V_\cA = & \Big( & V_\ra & V_\alpha & V^\alpha & \Big\vert & V_\rba & V_\balpha & V^\balpha &\Big)~. \\
    \text{relative dimension} & & 0 & -\tfrac12 & \tfrac12 & & 0 & -\tfrac12 & \tfrac12 &
\end{array}
\end{equation}
We have included above the \emph{relative dimension} of these various components. These dimensions can be understood as arising from the $\mathbb R_+$ factor in the decomposition
$\g{OSp}(10,10|64) \rightarrow \g{O}(9,1)_L \times \g{O}(1,9)_R \times \mathbb R_+$.
This dimension is one reason why we should not combine the two 16-component Weyl spinors $V_\alpha$ and $V^\alpha$ into a single 32-component Dirac spinor.

We have normalized the relative dimension so that it leads to the correct notion of engineering dimension for the flat derivatives $D_\cA = \cV_\cA{}^\cM \pa_\cM$,
\begin{equation}
\begin{array}{rccccccccc}
    D_\cA = & \Big( & D_\ra & D_\alpha & D^\alpha & \Big\vert & D_\rba & D_\balpha & D^\balpha &\Big)~. \\
    \text{engineering dimension} & & 1 & \tfrac12 & \tfrac32 & & 1 & \tfrac12 & \tfrac32 &
\end{array}
\end{equation}
At the component level in double field theory, $D_\ra$ and $D_\rba$ correspond to the two flat derivatives (built respectively with $e_\ra{}^m$ and $\bar e_\ra{}^m)$, while $D_\alpha$ and $D_\balpha$ correspond to the two supersymmetries. (The higher dimension $D^\alpha$ and $D^\balpha$ are discarded upon passing to component double field theory where one solves the section condition on the fermionic coordinates.)

Flat generalized vector indices are raised and lowered with
\begin{align}\label{E:eta.cAcB}
\eta_{\cA \cB} = 
\left(\begin{array}{ccc|ccc}
\eta_{\ra\rb} & 0 & 0 
    & 0 & 0 & 0\\
0 & 0 & \delta_\alpha{}^\beta 
    & 0 & 0 & 0\\
0 & -\delta^\alpha{}_\beta & 0
    & 0 & 0 & 0 \\ \hline
0 & 0 & 0
    & \eta_{\ol{\ra \rb}} & 0 & 0\\
0 & 0 & 0
    & 0 & 0 & \delta_\balpha{}^\bbeta \\
0 & 0 & 0
    & 0 & -\delta^\balpha{}_\bbeta & 0
\end{array}\right)~, \quad
%
\eta^{\cA \cB} = 
\left(\begin{array}{ccc|ccc}
\eta^{\ra\rb} & 0 & 0 
    & 0 & 0 & 0\\
0 & 0 & \delta^\alpha{}_\beta 
    & 0 & 0 & 0\\
0 & -\delta_\alpha{}^\beta & 0
    & 0 & 0 & 0 \\ \hline
\phantom{{}^{A^{A^A}}\!\!\!\!\!\!\!\!\!\!} 0 & 0 & 0
    & \eta^{\ol{\ra \rb}} & 0 & 0\\
0 & 0 & 0
    & 0 & 0 & \delta^\balpha{}_\bbeta \\
0 & 0 & 0
    & 0 & -\delta_\balpha{}^\bbeta & 0
\end{array}\right)~.
\end{align}
These matrices (and their chiral subblocks) are invariant under the double Lorentz group.

As in the bosonic case, there are unphysical ingredients present in the supervielbein, which are associated with local symmetry transformations
\begin{align}
\delta \cV_\cA{}^\cM = \lambda_\cA{}^\cB \cV_\cB{}^\cM~, \qquad 
\lambda_{\cA \cB} = -\lambda_{\cB \cA} (-)^{\grad{\cA} \grad{\cB}}~.
\end{align}
In the bosonic case, the local symmetry group is the double Lorentz group $\g{O}(D-1,1)_L \times \g{O}(1,D-1)_R$ with commuting left and right factors. In the supersymmetric case, this group is larger, although it still factors into two commuting chiral pieces. We denote it $\g{H}_L \times \g{H}_R$. The generators $\lambda_{\cA \cB}$ of $\g{H}_L$ are constrained as in Table \ref{T:lambdaConstraints}.
Unlike the bosonic case, there is no simple prescription whereby some invariant $\cH_{\cA \cB}$ determines $\lambda$; instead, one needs to take into account the constraint structure on the supersymmetric worldsheet \cite{Siegel:1993th}. For further details of this symmetry group, we refer to \cite{Butter:2021dtu,Butter:2022gbc}.

\begin{table}[t]
\centering
\renewcommand{\arraystretch}{1.5}
\begin{tabular}{c|c|c}
\toprule
dimension &  $\lambda_{\cB \cA}$ & constraint\\ 
\hline
$+1$ & $\lambda^{\beta \alpha}$ & $-$ \\
$+\tfrac{1}{2}$
    & $\lambda_\rb{}^{\alpha}$ 
    & $(\gamma^\rb)_{\beta \alpha} \lambda_\rb{}^\alpha = 0$ \\
$0$ 
    & $\lambda_{\rb \ra}$, $\lambda_{\beta}{}^{\alpha}$ 
    & $\lambda_{\beta}{}^{\alpha} = \tfrac{1}{4} \lambda_{\rb \ra} (\gamma^{\rb\ra})_\beta{}^\alpha$ \\
$-\tfrac{1}{2}$ & $\lambda_{\rb \alpha}$ & vanishing\\
$-1$ & $\lambda_{\beta \alpha}$ & vanishing \\
\bottomrule
\end{tabular}
\captionsetup{width=0.6\textwidth}
\caption{Constraints on $\g{H}_L$ parameters. $\g{H}_R$ parameters are analogous.}
\label{T:lambdaConstraints}
\end{table}

There are competing ways of parametrizing a generic supervielbein, depending on whether one wishes to make contact with component double field theory or with type II superspace. In this paper, we will only be concerned with the latter. Then as shown in \cite{Butter:2022gbc}, a generic supervielbein can be decomposed as a product of three simple factors:
\begin{align}\label{E:V.BELambdaS}
\cV_\cM{}^\cA
    = (\cV_B)_\cM{}^\cN \times (\cV_{E\Lambda})_\cN{}^\cB \times (\cV_S)_\cB{}^\cA
\end{align}
The first is built out of the Kalb-Ramond super two-form,
\begin{align}
(\cV_B)_\cM{}^\cN =
\begin{pmatrix}
\delta_M{}^N & B_{MN} (-)^n \\
0 & \delta^N{}_M
\end{pmatrix}~, 
\end{align}
just as in the bosonic case.
The second factor $\cV_{E\Lambda}$ is written, in a chiral decomposition of the $\cA$ index, as
\begin{align}
(\cV_{E\Lambda})_\cM{}^\cA =
\renewcommand{\arraystretch}{1.5}
\left(\begin{array}{ccc|ccc}
\frac{1}{\sqrt 2} E_M{}^\ra & \quad E_M{}^\alpha & 0 
    & \frac{1}{\sqrt 2} E_M{}^\rba & \quad E_M{}^\balpha & 0 \\ 
\frac{1}{\sqrt 2} E^\ra{}^M & \quad 0 & - E_\alpha{}^M (-)^m 
    & \frac{1}{\sqrt 2} E^\rba{}^M & \quad 0 & - E_\balpha{}^M (-)^m
\end{array}\right)~.
\end{align}
The two superfields $E_M{}^\ra$ and $E_M{}^\rba$ (along with their inverses) are related by a Lorentz transformation,
\begin{align}\label{E:LambdaRels}
E_M{}^\rba = E_M{}^\rb \Lambda_\rb{}^\rba~,\qquad
E_\rba{}^M = \Lambda_\rba{}^\rb E_\rb{}^M~.
\end{align}
We may think of $\cV_{E \Lambda}$ as being composed of a square invertible matrix $E_M{}^A = (E_M{}^\ra, E_M{}^\alpha, E_M{}^\balpha)$ and an additional Lorentz transformation $\Lambda$ with which we can define $E_\rba{}^M$ and $E_M{}^\rba$ by the relations \eqref{E:LambdaRels}. The $\cV_S$ factor is given, also in a chiral decomposition, as
\begin{align}
(\cV_S)_\cA{}^\cB &=
\renewcommand{\arraystretch}{1.5}
\left(\begin{array}{ccc|ccc}
\delta_\ra{}^\rb & \sqrt{2} S_\ra{}^\beta & 0 
    & 0 & 0 & 0 \\
0 & \delta_\alpha{}^\beta & 0
    & 0 & 0 & 0 \\
-\sqrt{2} S^{\rb \alpha} & \quad S^{\alpha\beta} - S^{\rc\alpha} S_\rc{}^\beta & \quad  \delta^\alpha{}_\beta
    & 0 & S^{\alpha \bbeta} & 0 \\ \hline
0 & 0 & 0
    &\delta_\rba{}^\rbb & \sqrt{2} S_\rba{}^\bbeta & 0  \\
0 & 0 & 0
    & 0 & \delta_\balpha{}^\bbeta & 0 \\
0 & S^{\balpha \beta}  & 0 
    & -\sqrt{2} S^{\ol{\rb \alpha}} & \quad S^{\ol{\alpha\beta}} - S^{\rbc\balpha} S_\rbc{}^\bbeta & \quad \delta^\balpha{}_\bbeta
\end{array} \right).
\end{align}
It consists of fermionic superfields $S_\ra{}^\beta$ and $S_\rba{}^{\bbeta}$ as well as the symmetric bosonic superfields $S^{\alpha\beta}$, $S^{\ol{\alpha\beta}}$, and $S^{\alpha \bbeta}$. All these constituents transform as their indices imply under double Lorentz transformations, while only $\cV_S$ transforms under the higher dimension $\g{H}_L \times \g{H}_R$ transformations:
\begin{subequations}
\begin{alignat}{2}
\delta S_\ra{}^\alpha &= -\frac{1}{\sqrt2} \lambda_\ra{}^\alpha~, 
&\qquad
\delta S_\rba{}^\balpha &= -\frac{1}{\sqrt2} \lambda_\rba{}^\balpha~,
\\
\delta S^{\alpha \beta} &= -\lambda^{\alpha\beta}
    + \sqrt{2}\, S^{c (\alpha} \lambda_\rc{}^{\beta)}~, 
&\qquad
\delta S^{\ol{\alpha \beta}} &= -\lambda^{\ol{\alpha\beta}}
    + \sqrt{2}\, S^{\rbc (\balpha} \lambda_\rbc{}^{\bbeta)}~.
\end{alignat}
\end{subequations}
The last condition implies that $S^{\alpha\beta}$ and $S^{\ol{\alpha\beta}}$ are pure
gauge, while the spin-1/2 parts of $S_\ra{}^\alpha$ and $S_\rba{}^\balpha$ are invariant
and constitute the dilatini
\begin{align}\label{E:dilatini.defs}
\chi_\alpha := -i S^{\ra \beta} (\gamma_{\ra})_{\beta\alpha}~, \qquad
\chi_\balpha := -i S^{\ol{\ra \beta}} (\gamma_{\rba})_{\ol{\beta\alpha}}~.
\end{align}
The invariant components $S^{\alpha\bbeta}$ contain the Ramond-Ramond bispinor 
field strength \eqref{E:RRtoF}.

The precise dictionary between these constituents and those of type II supergravity are reviewed in Appendix \ref{A:DemoTypeII}. It is instructive to compare the numbers of independent bosonic and fermionic components of these constituents with those of a generic $\g{OSp}(D,D|2s)$ 
element.\footnote{This is a count of \emph{superfields} rather than component fields. The number of bosonic and fermionic \emph{superfields} need not match.}
Taking into account the range of the indices $a=0,\dots,D-1$ and $\alpha=1,\dots,s/2$, we count
\begin{equation}
    \renewcommand{\arraystretch}{1.5}
    \begin{tabular}{l|l|l}
    object & bosonic & fermionic\\\hline
    $(\cV_B)_\cM{}^\cN$ & $\frac12 D(D-1)+ \frac12 s(s+1)$ & $Ds$\\
    $(\cV_{E\Lambda})_\cN{}^\cB$ & $\frac12 D(3D-1) + s^2$ & $2Ds$\\
    $(\cV_S)_\cB{}^\cA$ & $\frac12 s(s+1)$ & $Ds$\\\hline
    $\g{OSp}(D,D|2s)$ & $D(2D-1)+s(2s+1)$ & $4Ds$
    \end{tabular}
\end{equation}
In the same vein, we find that $\g{H}_L\times\g{H}_R$ gauge fixing gives rise to the physically relevant fields $E_M{}^A$ (modulo Lorentz transformations $\lambda_a{}^b$), $B_{MN}$, $\chi_\alpha$, $\chi_\balpha$ and the Ramond-Ramond bispinor $S^{\alpha\bbeta}$:
\begin{equation}
    \renewcommand{\arraystretch}{1.5}
    \begin{tabular}{l|l|l}
    object & bosonic & fermionic\\\hline
    $B_{MN}$ & $\frac12 D(D-1) + \frac12 s(s+1)$ & $Ds$ \\
    $E_M{}^A \, / \,\lambda_a{}^b$ & $D^2 - \frac12 D (D-1) + s^2$ & $2 Ds$ \\
    $\chi_{\alpha}$, $\chi_{\balpha}$ & $0$ & $s$ \\
    $S^{\alpha\bbeta}$ & $\frac14 s^2$ & $0 $\\ \hline
    $\g{OSp}(D,D|2s) \,/ \,\g{H}_L\times\g{H}_R$ & $D^2 + \frac74 s^2 +\frac12 s$ & $(3D+1)s$\\\hline
    $\g{H}_L\times\g{H}_R$ & $D(D-1) + \frac12 s (\frac12 s + 1)$ & $(D-1)s$ \\
    \end{tabular}
\end{equation}

\subsection{The structure of \texorpdfstring{$\g{OSp}(D,D|2s)$}{OSp(D,D|2s)} transformations}
\label{S:SDFT.OSp}
From their embedding in double field theory, we will be able to derive the generic transformations of the supervielbein, dilatini, Ramond-Ramond sector, and dilaton under $\g{OSp}(D,D|2s)$ transformations. For now, we will not concern ourselves with the precise form of these transformations. As we will discuss in the next sections, these encompass both bosonic T-duality \cite{Cvetic:1999zs,Kulik:2000nr,Bandos:2003bz, Benichou:2008it} and fermionic T-duality \cite{Berkovits:2008ic,Beisert:2008iq} (see also \cite{Bakhmatov:2009be,Sfetsos:2010xa,Grassi:2011zf,OColgain:2012si}), as well as more general non-abelian dualities involving a supergroup $G$ \cite{Borsato:2018idb}.

The key first step in uncovering the $\g{OSp}$ structure is to introduce square matrices $\cE_A{}^M$ and $\bar \cE_A{}^M$ defined by\footnote{These definitions serve as the starting point of the generalized supervielbein analysis, see appendix B of \cite{Butter:2022gbc}. Choosing these quantities to furnish two invertible supervielbeins leads to the solution discussed here. This is closely analogous to the bosonic analysis where one poses $V_\ra{}^m$ and $V_\rba{}^m$ to be invertible. These two vielbeins $\cE_A{}^M$ and $\bar \cE_A{}^M$ will end up being proportional to the operators $\cO_\pm$ discussed in the context of the $\eta$ and $\lambda$ deformations \cite{Borsato:2016ose}.}
\begin{alignat}{2}\label{E:cV1}
\cV_\ra{}^M &=: \frac{1}{\sqrt 2} \cE_\ra{}^M~, &\qquad
\cV_\rba{}^M &=: \frac{1}{\sqrt 2} \bar\cE_\rba{}^M~, \eol
\cV_\alpha{}^M &=: \cE_\alpha{}^M \equiv \bar \cE_\alpha{}^M ~, &\qquad
\cV_\balpha{}^M &= \cE_\balpha{}^M \equiv \bar \cE_\balpha{}^M ~.
\end{alignat}
These quantities are presumed invertible and related to $E_A{}^M$ by
\begin{subequations}
\begin{alignat}{2}
\cE_\alpha{}^M &= E_\alpha{}^M 
& \qquad
\bar\cE_\alpha{}^M &= E_\alpha{}^M ~,\\
\cE_\balpha{}^M &= E_\balpha{}^M 
& \qquad
\bar\cE_\balpha{}^M &= E_\balpha{}^M ~,\\
\cE_\ra{}^M &= E_\ra{}^M -2 S_\ra{}^\beta E_\beta{}^M 
& \qquad
\bar\cE_\rba{}^M &= E_\rba{}^M -2 S_\rba{}^\bbeta E_\bbeta{}^M ~.
\end{alignat}
\end{subequations}
For reference, the inverse relations are
\begin{subequations}
\label{E:cE-E.dictionary}
\begin{alignat}{2}
\cE_M{}^\ra &= E_M{}^{\ra} ~, 
& \qquad
\bar \cE_M{}^\rba &= E_M{}^{\rba} ~, \\
\cE_M{}^\alpha &= E_M{}^\alpha + 2\,E_M{}^\rb S_\rb{}^\alpha~,
& \qquad
\bar \cE_M{}^\alpha &= E_M{}^\alpha~, \\
\cE_M{}^\balpha &= E_M{}^\balpha~, 
& \qquad
\bar \cE_M{}^\balpha &= E_M{}^\balpha + 2\,E_M{}^\rbb S_\rbb{}^\balpha~.
\end{alignat}
\end{subequations}
Note that while $\bar\cE_M{}^\rba = \cE_M{}^\rb \Lambda_\rb{}^\rba$, this \emph{does not} hold for their inverses. A useful result is
\begin{align}\label{E:sdetE.same}
|\sdet E_M{}^A| = |\sdet \cE_M{}^A| = |\sdet \bar\cE_M{}^A|
\end{align}
since the matrices themselves differ only by Lorentz transformations on some of the elements.

In analogy to the bosonic case, we introduce
\begin{align}
G_{MN} &:= \cE_M{}^\ra \cE_{N}{}^\rb \eta_{\rb \ra} 
    = -\bar\cE_M{}^\rba \bar\cE_{N}{}^\rbb \eta_{\ol{\rb \ra}}~, \eol
\EE_{MN} &:= G_{MN} - B_{MN}~, \quad
\bar \EE_{MN} := G_{MN} + B_{MN}~,
\end{align}
in terms of which we find
\begin{alignat}{2}\label{E:cV2}
\cV_{\ra M} &= \frac{1}{\sqrt 2} \cE_\ra{}^N \EE_{NM} (-)^{\grad{M}}~,
&\qquad \qquad
\cV_{\rba M} &= -\frac{1}{\sqrt 2} \bar \cE_\rba{}^N \bar \EE_{NM} (-)^{\grad{M}} ~, \eol
\cV_{\alpha M} &= \cE_\alpha{}^N \EE_{NM} (-)^{\grad{M}}
&\quad 
\cV_{\alpha M} &= -\bar\cE_\alpha{}^N \bar \EE_{NM} (-)^{\grad{M}}~, \eol
\cV_{\balpha M} &= \cE_\balpha{}^N \EE_{NM} (-)^{\grad{M}}
&\quad 
\cV_{\balpha M} &= -\bar\cE_\balpha{}^N \bar \EE_{NM} (-)^{\grad{M}}~. 
\end{alignat}
A generic orthosymplectic transformation can be written as
$\cV'_\cA{}^\cM = \cV_\cA{}^\cN \cU_\cN{}^\cM$
where
\begin{align}
\cU_{\cM}{}^\cN =
\begin{pmatrix}
U_M{}^N & U_{MN} (-)^{\grad{N}} \\
U^{M N} & U^M{}_N (-)^{\grad{N}}
\end{pmatrix}~.
\end{align}
Defining
\begin{alignat}{2}
X_M{}^N &:= U_M{}^N + \EE_{M P} U^{P N} (-)^{\grad{P}} ~, &\qquad\qquad
\bar X_M{}^N &:= U_M{}^N - \bar \EE_{M P} U^{P N} (-)^{\grad{P}}~, \eol
Y_{MN} &:= U_{MN} + \EE_{M P} U^P{}_N (-)^{\grad{P}}& \quad
\bar Y_{MN} &:= U_{MN} - \bar \EE_{M P} U^P{}_N (-)^{\grad{P}}~,
\end{alignat}
one can show that
\begin{alignat}{2}
\cE'_A{}^M &= \cE_A{}^N X_N{}^M ~, &\qquad\qquad
\bar \cE'_A{}^{M} &= \bar \cE_A{}^N \bar X_N{}^M, \eol
\EE'_{MN} &= (X^{-1})_M{}^P Y_{P N}~, &\qquad
\bar \EE'_{MN} &= (\bar X^{-1})_M{}^P \bar Y_{P N}~.
\label{E:FormalOSpTrafos}
\end{alignat}
From these equations one can read off the transformations of $B_{MN}$ and $G_{MN}$.

Similarly, from 
$\cE'_M{}^A = (X^{-1})_M{}^N \cE_N{}^A$ and
$\bar \cE'_M{}^A = (\bar X^{-1})_M{}^N \cE_N{}^A$, we deduce the
transformations for the graviton one-form
\begin{align}
E'_M{}^\ra = (X^{-1})_M{}^N E_N{}^\ra~, \qquad
E'_M{}^\rba = (\bar X^{-1})_M{}^N E_N{}^\rba~,
\end{align}
and these are related by the Lorentz transformation
\begin{align}
\Lambda'_\ra{}^\rbb 
    &= \underbrace{E_\ra{}^M (X \bar X^{-1})_M{}^N E_N{}^\rb}_{{\bm\Lambda}(U)_\ra{}^\rb}
    \times \Lambda_\rb{}^\rbb
    = \Lambda_\ra{}^\rba \times 
    \underbrace{E_\rba{}^M (X \bar X^{-1})_M{}^N E_N{}^\rbb}_{{\bm\Lambda}(U)_\rba{}^\rbb}~.
\end{align}
Some useful identifies are
\begin{align}
U^{M P} (X^{-1})_P{}^N &= -U^{N P} (\bar X^{-1})_P{}^M (-)^{\grad{M} \grad{N}}~, \eol
(X \bar X^{-1})_M{}^N 
    &= \delta_M{}^N + 2 \,G_{MP} \,U^{PQ} (\bar X^{-1})_Q{}^N (-)^{\grad{P}}~, \eol
{\bm\Lambda}(U)_\ra{}^\rb &= 
    \delta_\ra{}^\rb + 2  \,U^{MP} (\bar X^{-1})_P{}^N E_N{}^\rb \,E_M{}_\ra~.
\label{E:OSp.Identities}
\end{align}

The gravitini are identified in Dirac spinor language using \eqref{E:Gravitini.Dirac}.
Applying this result gives the transformations
\begin{align}
E'_M{}^{1 \hbeta} 
    = (\bar X^{-1})_M{}^N E_N{}^{1\hbeta}~, \qquad
E'_M{}^{2 \hbeta} 
    = (X^{-1})_M{}^N E_N{}^{2\hbeta}
        (\bm{\slashed\Lambda}(U)^{-1})_{\hbeta}{}^{\halpha} 
\end{align}
where $\bm{\slashed\Lambda}(U)$ is the spinorial version of ${\bm\Lambda}(U)$.

The transformations for the dilatini \eqref{E:dilatini.defs} are a bit more involved.
From $S_\rb{}^{\alpha} = -\frac{1}{2} \cE_\rb{}^M \bar \cE_M{}^\alpha$, we can show
\begin{align}
S'{}^{\rb \alpha}
    &= S^{\rb \alpha} - E_N{}^\rb U^{N M} (\bar X^{-1})_M{}^P E_P{}^\alpha (-)^{\grad{N}}
    \quad \implies \eol
\chi'_{\alpha} &= \chi_\alpha - i \,U^{N M} (X^{-1})_M{}^P E_P{}^\rb 
(\gamma_\rb)_{\alpha \beta} E_N{}^\beta
\end{align}
where we have used the first identity in \eqref{E:OSp.Identities} to replace $\bar X$ with $X$.
A similar expression holds for $\chi_\balpha$. Converting to Dirac notation gives\footnote{Several sign factors factors appear in the second term of $\chi'_{2\halpha}$ relative to $\chi'_{1\halpha}$. A relative minus sign comes about essentially from converting $\bar\gamma_\rbb$ to $-\gamma_* \gamma_\rb$ after conjugating by all the $\Lambda$ factors. A factor of $\alpha_\Lambda$ comes from converting $\bar C^{-1}$ to $C^{-1}$. Finally, a factor of $\alpha_\Lambda \beta_\Lambda$ appears after eliminating the $\gamma_*$.}
\begin{align}
\chi'_{1\halpha} &= \chi_{1\halpha} - i \,U^{N M} (X^{-1})_M{}^P E_P{}^\rb 
(\gamma_\rb C^{-1})_{\halpha \hbeta} E_{N}{}^{1\hbeta}~, \eol
\chi'_{2\halpha}
    &= {\bm{\slashed{\Lambda}}}(U)_{\halpha}{}^{\hbeta} \Big(
    \chi_{2\hbeta} + i \,\beta_\Lambda \,U^{N M} (\bar X^{-1})_M{}^P E_P{}^\rb
    (\gamma_\rb C^{-1})_{\hbeta\hgamma} E_{N}{}^{2\hgamma}
    \Big)
\end{align}
The $\beta_\Lambda$ factor is $+1$ for IIB/IIA$^*$ and $-1$ for IIA/IIB$^*$.

The Ramond-Ramond bispinor in Weyl notation is 
$S^{\alpha \bbeta} = -\cV^{\alpha M} E_M{}^{\bbeta}$.
This transforms as
\begin{align}
S'{}^{\alpha \bbeta} 
    &= 
-\Big(\cV^{\alpha N} X_N{}^M
+ (\cV^\alpha{}_N - \cV^{\alpha P} \EE_{P N} )\,U^{N M} (-)^{\grad{N}} \Big)
(X^{-1})_M{}^P E_P{}^\bbeta~.
\end{align}
One can show that $\cV^\alpha{}_N - \cV^{\alpha P} \EE_{PN} = \bar \cE_N{}^\alpha$
and translating this to Dirac form gives
\begin{align}
S'{}^{1\halpha \, 2\hbeta}
    = 
    \Big( S^{1\halpha \, 2\hgamma} 
    - E_N{}^{1\halpha} \,U^{N M} (X^{-1})_M{}^P E_P{}^{2\hgamma}
    \Big) ({\bm{\slashed\Lambda}}(U)^{-1})_{\hgamma}{}^{\hbeta} ~.
\end{align}
In the democratic formulation of type II supergravity, we define
\begin{align}\label{E:cF12}
\widehat \cF^{1\halpha \, 2\hbeta} := 
\begin{cases}
\sum_{p \,\text{odd}} \frac{1}{p!} \widehat \cF_{a_1 \cdots a_p} (C P_R \gamma^{a_1 \cdots a_p})^{\halpha \hbeta}
& \text{IIB/IIB$^*$} \\
\sum_{p \,\text{even}} \frac{1}{p!} \widehat \cF_{a_1 \cdots a_p} (C P_R \gamma^{a_1 \cdots a_p})^{\halpha \hbeta}
& \text{IIA/IIA$^*$}
\end{cases}
\end{align}
From \eqref{E:RRtoF}, we deduce the transformation
\begin{align}
e^{\varphi'} \widehat \cF'{}^{1\halpha \, 2\hbeta}
    = 
    \Big( e^{\varphi} \widehat \cF^{1\halpha \, 2\hgamma} 
    - 32i\, E_N{}^{1\halpha} \,U^{N M} (X^{-1})_M{}^P E_P{}^{2\hgamma}
    \Big) ({\bm{\slashed\Lambda}}(U)^{-1})_{\hgamma}{}^{\hbeta} ~.
\end{align}

The above requires the transformation of the dilaton, which is our last field to discuss.
Its behavior in super-DFT mirrors its bosonic cousin. It is a superfield $\Phi$ that transforms as a scalar density under generalized Lie derivatives
\begin{align}
\dLie_\xi \Phi = \xi^\cM \pa_\cM \Phi + \pa_\cM \xi^\cM \Phi
    = \pa_\cM (\xi^\cM \Phi)~.
\end{align}
The generalized superdilaton $\Phi$ is related to the supergravity dilaton $\varphi$ by
$\Phi = e^{-2\varphi} \sdet E_M{}^A$.\footnote{Note that $\Phi$ is \emph{not} simply related to the component dilaton $e^{-2d}$. They differ by a factor of $\sdet E_M{}^A / \det e_m{}^\ra$.}
Presuming the superdilaton to transform by a scalar factor
$\Phi' = \Phi \,\cU_\Delta$,
it follows that
\begin{align}
e^{-2\varphi'} = e^{-2 \varphi} \times \sdet X_M{}^N \times \cU_\Delta~.
\end{align}
The factor $\cU_\Delta$ is \emph{a priori} independent of $\cU_\cM{}^\cN$.

\section{Super non-abelian T-duality}
\label{S:SNATD}

The simplest and most direct situation where we can explicitly see how the $\g{OSp}$ transformations of double field theory come about is in the context of T-duality for a supersymmetric $\sigma$-model, namely the Green-Schwarz superstring with a non-abelian (or abelian) isometry supergroup $G$. This situation was fully analyzed by Borsato and Wulff a few years ago \cite{Borsato:2018idb}. We first summarize their construction and then reinterpret their results in the language of double field theory.

\subsection{Worldsheet formulation of non-abelian T-duality}
\label{S:SNATD.worldsheet}
Following \cite{Borsato:2018idb}, the starting point is a worldsheet Lagrangian
\begin{align}
\cL = -\frac{1}{2} \sqrt{-h} \,h^{ij} \,\pa_i Z^M \pa_j Z^N G_{NM}
    - \frac{1}{2} \veps^{ij} \,\pa_i Z^M \pa_j Z^N B_{NM}
\end{align}
The supercoordinates $Z^M = (X^m, \Theta^\mu)$ parametrize a target superspace.
The worldsheet metric $h_{ij}$ is presumed to have Lorentzian signature $(-,+)$ and
the worldsheet antisymmetric tensor density $\veps^{ij}$ obeys $\veps^{01} = +1$.
The target space tensors $G_{MN}(Z)$ and $B_{MN}(Z)$ are graded symmetric and antisymmetric
respectively.\footnote{One may refer to $G_{MN}$ as the supermetric but this is something of a misnomer as it need not be invertible and the usual considerations of Riemannian geometry do not apply. For the Green-Schwarz superstring, $G_{MN}$ is built from a rectangular piece $E_M{}^a$ of the supervielbein $E_M{}^A$ as $G_{MN} = E_M{}^a E_N{}^b \eta_{ab}$.}

Let the $\sigma$-model admit a supergroup $G$ of isometries described by supervectors $k_{\Iso1}$ obeying
$[k_{\Iso1}, k_{\Iso2}] = f_{\Iso1 \Iso2}{}^{\Iso3} k_{\Iso3}$.
This is a graded commutator, and the isometry label $\Iso1$ should be understood to decompose into bosonic and fermionic isometries, $\Iso1 = (\iso1, \fIso1)$. We presume that we can adopt a coordinate system where the coordinates $Z^M$ factorize into coordinates $Y^{\cIso1}$ on which the isometries act and spectator coordinates $Z^{\uM}$, so that $k_{\Iso1} = k_{\Iso1}{}^{\cIso1} \pa_{\cIso1}$. The superfields $G$ and $B$ decompose as
\begin{subequations}\label{E:FlattenedGB}
\begin{align}
G &= 
    e^{\Iso1} \otimes e^{\Iso2} G_{\Iso2 \Iso1}(\ul Z)
    + 2 \,e^{\Iso1} \otimes \rd Z^\uN G_{\uN \Iso1}(\ul Z)
    + \rd Z^\uM \otimes \rd Z^\uN G_{\uN \uM}(\ul Z)
    ~,\\
B &= \frac{1}{2} e^{\Iso1} \wedge  e^{\Iso2} B_{\Iso2 \Iso1}(\ul Z)
    + e^{\Iso1} \wedge \rd Z^\uN B_{\uN \Iso1}(\ul Z)
    + \frac{1}{2} \rd Z^\uM \wedge \rd Z^\uN B_{\uN \uM}(\ul Z)~.
\end{align}
\end{subequations}
All the dependence on the coordinates $Y^{\cIso1}$ is sequestered in the left-invariant vector fields $e^{\Iso1}$ in the usual manner, $e^{\Iso1} t_{\Iso1} = g^{-1} \rd g$ for $g(Y)\in G$.
We review in Appendix \ref{A:GaugedSigmaModels} how the above conditions come about.
The generators $t_{\Iso1}$ obey the algebra
\begin{align}\label{E:superLieAlgebra}
[t_{\Iso1}, t_{\Iso2}] = - f_{\Iso1 \Iso2}{}^{\Iso3} t_{\Iso3}~.
\end{align}
Our supergroup conventions are given in Appendix \ref{A:Supergroups}. 

When the isometries act freely (that is, without isotropy), the above has a clear geometric interpretation: the coordinates $Y^{\cIso1}$ parametrize the orbits of $G$ on the manifold. When the isometries act with an isotropy group $H$, then we can (at least locally) take the coordinates $Y^{\cIso1}$ to parametrize the orbits of $G/H$.\footnote{The strategy reviewed here follows \cite{Borsato:2018idb} and is equivalent to extending the coordinates $\dot Z$ by additional $H$ coordinates so that the full group $G$ acts freely. The conditions \eqref{E:G/H_condition1} and \eqref{E:G/H_condition2} guarantee that the additional degrees of freedom drop out.} The isotropy condition amounts to invariance under $g \rightarrow g h$ for $h \in H$, meaning that $G_{MN}$ (and similarly for $B_{MN}$) must be invariant under the adjoint action of $H$,
\begin{align}\label{E:G/H_condition1}
(\Adj h)_{\Iso1}{}^{\Iso1'} G_{\Iso1' \Iso2'} (\Adj h)_{\Iso2}{}^{\Iso2'} 
    \, (-)^{\grad{\Iso2} \grad{\Iso2'} + \grad{\Iso2'}}
= G_{\Iso1 \Iso2}~, \qquad
(\Adj h)_{\Iso1}{}^{\Iso1'} G_{\Iso1' \uN} 
= G_{\Iso1 \uN}~.
\end{align}
It must also project out the Lie algebra $\mathfrak{h}$,
\begin{align}\label{E:G/H_condition2}
\zeta^{\Iso1} G_{\Iso1 \Iso2} = \zeta^{\Iso1} G_{\Iso1 \uN} = 0~, \qquad \zeta \in \mathfrak{h}~.
\end{align}

Non-abelian T-duality is effected by replacing $\pa_i Y^{\cIso1} e_{\cIso1}{}^{\Iso1}$ with a $\mathfrak{g}$-valued worldsheet one-form $\tilde A_i{}^{\Iso1}$, and adding a term $\veps^{ij} F(\tilde A)_{ij}{}^{\Iso1} \nu_{\Iso1}$ where $F(\tilde A)_{ij}{}^{\Iso1}$ is the worldsheet $G$-curvature built from $\tilde A$. Treating $\nu_{\Iso1}$ as a Lagrange multiplier, one recovers the original action where $\tilde A = g^{-1} \rd g$ is pure gauge. The T-dual model arises if we instead integrate out the one-form $\tilde A$. Working in lightcone coordinates for simplicity, the Lagrangian becomes
\begin{align}
\cL &=
    \pa_+ Z^\uM \,\EE_{\uM \uN} \, \pa_- Z^\uN \,(-)^{\grad{\uN}}
    + \tilde A_+^{\Iso1} \,\widehat \EE_{\Iso1 \Iso2} \,\tilde A_-^{\Iso2} \, (-)^{\grad{\Iso2}}
    \eol & \quad
    + \tilde A_+^{\Iso1} \Big(\pa_- \nu_{\Iso1} 
    + \EE_{\Iso1 \uM} \pa_- Z^\uM \,(-)^{\grad{\uM}} 
    \Big)
    + \Big(\pa_+ Z^\uM \EE_{\uM \Iso1} - \pa_+ \nu_{\Iso1} 
    \Big) \tilde A_-^{\Iso1} \, (-)^{\grad{\Iso1}}
\end{align}
where we have introduced
\begin{subequations}
\begin{alignat}{2}
\EE_{\uM \uN} &= G_{\uM \uN} - B_{\uM \uN}~, \\
\EE_{\Iso1 \uM} &= G_{\Iso1 \uM} - B_{\Iso1 \uM}~, &\quad
\EE_{\uM \Iso1 } &= G_{\uM \Iso1 } - B_{\uM \Iso1}~, \\
\widehat \EE_{\Iso1 \Iso2} &= G_{\Iso1 \Iso2} - B_{\Iso1 \Iso2} - f_{\Iso1 \Iso2}{}^{\Iso3} \nu_{\Iso3}~.
\label{E:Ehat}
\end{alignat}
\end{subequations}
The addition of the Lagrange multiplier to $\widehat \EE_{\Iso1 \Iso2}$ is the major difference with respect to abelian T-duality. Integrating out the worldsheet one-forms gives the dual model
\begin{align}
\cL = \pa_+ Z'{}^M \,\EE'_{M N} \,\pa_- Z'{}^N \, (-)^{\grad{N}}
\end{align}
where the new coordinates are $Z'{}^M = (Z^\uM, \tilde Y{}^{\Iso1})$ with
\begin{align}\label{E:dualYCoordinate}
\tilde Y^{\Iso1} = \nu_{\Iso2}  \,\delta^{\Iso2 \Iso1} (-)^{\grad{\Iso2}}
    = \nu_{\Iso1}  (-)^{\grad{\Iso1}} ~.
\end{align}
The choice of grading here may seem awkward, but it makes subsequent formulae simpler:
\begin{align}
\EE'_{\Iso1 \Iso2} &= \widehat \EE^{\Iso1 \Iso2} ~, \qquad
\EE'_{\Iso1 \uM} = \widehat \EE^{\Iso1 \Iso2} \EE_{\Iso2 \uM}~, \qquad
\EE'_{\uM \Iso1} = -\EE_{\uM \Iso2} \widehat \EE^{\Iso2 \Iso1} (-)^{\grad{\Iso2}}~, \eol
\EE'_{\uM \uN} &= \EE_{\uM \uN} - \EE_{\uM \Iso1} \widehat \EE^{\Iso1 \Iso2} \EE_{\Iso2 \uN} (-)^{\grad{\Iso1}}
\end{align}
where we define $\widehat \EE^{\Iso1 \Iso2}$ as the graded inverse, 
$\widehat \EE^{\Iso1 \Iso3} \widehat \EE_{\Iso3 \Iso2} 
    = \delta_{\Iso2}{}^{\Iso1}\, (-)^{\grad{\Iso2} \grad{\Iso1}}$.

Comparing the expressions for $\EE'_{MN}$ with the formal result \eqref{E:FormalOSpTrafos} for a generic $\g{OSp}(D,D|2s)$ transformation, we find $\cU$ can be written as a sequence of three orthosymplectic transformations, $\cU = \cU_{(0)} \cU_{(1)} \cU_{(2)}$, where
\begin{align}\label{E:NATD.U}
\cU_{(0)} &=
\left(
\begin{array}{cc|cc}
\delta_\uM{}^\uN & 0 & 0 & 0 \\
0 & e_{\cIso1}{}^{\Iso2} & 0 & 0 \\ \hline
0 & 0 & \delta^\uM{}_\uN & 0 \\
0 & 0 & 0 & e_{\Iso2}{}^{\cIso1} (-)^{\grad{\cIso1} \grad{\Iso2}+\grad{\Iso2}}
\end{array}\right)~, \quad
\cU_{(1)} = 
\left(
\begin{array}{cc|cc}
\delta_\uM{}^\uN & 0 & 0 & 0 \\
0 & \delta_{\Iso1}{}^{\Iso2} & 0 & -f_{\Iso1 \Iso2}{}^{\Iso3} \nu_{\Iso3} (-)^{\grad{\Iso2}}\\ \hline
0 & 0 & \delta^\uM{}_\uN & 0 \\
0 & 0 & 0 & \delta^{\Iso1}{}_{\Iso2}
\end{array}\right)~, \eol[2ex]
\cU_{(2)} &= 
\left(
\begin{array}{cc|cc}
\delta_\uM{}^\uN & 0 & 0 & 0 \\
0 & 0 & 0 & \delta_{\Iso1 \Iso2} (-)^{\grad{\Iso2}} \\ \hline
0 & 0 & \delta^\uM{}_\uN & 0 \\
0 & \delta^{\Iso1 \Iso2} & 0 & 0
\end{array}\right)~.
\end{align}
The factor $\cU_{(0)}$ flattens $G$ and $B$ in the isometric
directions with the left-invariant vielbein: this occurred in \eqref{E:FlattenedGB}. 
The factor $\cU_{(1)}$ gives the non-abelian factor that replaces
$\EE_{\Iso1 \Iso2}$ with $\widehat \EE_{\Iso1 \Iso2}$ in \eqref{E:Ehat}.
Finally, $\cU_{(2)}$ induces the familiar T-duality transformation \`a la Buscher.
Now one can use the results in section \ref{S:SDFT.OSp} to compute the new gravitini, dilatini, and Ramond-Ramond bispinors. (We will return to the question of the dilaton in due course.)
The additional ingredients we will need are
\begin{align}
X_M{}^N =
\begin{pmatrix}
\delta_{\uM}{}^{\uN} & \EE_{\uM \Iso2} (-)^{\grad{\Iso2}} \\[1ex]
0 & e_{\cIso1}{}^{\Iso1} \,\hat \EE_{\Iso1 \Iso2} (-)^{\grad{\Iso2}}
\end{pmatrix}~, \quad
(X^{-1})_M{}^N =
\begin{pmatrix}
\delta_{\uM}{}^{\uN} 
    & -\EE_{\uM \Iso1} \widehat \EE^{\Iso1 \Iso2} \,e_{\Iso2}{}^{\cIso2} 
    (-)^{\grad{\Iso1}}
    \\[1ex]
0 & \widehat \EE^{\Iso1 \Iso2} \,e_{\Iso2}{}^{\cIso2}
\end{pmatrix}~.
\end{align}
Now we can directly compute the new supervielbein
\begin{align}
E'{}^a &= 
    \rd Z^\uM \Big(
    E_\uM{}^a
    - \EE_{\uM \Iso1} (-)^{\grad{\Iso1}} \widehat{\EE}^{\Iso1 \Iso2} E_{\Iso2}{}^a
    \Big)
    + \rd \nu_{\Iso1} (-)^{\grad{\Iso1}} \widehat{\EE}^{\Iso1 \Iso2} E_{\Iso2}{}^a~, \\
E'{}^{1\halpha} 
    &= \rd Z^\uM \Big(
    E_\uM{}^{1\halpha}
    - \bar \EE_{\uM \Iso1} (-)^{\grad{\Iso1}} \widehat{\bar \EE}{}^{\Iso1 \Iso2} E_{\Iso2}{}^{1\halpha}
    \Big)
    - \rd \nu_{\Iso1} (-)^{\grad{\Iso1}} \widehat{\bar \EE}{}^{\Iso1 \Iso2} E_{\Iso2}{}^{1\halpha}~, \\
E'{}^{2\halpha} 
    &= \Big[\rd Z^\uM \Big(
    E_\uM{}^{2\hbeta}
    - \EE_{\uM \Iso1} (-)^{\grad{\Iso1}} \widehat{\EE}{}^{\Iso1 \Iso2} E_{\Iso2}{}^{2\hbeta}
    \Big)
    + \rd \nu_{\Iso1} (-)^{\grad{\Iso1}} \widehat{\EE}^{\Iso1 \Iso2} E_{\Iso2}{}^{2\hbeta}
    \Big] (\bm{\slashed\Lambda}^{-1})_{\hbeta}{}^\halpha~.
\end{align}
The Lorentz transformation $\bm\Lambda$ and its inverse are
\begin{align}
{\bm\Lambda}_a{}^b = \delta_a{}^b - 2 \,\widehat{\bar \EE}{}^{\Iso1 \Iso2} E_{\Iso2}{}^b E_{\Iso1 a}~, \qquad
(\bm\Lambda^{-1})_a{}^b = \delta_a{}^b - 2 \,\widehat{\EE}{}^{\Iso1 \Iso2} E_{\Iso2}{}^b E_{\Iso1 a}~.
\end{align}
It is difficult to characterize fully this Lorentz transformation, although one can show that
$\det {\bm\Lambda} = (-1)^{\dim_B G}$ where by $\dim_B$ we mean the bosonic dimension. This was proven in \cite{Borsato:2018idb} for a bosonic group. Adapting their proof for a supergroup is straightforward. In their eq. (3.10), promote traces and determinants to supertraces and superdeterminants, leading to
$\det {\bm\Lambda} = \sdet(-1) \times 
    \frac{\sdet \widehat {\bar\EE}_{\Iso1 \Iso2}}{\sdet \widehat \EE_{\Iso1 \Iso2}}$.
Because $\widehat {\bar \EE}$ is the supertranspose of $\widehat \EE$, their superdeterminants are related as
$\sdet \widehat{\bar \EE}_{\Iso1 \Iso2} 
= \sdet \widehat{\EE}_{\Iso1 \Iso2} \times (-1)^{\dim_F G}$ where $\dim_F$ denotes
the fermionic dimension. The result follows since $\sdet(-1) = (-1)^{\dim G}$.

The super two-form, covariant field strengths, and dilatini transform as
\begin{align}
B' &= \frac{1}{2} \rd Z^\uM \wedge \rd Z^\uN B_{\uN \uM}
    - \frac{1}{2} \widehat \EE^{\Iso1 \Iso2}
    \Big(
    \rd \nu_{\Iso2} + \rd Z^{\uN} \bar \EE_{\uN \Iso2}
    \Big)\wedge 
        \Big(
    \rd \nu_{\Iso1} - \rd Z^{\uM} \EE_{\uM \Iso1}
    \Big) ~, \\
e^{\varphi'} \widehat \cF'{}^{1\halpha \, 2\hbeta}
    &= 
    \Big( e^{\varphi} \widehat \cF^{1\halpha \, 2\hgamma} 
    -32i\, E_{\Iso1}{}^{1\halpha} \,\widehat \EE^{\Iso1 \Iso2} E_{\Iso2}{}^{2\hgamma}
    \Big) ({\bm{\slashed\Lambda}}^{-1})_{\hgamma}{}^{\hbeta} ~, \\
\chi'_{1\halpha} &= \chi_{1\halpha} - i \,
    \widehat{\EE}^{\Iso1 \Iso2} E_{\Iso2}{}^b E_{\Iso1}{}^{1\hbeta}
(\gamma_b C^{-1})_{\hbeta \halpha } ~, \\
\chi'_{2\halpha} &= \bm{\slashed\Lambda}_{\halpha}{}^{\hbeta} 
    \Big(\chi_{2\hbeta}
    + i \,\beta_\Lambda \,\widehat{\bar\EE}{}^{\Iso1 \Iso2} E_{\Iso2}{}^b 
    E_{\Iso1}{}^{2\hgamma} (\gamma_b C^{-1})_{\hgamma \hbeta} 
    \Big)~.
\end{align}
These results match those found by Borsato and Wulff \cite{Borsato:2018idb} subject to
the identifications
\begin{align}
\nu_I = -\nu_{\Iso1}~, \quad
f_{IJ}{}^K = -f_{\Iso1 \Iso2}{}^{\Iso3}~, \quad
N_+^{IJ} = (-)^{\grad{\Iso1}} \widehat \EE^{\Iso1 \Iso2}~, \quad
N_-^{IJ} = -(-)^{\grad{\Iso1}} \widehat{\bar \EE}{}^{\Iso1 \Iso2}~.
\end{align}

This argument is perhaps a bit too slick as it appears to ignore a key point: the transformation of $\EE_{MN}$ does not completely determine $\cU$. Put simply, there are as many degrees of freedom in $\cU$ as there are in $\cV_\cA{}^\cM$ itself, but only some of these appear in $\EE_{MN}$. The choice \eqref{E:NATD.U} was merely the simplest choice that reproduces $\EE'_{MN}$, but this is hardly conclusive. What actually singles it out (we will show) is that it leaves the generalized fluxes of double field theory invariant --- this has the crucial effect that it guarantees the dual theory will possess the proper supergravity constraints.

\subsection{Double field theory interpretation}
\label{S:SNATD.DFT}
In defining the dual coordinate $\tilde Y^{\Iso1}$ in \eqref{E:dualYCoordinate}, we have (as usual in bosonic T-duality) identified it with the Lagrange multiplier $\nu_{\Iso1}$ directly, swapping the index location by hand. This may not actually be the most natural choice; instead, what we can do is to think of $\nu_{\Iso1}$ as a function of the new coordinates, which we denote $\tilde Y_{\cIso1}$.\footnote{One could presumably also let $\nu_{\Iso1}$ depend on the spectator coordinates, but this muddies the water.} These can be interpreted as the natural DFT coordinates dual
to $Y^{\cIso1}$. Then in the $\sigma$-model action, we denote
\begin{align}\label{E:tilde_e.def}
\rd \nu_{\Iso1} = \tilde e_{\Iso1}{}^{\cIso1} \rd \tilde Y_{\cIso1} ~, \qquad
\tilde e_{\Iso1}{}^{\cIso1} := \pa^{\cIso1} \nu_{\Iso1}  (-)^{\grad{\cIso1} \grad{\Iso1}}
    = \tilde e^{\cIso2}{}_{\Iso1} (-)^{\grad{\cIso1} \grad{\Iso1}}
\end{align}
with $\tilde e_{\Iso1}{}^{\cIso1}$ is interpreted as the dual analogue of $e_{\cIso1}{}^{\Iso1}$.
The crucial feature is that while $e_{\cIso1}{}^{\Iso1}$ is the left-invariant vector field of the group $G$ and therefore carries a flux, the dual vielbein $\tilde e_{\Iso1}{}^{\cIso1}$ is purely flat. This slightly modifies $\cU_{(1)}$ to
\begin{align}
\cU_{(1)} &= 
\left(
\begin{array}{cc|cc}
\delta_\uM{}^\uN & 0 & 0 & 0 \\
0 & \tilde e_{\Iso1}{}^{\cIso2} & 
    0 & -f_{\Iso1 \Iso2}{}^{\Iso3} \nu_{\Iso3} \,\tilde e^{\Iso2}{}_{\cIso2}
        (-)^{\grad{\Iso2}+\grad{\cIso2}}
        \\ \hline
0 & 0 & \delta^\uM{}_\uN & 0 \\
0 & 0 & 0 & \tilde e^{\Iso1}{}_{\cIso2} (-)^{\grad{\cIso2}}
\end{array}\right)~.
\end{align}
For $\cU_{(2)}$, we simply replace $\Iso1$ with $\cIso1$ everywhere.
Now it will be convenient to denote the indices of these matrices as
\begin{align}\label{E:Umatrices}
(\cU_{(0)})_\cM{}^{\wtcN}~, \qquad
(\cU_{(1)})_\wtcM{}^{\cN}~, \qquad
(\cU_{(2)})_{\cM}{}^{\cN}
\end{align}
where $\wtcM$ is flattened in the isometry direction, i.e. it involves
${}^\uM, {}_\uM$ and ${}^{\Iso1},{}_{\Iso1}$. From the perspective of double field theory, we can dispense with $\cU_{(2)}$: this merely has the effect of swapping which coordinates we view as physical and which as winding, so we can think of it as a purely passive transformation. What interpretation do we give to $\cU_{(0)}$ and $\cU_{(1)}$?

Suppose we have a generalized vielbein depending on two sets of doubled coordinates, $Y^{\cIso1}$ and ${\tilde Y}_{\cIso1}$ as well as $Z^\uM$ and $\tilde Z_\uM$, in such a way that it decomposes into a product of two factors:
\begin{align}\label{E:VielbeinDecomposedDuality}
\cV_\cM{}^\cA
    = \mathring \cV_\cM{}^{\wtcN}(Y, \tilde Y) \times 
    \wtcV_{\wtcN}{}^{\cA}(Z, \tilde Z)~.
\end{align}
The first factor involves only the $Y$ coordinates and the second only the spectators. 
(We don't actually need the dual $\tilde Z_\uM$ coordinates, but we keep them for generality.) In the bosonic limit $s=0$ \eqref{E:VielbeinDecomposedDuality} reduced to the generalized Scherk-Schwarz ansatz \cite{Dabholkar:2002sy,Aldazabal:2011nj,Geissbuhler:2011mx} in DFT. Here, we study its natural supersymmetrization. The tilde index  ${}^{\wtcM} = ({}^{\wtM}, {}_{\wtM})$ decomposes as $\wtM = (\uM, {\bigIso1})$. We presume $\mathring \cV$ is chosen so that
\begin{align}\label{E:mathringcVcondition}
\mathring \cV_{\wtcN}{}^{\uM} = \delta_{\wtcN}{}^{\uM}~, \qquad
\mathring \cV_{\wtcN \,\uM} = \delta_{\wtcN \uM}~,
\end{align}
That is, $\mathring \cV$ is the identity in the non-isometric directions; this is the situation in the case at hand. The original model and its dual differ only in the choice of $\mathring \cV$ which in the two cases is
\begin{align}
\label{E:zV.original}
\underline{\text{original model}} \quad \mathring \cV_\cM{}^{\wtcN}
&=
\left(
\begin{array}{cc|cc}
\delta_\uM{}^\uN & 0 & 0 & 0 \\
0 & e_{\cIso1}{}^{\Iso2} & 0 & 0 \\ \hline
0 & 0 & \delta^\uM{}_\uN & 0 \\
0 & 0 & 0 & e^{\cIso1}{}_{\Iso2} (-)^{\grad{\Iso2}}
\end{array}\right)
= (\cU_{(0)})_\cM{}^{\wtcN}~, \\[3ex]
\label{E:zV.dual}
\underline{\text{dual model}} \quad \mathring \cV_\cM{}^{\wtcN}
&=
\left(
\begin{array}{cc|cc}
\delta_\uM{}^\uN & 0 & 0 & 0 \\
0 & \tilde e_{\cIso1}{}^{\Iso2} & 0 & 
    \tilde e_{\cIso1}{}^{\Iso1} f_{\Iso1 \Iso2}{}^{\Iso3} \nu_{\Iso3} 
    (-)^{\grad{\Iso2}} \\ \hline
0 & 0 & \delta^\uM{}_\uN & 0 \\
0 & 0 & 0 & \tilde e^{\cIso1}{}_{\Iso2} (-)^{\grad{\Iso2}}
\end{array}\right)
= (\cU_{(1)}^{-1})_\cM{}^{\wtcN}
\end{align}
Here one should think of $\nu_{\Iso1}(\tilde Y)$ as the potential for $\tilde e_{\Iso1}{}^{\cIso1}$ as in \eqref{E:tilde_e.def}. 

The first generalized vielbein depends on $Y$  but not $\tilde Y$, and vice-versa for the second. Both of these, viewed as generalized vielbeins, \emph{involve the same flux tensor}.
Recall that in double field theory, one can build a generalized flux tensor $\cF_{\cA \cB \cC}$
from the generalized vielbein,
\begin{align}\label{E:GenFluxTensor}
\dLie_{\cV_\cA} \cV_\cB{}^\cM = -\cF_{\cA \cB}{}^\cC \cV_{\cC}{}^\cM~, \qquad
\cF_{\cA \cB \cC} := -3\, \cV_{[\cA}{}^\cM \pa_\cM \cV_\cB{}^{\cN} \cV_{\cN \cC]}
\end{align}
with the sign here chosen so that the definition of the flux tensor matches that of the torsion tensor in conventional (undoubled) superspace. Using the decomposition \eqref{E:VielbeinDecomposedDuality}, one finds
\begin{align}
\cF_{\cA \cB \cC} &= \widetilde \cF_{\cA \cB \cC } 
    + \wtcV_\cA{}^{\wtcM}
    \wtcV_\cB{}^{\wtcN}
    \wtcV_\cC{}^{\wtcP}
    \mathring \cF_{\wtcM \wtcN \wtcP } \qquad \text{(gradings suppressed)}
\end{align}
where $\widetilde \cF_{\cC \cB \cA}$ is built purely from $\widetilde \cV$ (which is unchanged under duality) and
\begin{align}\label{E:NATD.fluxes}
\mathring \cF_{\wtcM \wtcN \wtcP}
:= -3 \,\mathring \cV_{[\wtcM|}{}^{\cM} \pa_\cM \mathring \cV_{|\wtcN|}{}^{\cN} \mathring \cV_{\cN |\wtcP]}
= 
\begin{cases}
f_{\Iso1 \Iso2}{}^{\Iso3} & {}_{\wtcM\wtcN\wtcP}={}_{\Iso1\Iso2}{}^{\Iso3} \\
0 & \text{otherwise}
\end{cases}
\end{align}
for \emph{both} the original and dual models.

This suggests an alternative way of seeing that the class of Green-Schwarz superstrings obeying the $\kappa$-symmetry constraints \eqref{E:FluxConstraint1} and \eqref{E:FluxConstraint2} is closed under super non-abelian T-duality, a result established explicitly in \cite{Borsato:2018idb}. Let's begin with two observations:
\begin{itemize}
\item 
The Green-Schwarz action on its own does not contain all of the physical data --- it contains only $G_{MN} = E_M{}^a E_N{}_a$ and $B_{MN}$.  However, if it obeys the $\kappa$-symmetry constraints, then one can uniquely identify the gravitini $E_M{}^{1\halpha}$ and $E_M{}^{2\halpha}$, as well as the dilatini and Ramond-Ramond bispinor by imposing various purely conventional constraints on top of $\kappa$-symmetry \cite{Wulff:2016tju}. From these data, one can identify the generalized supervielbein up to its local tangent space symmetries (which include the double Lorentz group).

\item The duality transformations from the GS action determine $\EE'_{MN}$ from $\EE_{MN}$, 
but this does not allow one to completely determine the orthosymplectic element $\cU$.
There is residual ambiguity corresponding \emph{precisely} to the elements not appearing
explicitly in $\EE_{MN}$ (and thus the GS action) --- the gravitini, dilatini, Ramond-Ramond bispinor (plus the extra local gauge symmetries). We merely guessed the simplest form of $\cU$.
\end{itemize}
But these issues are related! The simple choice of $\cU$ turns out to leave the generalized flux unchanged. Since the $\kappa$-symmetry constraints are \emph{already} encoded in the fluxes, these are maintained as well. Hence, $\kappa$-symmetry is preserved under non-abelian T-duality.\footnote{To put it another way, the simple choice of $\cU$ turns out to be the one that leads to the same choices of gravitini, dilatini, and Ramond-Ramond bispinor made in \cite{Wulff:2016tju}.}

\subsection{The role of the dilaton and modified / generalized double field theory}
\label{S:SNATD.dilaton}
We have not addressed how the dilaton changes under the duality.\footnote{From the perspective of the $\sigma$-model, the dilaton is an additional field added in order to restore Weyl invariance at the one-loop level. From the perspective of supergravity, the dilaton is a scalar field whose supersymmetry variation gives the dilatini. The perspective here is analogous to the supergravity point of view.} We will do this momentarily, but first, let us make a brief digression on the subject of what we call \emph{generalized double field theory}.
Recall that the DFT dilaton $\Phi$ (here a scalar density of weight 1) can be used to construct a flux
\begin{align}\label{E:DilatonFlux}
\cF_\cA = \cV_\cA{}^\cM \pa_\cM \log\Phi 
    + \pa^\cM \cV_{\cM \cA}~.
\end{align}
Upon solving the section condition, the generalized dilaton is related to a conventional dilaton $e^{-2\varphi}$ via the superspace measure, $\Phi = e^{-2 \varphi} \sdet E_M{}^A$.
Just as generalized supergravity \cite{Arutyunov:2015mqj, Wulff:2016tju} relaxes the assumption that a dilaton exists, one can define generalized double field theory by relaxing the assumption that a generalized dilaton exists. Then one replaces the flux \eqref{E:DilatonFlux} with
\begin{align}\label{E:GenDilatonFlux}
\cF_\cA = \cV_\cA{}^\cM \cX_\cM
    + \pa^\cM \cV_{\cM \cA}~.
\end{align}
This is written in terms of a vector field $\cX_\cM$ which \emph{a priori} obeys no particular
constraints. In order for $\cF_\cA$ to be a scalar under generalized diffeomorphisms, $\cX_\cM$ should transform as
\begin{align}\label{E:delta.cX}
\delta_\xi \cX_\cM = \dLie_\xi \cX_\cM + \pa_\cM \pa^\cN \xi_\cN~.
\end{align}
What distinguishes the choice of $\cX_\cM$ is that one requires $\cF_\cA$ to obey the same properties as it did when the dilaton existed. That is, we impose the same constraints and the same Bianchi identities. Viewed in this way, \emph{$\cX_\cM$ is defined in terms of $\cF_\cA$.}

What exactly does this mean? The flux tensors $\cF_{\cA \cB \cC}$ and $\cF_\cA$ obey the Bianchi identities
\begin{subequations}
\label{E:FluxBianchi}
\begin{alignat}{2}
\label{E:FluxBianchi.1}
\cZ_{\cA \cB \cC \cD} &:=
    4 \,D_{[\cA} \cF_{\cB \cC \cD]} + 3 \cF_{[\cA \cB|}{}^{\cE} \cF_{\cE |\cC \cD]} &\quad&= 0~, \\
\label{E:FluxBianchi.2}
\cZ_{\cA \cB} &:= 2 D_{[\cA} \cF_{\cB]} + \cF_{\cA \cB}{}^\cC \cF_\cC + D^\cC \cF_{\cC \cA \cB} &\quad&= 0~, \\
\label{E:FluxBianchi.3}
\cZ &:= D^\cA \cF_{\cA} + \frac{1}{2} \cF^\cA \cF_\cA + \frac{1}{12} \cF^{\cA \cB \cC} \cF_{\cC \cB \cA} &\quad&= 0~.
\end{alignat}
\end{subequations}
The expression for $\cZ_{\cA\cB}$ \eqref{E:FluxBianchi.2} can be rewritten in two equivalent ways
\begin{align}
\cZ_{\cM \cN} = 2\, \pa_{[\cM} \cX_{\cN]} + \cX^\cP \pa_\cP \cV_{\cM}{}^\cA \cV_{\cA \cN} \quad
\implies \quad
\dLie_\cX \cV_\cM{}^\cA = \cV_\cM{}^\cB \cZ_\cB{}^\cA~,
\end{align}
while $\cZ$ in \eqref{E:FluxBianchi.3} can be rewritten as
\begin{align}
\cZ = \pa^\cM \cX_\cM + \frac{1}{2} \cX^\cM \cX_\cM  \quad \implies \quad
\dLie_\cX \cX^\cM = \pa^\cM \cZ~.
\end{align}
When $\cZ_{\cA \cB}$ and $\cZ$ vanish, $\cX^\cM$ has an obvious interpretation as a generalized Killing vector.

Generalized double field theory is nearly (perhaps completely) equivalent to \emph{modified double field theory (mDFT)} \cite{Sakatani:2016fvh}. The distinction is that mDFT imposes the section condition on the index of $\cX^\cM$, so that $\cX^\cM \cX_\cM = 0$ and $\cX^\cM \otimes \pa_\cM = 0$. Upon doing so, $\cZ$ vanishes and $\cZ_{\cM \cN}$ vanishes only if $\cX_\cM$ is the gradient of some other field. It is unclear to us whether the reverse is true, whether imposing $\cZ = \cZ_{\cM \cN} = 0$ necessarily implies the section condition on $\cX_\cM$. If it is, then mDFT and generalized DFT should be identical.\footnote{In mDFT, one has both a vector $\cX_\cM$ and the dilaton gradient $\pa_\cM \Phi$, but in principle one could just absorb the latter into the former to arrive at the formulation discussed here.  It was argued in \cite{Sakamoto:2017wor} that mDFT can always be interpreted as conventional DFT where the dilaton carries a linear dependence on some of the winding coordinates, while still satisfying the section condition. This forces the generalized vielbein to be independent of those winding coordinates.}

Both generalized DFT and mDFT lead to generalized supergravity upon solving the section condition, where we define
\begin{align}
\cX_M = -2 (X_M - K^N B_{NM}) + \pa_M \log\sdet E_N{}^A~, \qquad
\cX^M = -2 K^M~.
\label{E:XKdefs}
\end{align}
The measure factor in the first equation accounts for the inhomogeneous term in \eqref{E:delta.cX} so that both $X_M$ and $K^M$ are a conventional one-form and vector respectively.
The explicit factor of the $B$-field ensures that $X_M$ is inert under the $B$-field gauge transformations. The factors of $-2$ are chosen so that $X_M = \pa_M \varphi$ when a dilaton exists. Now one can show that if the modified flux tensor $\cF_\cA$ obeys the same Bianchi identities and same constraints as before, then the vector $K^M$ turns out to be a Killing vector in conventional superspace and $X_M$ is a one-form whose spinorial components are the dilatini. The other relations discussed in generalized supergravity \cite{Arutyunov:2015mqj, Wulff:2016tju} can be derived in like manner from generalized / modified DFT. We hope to elaborate on this in superspace in the future; the bosonic proof of this was given in \cite{Sakatani:2016fvh}.

Returning to the original question: how does the dilaton or, more generally, $X$ and $K$ change under duality? Factorizing the supervielbein as in \eqref{E:VielbeinDecomposedDuality}, the dilaton flux becomes
\begin{align}
\cF_\cA &= \wtcV_\cA{}^\wtcM 
    \Big(
    \mathring \cV_{\wtcM}{}^{\cN} \cX_\cN
    + \pa^\cN \mathring \cV_{\cN\wtcM}
    \Big)
    + \pa^\wtcM \wtcV_{\wtcM\cA}~.
\end{align}
We posit that the dilaton flux should remain unchanged. If so, then the element in parentheses must be fixed. In the spectator directions, we have simply
$\cX'_\uM = \cX_\uM$ and 
$\cX'{}^\uM = \cX^\uM$.
In the isometry directions, we find more intricate relations
\begin{align}
\cX'{}^{\Iso1}
    = \cX^{\Iso1} + \tilde D^{\Iso1} \sdet \tilde e_{\Iso2}{}^{\cIso2}
~, \qquad
\cX'_{\Iso1} 
    = \cX_{\Iso1} + D_{\Iso1} \sdet e_{\Iso2}{}^{\cIso2}
    + 2 f_{\Iso1 \Iso2}{}^{\Iso2} (-)^{\grad{\Iso2}}
    - \cX^{\Iso2}  f_{\Iso2\Iso1}{}^{\Iso3} \nu_{\Iso3}
\end{align}
where for convenience we have defined
\begin{alignat}{3}
D_{\Iso1} &= e_{\Iso1}{}^{\cIso1} \pa_{\cIso1}~, 
&\qquad
\cX_{\Iso1} &= e_{\Iso1}{}^{\cIso1} \cX_{\cIso1}~, 
&\qquad
\cX'_{\Iso1} &= \tilde e_{\Iso1}{}^{\cIso1} \cX'_{\cIso1}~, 
\eol
D^{\Iso1} &= \pa^{\cIso1}\times \tilde e_{\cIso1}{}^{\Iso1} ~,
&\qquad
\cX^{\Iso1} &= \cX^{\cIso1} e_{\cIso1}{}^{\Iso1}~,
&\qquad
\cX'{}^{\Iso1} &= \cX'{}^{\cIso1} \tilde e_{\cIso1}{}^{\Iso1}~.
\end{alignat}
The next step is to strip out the density behavior of $\cX_\cM$ and $\cX'_\cM$ by subtracting
factors of $\pa_\cM \log E_M{}^A$ and $\pa_\cM \log \sdet E'_{M'}{}^A$.
We explicitly prime the index $M$ of the dual model to emphasize that it involves a different
coordinate set $(Z, \tilde Y)$ from the original model $(Z, Y)$. Here we will need the explicit transformation of the supervielbein in terms of $X_M{}^N$. This leads to
\begin{align}\label{E:sdetE'}
\sdet E'_A{}^{M'}(Z,\tilde Y)
    &= \sdet E_A{}^M(Z,Y) \times \sdet \widehat{\EE}_{\Iso1 \Iso2}(Z, \tilde Y)
    \times \sdet e_{\cIso1}{}^{\Iso1}(Y) \times \sdet \tilde e_{\cIso1}{}^{\Iso1}(\tilde Y)
\end{align}
where we have exhibited the dependence on the coordinates $Y$, $\tilde Y$, and the spectator coordinates $Z$. From these relations, we find
\begin{subequations}
\begin{align}
\Big(\cX'_\uM - \pa_\uM \log E'\Big) &= \Big(\cX_\uM - \pa_\uM \log E\Big) 
    + \pa_\uM \log \sdet \widehat{\EE}_{\Iso1 \Iso2}~, \\
\Big(\cX'{}^{\Iso1} - D^{\Iso1} \log E'\Big)
    &= \cX^{\Iso1} + D^{\Iso1} \log\sdet \widehat{\EE}_{\Iso2 \Iso3}
~, \\
\cX'{}^{\uM} &= \cX^{\uM}~, \\
\cX'_{\Iso1} 
    &= \Big(\cX_{\Iso1} - D_{\Iso1} \log E\Big)
    + 2 f_{\Iso1 \Iso2}{}^{\Iso2} (-)^{\grad{\Iso2}}
    - \cX^{\Iso2} f_{\Iso2 \Iso1}{}^{\Iso3} \nu_{\Iso3}
\end{align}
\end{subequations}
Now we may identify the $X$ and $K$ fields. In the original model, we take
\begin{alignat}{2}
\cX_\uM - \pa_\uM \log E &= -2 \widehat X_{\uM}~, 
&\qquad
\cX^{\uM} &= -2 K^{\uM}~, \\
\cX_{\Iso1} - D_{\Iso1} \log E &= -2 \widehat X_{\Iso1}~, 
&\qquad
\cX^{\Iso1} &= -2 K^{\Iso1}
\end{alignat}
where we denote $\widehat X_M = X_M - K^N B_{NM}$ for convenience. The indices $\Iso1$ are flattened
with $e_{\cIso1}{}^{\Iso1}$. In the dual model, we have the somewhat more complicated expressions
\begin{alignat}{2}
\cX'_\uM - \pa_\uM \log E &= -2 \widehat X'_{\uM}~, 
&\qquad
\cX'{}^{\uM} &= -2 K'{}^{\uM}~, \\
\cX'{}^{\Iso1} - D'{}^{\Iso1} \log E' &= -2 \widehat X'{}^{\Iso1}~, 
&\qquad
\cX'_{\Iso1} &= -2 K'_{\Iso1}~.
\end{alignat}
Here we must remember that the duality involves a passive coordinate transformation so
\begin{align}
X'_M 
= 
\begin{pmatrix}
X'_\uM \,&
X'{}^{\cIso1}
\end{pmatrix}~, \qquad
K'{}^M 
=
\begin{pmatrix}
K'{}^\uM \,&
K'_{\cIso1} (-)^{\grad{\cIso1}} 
\end{pmatrix}~.
\end{align}
Rather then perform index gymnastics with the isometry coordinates, we will simply express
relations in terms of the flattened isometry index, even though it is in the ``wrong'' position:
\begin{subequations}
\label{E:XK.trafo}
\begin{align}
\widehat X'_\uM &= \widehat X_\uM -\frac{1}{2} \pa_\uM \log \sdet \widehat{\EE}_{\Iso2 \Iso3}~, \\
\widehat X'{}^{\Iso1}
    &= K^{\Iso1} - \frac{1}{2} D^{\Iso1} \log\sdet \widehat{\EE}_{\Iso2 \Iso3}
~, \\
K'{}^{\uM} &= K^{\uM}~, \\
K'_{\Iso1}
    &= \widehat X_{\Iso1}
    - f_{\Iso1 \Iso2}{}^{\Iso2} (-)^{\grad{\Iso2}}
    - K^{\Iso2} f_{\Iso2 \Iso1}{}^{\Iso3} \nu_{\Iso3}~.
\end{align}
\end{subequations}

A rather strict check of these relations is this: $K \lrcorner \widehat X$ should vanish in the dual model when it vanishes in the original model. This is a consequence of T-duality preserving $\kappa$-symmetric Green-Schwarz actions. We find
\begin{align}\label{E:KX.calc1}
K' \lrcorner \widehat X'
    - K \lrcorner \widehat X 
    &= - \frac{1}{2} K^{\uM} \pa_\uM \log \sdet \widehat{\EE}_{\Iso2 \Iso3}
    - K^{\Iso1} f_{\Iso1 \Iso2}{}^{\Iso2} (-)^{\grad{\Iso2}}
    \eol & \quad
    + \frac{1}{2} D^{\Iso1} \log\sdet \widehat{\EE}_{\Iso2 \Iso3} \times
        \Big(f_{\Iso1 \Iso4}{}^{\Iso4} (-)^{\grad{\Iso4}} 
            + K^{\Iso4} f_{\Iso4 \Iso1}{}^{\Iso5} \nu_{\Iso5} 
            - \widehat X_{\Iso1} \Big)
\end{align}
The second line can be rewritten as
\begin{align}\label{E:KX.calc2}
\frac{1}{2} \widehat{\EE}^{\Iso3 \Iso2} 
        \Big(- f_{\Iso2 \Iso3}{}^{\Iso1} f_{\Iso1 \Iso4}{}^{\Iso4} (-)^{\grad{\Iso4}} 
            - f_{\Iso2 \Iso3}{}^{\Iso1} K^{\Iso4} f_{\Iso4 \Iso1}{}^{\Iso5} \nu_{\Iso5} 
            + f_{\Iso2 \Iso3}{}^{\Iso1} \widehat X_{\Iso1} \Big)~.
\end{align}
The first term drops out immediately using the Jacobi identity.
To evaluate the remaining terms requires a few features of $K$ and $X$ that arise in generalized supergravity. First, $G_{MN}$ and $B_{MN}$, once flattened as in \eqref{E:FlattenedGB}, are independent of $Y^{\cIso1}$. The same should be true of $K$ and $X$ in generalized supergravity, because their various components appear in the torsion and curvatures.\footnote{We are speaking here of the flattened versions $K^{\Iso1}$ and $X_{\Iso1}$.} This means that the isometry condition on $G_{MN}$ implies in particular that
\begin{align}
0 = K^\uM \pa_\uM G_{\Iso1 \Iso2}
    + 2 K^{\Iso3} f_{\Iso3 (\Iso1}{}^{\Iso4} G_{\Iso4 \Iso2)}~.
\end{align}
For the $B$-field, the relevant relation we need is $\rd X = - K \lrcorner H$. From this, one can show that
\begin{align}
-f_{\Iso1 \Iso2}{}^{\Iso3} \widehat X_{\Iso3}
    = K^\uM \pa_\uM B_{\Iso1 \Iso2}
    + 2 K^{\Iso3} f_{\Iso3 [\Iso1}{}^{\Iso4} B_{\Iso4 \Iso2]}~.
\end{align}
Taking the difference between these two relations lets one rewrite the right-hand side in terms of
$\EE = G-B$. Introducing the Lagrange multiplier field converts $\EE_{\Iso1 \Iso2}$ to $\widehat \EE_{\Iso1 \Iso2}$. Using the Jacobi identity simplifies the result to
\begin{align}
f_{\Iso1 \Iso2}{}^{\Iso3} \Big(\hat X_{\Iso3} - K^{\Iso4} f_{\Iso4 \Iso3}{}^{\Iso5} \nu_{\Iso5}\Big)
        = K^\uM \pa_\uM \widehat{\EE}_{\Iso1 \Iso2}
    + K^{\Iso3} \Big(f_{\Iso3 \Iso1}{}^{\Iso4} \widehat{\EE}_{\Iso4 \Iso2} 
        + f_{\Iso3 \Iso2}{}^{\Iso4} \widehat\EE_{\Iso1 \Iso4} \Big)
\end{align}
where we have suppressed gradings in the final term for readability. The term on the left-hand side is exactly what remains in \eqref{E:KX.calc2}. Substituting this expression, we find the complete cancellation of the remainder of the right-hand side of \eqref{E:KX.calc1}.

A specific case of interest is when we start with a model with a dilaton, a case analyzed in
\cite{Borsato:2018idb}. Then $K=0$ and $\widehat X = X = \rd \varphi$. The equations \eqref{E:XK.trafo} can be rewritten
\begin{alignat}{2}
\widehat X'_\uM &= \pa_\uM \Big( \varphi -\frac{1}{2} \log \sdet \widehat \EE_{\Iso2 \Iso3} \Big)~,
&\qquad \qquad
K'{}^{\uM} &=0 ~,
\eol
\widehat X'{}^{\Iso1}
    &= D^{\Iso1} \Big( \varphi - \frac{1}{2}  \log\sdet \widehat \EE_{\Iso2 \Iso3} \Big)~,
&\qquad
K'_{\Iso1}
    &= D_{\Iso1} \varphi
    - f_{\Iso1 \Iso2}{}^{\Iso2} (-)^{\grad{\Iso2}}~,
\end{alignat}
where we have used $D^{\Iso1} \varphi = 0$.
Now the dual theory satisfies the conventional supergravity constraints when $K'=0$, so
$D_{\Iso1} \varphi = f_{\Iso1 \Iso2}{}^{\Iso2} (-)^{\grad{\Iso2}}$.
This \emph{imposes} a requirement for how the dilaton should depend on the coordinates we
are dualizing. To solve this, we could extract from the dilaton a purely $Y$-dependent piece that generates this term, i.e.
\begin{align}
\varphi(Z, Y) = \varphi_0(Z) + \Delta(Y)~, \qquad
D_{\Iso1} \Delta = f_{\Iso1 \Iso2}{}^{\Iso2} (-)^{\grad{\Iso2}}~.
\end{align}
In general there is no local obstruction to the existence of $\Delta$, since it obeys the consistency condition $[D_{\Iso1}, D_{\Iso2}] \Delta = -f_{\Iso1 \Iso2}{}^{\Iso3} D_{\Iso3} \Delta$ by virtue of the Jacobi identity.  Now the dual dilaton can be identified as
\begin{align}
\varphi'(Z, \tilde Y) = \varphi_0(Z) - \frac{1}{2} \log \sdet \widehat \EE_{\Iso2 \Iso3}(Z, \tilde Y)
\end{align}
so that $\widehat X' = X' = \rd \varphi'$.

\subsection{Component description}
The previous discussion has been at the level of superspace. In order to make contact with the literature on fermionic and bosonic T-dualities of bosonic backgrounds, we should rewrite our expressions at the component level. Here we must already make a distinction between bosonic and fermionic isometries that arise from the algebra of supervectors 
\begin{equation}\label{E:SuperKillingAlgebra}
    [k_{\Iso1}, k_{\Iso2}] = f_{\Iso1 \Iso2}{}^{\Iso3} k_{\Iso3}\,:
\end{equation}%
\begin{itemize}
\item Bosonic isometries are treated as conventional vectors $k_{\iso1} = k_{\iso1}{}^{m} \pa_m$ acting on bosonic coordinates. These arise by taking the $\theta=0$ parts of the bosonic supervectors $k_{\iso1} = k_{\iso1}{}^M \pa_M$. Since $k_{\iso1}{}^\mu$ is fermionic, it must be at least linear in $\theta$, and so can be discarded.
\item Fermionic isometries are described by \emph{commuting spinors} $\veps_{\fIso1}{}^{i \halpha}$
with $i=1,2$. These arise by flattening the fermionic isometries $k_{\fIso1}$ with the gravitino one-forms and setting $\theta=0$:
\begin{align}
\veps_{\fIso1}{}^{i\halpha} = k_{\fIso1}{}^{M} E_M{}^{i \halpha} \vert_{\theta=0}
    = k_{\fIso1}{}^{\mu} E_\mu{}^{i \halpha} \vert_{\theta=0}~.
\end{align}
Since $k_{\fIso1}{}^m$ is fermionic (being linear in $\theta$), it can be discarded.
\end{itemize}

As is well known, bosonic isometries can arise as bilinears of fermionic ones. To describe this, we first rewrite \eqref{E:SuperKillingAlgebra} with flat indices. Under a covariant Lie derivative generated by $k_{\Iso1}$, the supervielbein is merely rotated,
\begin{align}
\cL^{\rm cov}_{\Iso1} E_M{}^A 
:= k_{\Iso1}{}^N \cD_N E_M{}^A + \pa_M k_{\Iso1}{}^N E_N{}^A
= - E_M{}^B (\lambda_{\Iso1})_B{}^A
\end{align}
where $\lambda_{\Iso1}$ is a Lorentz transformation.
This follows from the Green-Schwarz action, since invariance of $G_{MN}$ implies the result for $E_M{}^a$; for $E_M{}^{i \halpha}$, one must employ the torsion constraints (which arise from $\kappa$-symmetry). This expression may equivalently be written
\begin{align}
\cD_B k_{\Iso1}{}^{A} (-)^{\grad{B} \grad{\Iso1}} 
    + k_{\Iso1}{}^C T_{C B}{}^A = -(\lambda_{\Iso1})_B{}^A~.
\end{align}
The algebra of Killing supervectors can then be rewritten (with gradings suppressed)
\begin{align}\label{E:FlatKillingAlgebra}
f_{\Iso1 \Iso2}{}^{\Iso3} k_{\Iso3}{}^A
    = k_{\Iso1}{}^B k_{\Iso2}{}^C T_{B C}{}^A
    - k_{\Iso1}{}^B (\lambda_{\Iso2})_B{}^A
    + k_{\Iso2}{}^B (\lambda_{\Iso1})_B{}^A~.
\end{align}
These expressions lead immediately to several useful results. First,
taking \eqref{E:FlatKillingAlgebra} with $A=a$ and $\Iso1 \Iso2 = \fIso1 \fIso2$, we find
how a bosonic Killing vector is generated from two Killing spinors:
\begin{align}\label{E:SUSYIsometries}
i f_{\fIso1 \fIso2}{}^{\iso3} k_{\iso3}{}^a
    = \bar \veps^1_{\fIso1} \gamma^a \veps_{\fIso2}^1
    + \beta_\Lambda \,\bar \veps^2_{\fIso1} \gamma^a \veps_{\fIso2}^{2}
    = 
\begin{cases}
    \bar \veps^1_{\fIso1} \gamma^a \veps_{\fIso2}^1
    + \bar \veps^2_{\fIso1} \gamma^a \veps_{\fIso2}^{2}
    & \text{IIB/IIA$^*$} \\[2ex]
    \bar \veps^1_{\fIso1} \gamma^a \veps_{\fIso2}^1
    - \bar \veps^2_{\fIso1} \gamma^a \veps_{\fIso2}^{2}
    & \text{IIA/IIB$^*$} 
\end{cases}
\end{align}
where $k_{\iso3}{}^a = k_{\iso3}{}^m e_m{}^a$. The chirality of $\veps^1$ is fixed while that of $\veps^2$ depends on whether one lies in a IIB/IIB$^*$ or IIA/IIA$^*$ duality frame. A crucial point is that the fermionic indices appear \emph{symmetrically} in \eqref{E:SUSYIsometries} and the two commuting spinors may be taken to be the same:
\begin{align}
i f_{\fIso1 \fIso1}{}^{\iso3} k_{\iso3}{}^a
    = \bar \veps_{\fIso1}^1 \gamma^a \veps_{\fIso1}^1
    + \beta_\Lambda \bar \veps_{\fIso1}^2 \gamma^a \veps_{\fIso1}^{2}~.
\end{align}
Taking $A$ to be spinorial in \eqref{E:FlatKillingAlgebra}, we find the other useful relations
\begin{align}
f_{\iso1 \fIso1}{}^{\fIso2} \veps_{\fIso2}^1
    &= 
    -\frac{1}{8} k_{\iso1}{}^b H_{bcd}  \,\gamma^{cd} \veps_{\fIso1}^1
    + \frac{\beta_\Lambda}{16} e^{\varphi} k_{\iso1}{}^b 
    C^{-1} \widehat{\slashed\cF} \gamma_b \veps_{\fIso1}^{2} 
    - \slashed{\lambda}_{\iso1} \veps_{\fIso1}^{1}~, \\
f_{\iso1 \fIso1}{}^{\fIso2} \veps_{\fIso2}^2
    &= 
    + \frac{1}{8} k_{\iso1}{}^b H_{bcd} \gamma^{cd} \veps_{\fIso1}^2
    - \frac{1}{16} e^{\varphi}\, k_{\iso1}{}^b 
    C^{-1} \widehat{\slashed\cF}{}^T \gamma_b \veps_{\fIso1}^{1} 
    - \slashed{\lambda}_{\iso1} \veps_{\fIso1}^{2}
\end{align}
with $\widehat{\slashed\cF}$ given by \eqref{E:cF12}.

The vielbein and $B$-field expressions match what the computations purely from 
the bosonic dualities would give,
\begin{align}\label{E:dualEandB}
e'{}^a &= 
    \rd x^\um \Big(
    e_\um{}^a
    - \EE_{\um \iso1} \widehat{\EE}^{\iso1 \iso2} e_{\iso2}{}^a
    \Big)
    + \rd \nu_{\iso1} \widehat{\EE}^{\iso1 \iso2} e_{\iso2}{}^a~, \\
B' &= \frac{1}{2} \rd x^\um \wedge \rd x^\un B_{\un \um}
    - \frac{1}{2} \widehat{\EE}^{\iso1 \iso2}
    \Big(
    \rd \nu_{\iso2} + \rd x^{\un} \bar \EE_{\un \iso2}
    \Big)\wedge 
        \Big(
    \rd \nu_{\iso1} - \rd x^{\um} \EE_{\um \iso1}
    \Big) ~,
\end{align}
indicating that the fermionic T-dualities have no effect on them \cite{Berkovits:2008ic}.
Where the fermionic T-dualities matter is for the Ramond-Ramond background and for the dilaton, where we find
\begin{align}
\varphi' &= \varphi_0 - \frac{1}{2} \log \det \widehat{\EE}_{\iso1 \iso2}
    + \frac{1}{2} \log \det \widehat{\EE}_{\fIso1 \fIso2}~, \\
e^{\varphi'} \widehat \cF'{}^{1\halpha \, 2\hbeta}
    &= 
    \Big( e^{\varphi} \widehat \cF^{1\halpha \, 2\hgamma} 
    - 32i\, E_{\fIso1}{}^{1\halpha} \,\widehat\EE{}^{\fIso1 \fIso2} E_{\fIso2}{}^{2\hgamma}
    \Big) ({\bm{\slashed\Lambda}}^{-1})_{\hgamma}{}^{\hbeta}~.
\end{align}
The additional Lorentz transformation above is given in vectorial form as
\begin{align}
{\bm\Lambda}_a{}^b = \delta_a{}^b - 2 \,\widehat{\bar \EE}{}^{\iso1 \iso2} e_{\iso2}{}^b e_{\iso1 a}
\end{align}
and depends purely on the bosonic isometries.

The expression for the Ramond-Ramond bispinor involves 
$E_{\fIso1}{}^{1 \halpha} = e_{\fIso1} \lrcorner E^{1 \halpha}$, but it would be more useful
to rewrite this in terms of $k_{\fIso1} \lrcorner E^{1 \halpha} = \veps_{\fIso1}^{1\halpha}$.
To do that, we need to apply the adjoint action of $g$ to the isometry indices.
Recall we have
$\widehat \EE_{\Iso1 \Iso2} = E_{\Iso1}{}^a E_{\Iso2}{}^b \eta_{ab}
    - B_{\Iso1 \Iso2}
    - f_{\Iso1 \Iso2}{}^{\Iso3} \nu_{\Iso3}
$ where we have expanded the supervielbein in the original model as
$E^a = \rd Z^\uM E_\uM{}^a + e^{\Iso1} E_{\Iso1}{}^a$.
In choosing the original coordinate system \eqref{E:FlattenedGB}, we
expanded in terms of the left-invariant vector fields. The right-invariant vector
fields are
$\rd g g^{-1} = \rd Y^{\cIso1} k_{\cIso1}{}^{\Iso1} t_{\Iso1}$
and these are related to $e_{\cIso1}{}^{\Iso1}$ by
\begin{align}
k_{\Iso1} \lrcorner e^{\Iso2} = (\Adj g^{-1})_{\Iso1}{}^{\Iso2} \qquad
\text{where} \quad g\,\xi^{\Iso1} t_{\Iso1} \,g^{-1}  = \xi^{\Iso1} (\Adj g)_{\Iso1}{}^{\Iso2} t_{\Iso2}~.
\end{align}
Applying the adjoint action to $\widehat \EE_{\Iso1 \Iso2}$ gives
\begin{align}
\cQ_{\Iso1 \Iso2} &:= (\Adj g)_{\Iso1}{}^{\Iso1'} (\Adj g)_{\Iso2}{}^{\Iso2'} \widehat \EE_{\Iso1' \Iso2'} 
    = k_{\Iso1}{}^a k_{\Iso2}{}^b \eta_{ab}
    + k_{\Iso1} \lrcorner k_{\Iso2} \lrcorner B
    - f_{\Iso1 \Iso2}{}^{\Iso3} (\Adj g^{-1})_{\Iso3}{}^{\Iso4} \nu_{\Iso4}
\end{align}
where we suppressed gradings in the first equality.
Since we only care about purely bosonic expressions, we have simply
\begin{align}
\cQ_{\iso1 \iso2} &= k_{\iso1}{}^m k_{\iso2}{}^n (g_{mn} - B_{mn}) 
    - f_{\iso1 \iso2}{}^{\iso3} (\Adj g^{-1})_{\iso3}{}^{\iso4} \nu_{\iso4}~, \\
\cQ_{\fIso1 \fIso2} &= -\Lambda_{\fIso1 \fIso2} 
    - f_{\fIso1 \fIso2}{}^{\iso3} (\Adj g^{-1})_{\iso3}{}^{\iso4} \nu_{\iso4}
\end{align}
where $\Lambda_{\fIso1 \fIso2} := -k_{\fIso1} \lrcorner k_{\fIso2} \lrcorner B = k_{\fIso1}{}^M B_{M N} k_{\fIso2}{}^N$.
The dilaton can be written
\begin{align}
\varphi' = \varphi_0 - \frac{1}{2} \log \det \widehat \EE_{\iso1 \iso2} 
    + \frac{1}{2} \log\det \cQ_{\fIso1 \fIso2}
    + \log \det (\Adj g)_{\fIso1}{}^{\fIso2}
\end{align}
The Ramond-Ramond bispinor becomes
\begin{align}
e^{\varphi'} \widehat \cF'{}^{1\halpha \, 2\hbeta}
    = 
    \Big( e^{\varphi} \widehat \cF^{1\halpha \, 2\hgamma} 
    + 32i\, \veps_{\fIso1}^{1\halpha} \,(\cQ^{-1})^{\fIso1 \fIso2} \veps_{\fIso2}^{2\hgamma}
    \Big) ({\bm{\slashed\Lambda}}^{-1})_{\hgamma}{}^{\hbeta}~.
\end{align}
An extra sign has appeared because we use the inverse $(\cQ^{-1})^{\fIso1 \fIso2}$  rather than the graded inverse. 

What can we say about $\cQ_{\fIso1 \fIso2}$? While it is fully characterized in superspace, on the bosonic background it can really only be described by its derivative. From the definition of $\Lambda_{\fIso1 \fIso2}$ in superspace, one can show that
\begin{align}
\rd \Lambda_{\fIso1 \fIso2} &= 
    - k_{\fIso1} \lrcorner k_{\fIso2} \lrcorner H
    - f_{\fIso1 \fIso2}{}^{\iso3} k_{\iso3} \lrcorner B~.
\end{align}
This follows because the $B$-field in \eqref{E:FlattenedGB} (like the metric) obeys $\cL_{k_{\Iso1}} B = 0$. The more general case is discussed in Appendix \ref{A:GaugedSigmaModels}.
This leads to
\begin{align}\label{E:dQ}
\rd \cQ_{\fIso1 \fIso2}
    = 
    k_{\fIso1} \lrcorner k_{\fIso2} \lrcorner H
    + f_{\fIso1 \fIso2}{}^{\iso3} \Big(k_{\iso3} \lrcorner B
    - (\Adj g^{-1})_{\iso3}{}^{\iso4} \rd \nu_{\iso4}
    - k^{\iso4} f_{\iso4 \iso3}{}^{\iso3'} (\Adj g^{-1})_{\iso3'}{}^{\iso4} \nu_{\iso4}
    \Big)~.
\end{align}
The quantity $\cQ_{\fIso1 \fIso2}$ should depend on the spectator coordinates,
the dual $\tilde y$ coordinate (via $\nu_{\iso4}$), and on the coordinates $y$ only
via the adjoint action (since $\widehat \EE_{\fIso1 \fIso2}$ was $y$-independent). This means
\begin{align}
k_{\iso1} \lrcorner \rd \cQ_{\fIso1 \fIso2}
    = -f_{\iso1 \fIso1}{}^{\fIso3} \cQ_{\fIso3 \fIso2}
    -f_{\iso1 \fIso2}{}^{\fIso3} \cQ_{\fIso1 \fIso3}~.
\end{align}
The terms involving $\nu$ already have this form, since
$k_{\iso1} \lrcorner \rd \nu_{\iso4} = 0$ 
and the the Jacobi identity allows one to rewrite the pair of structure constants appropriately.
For the terms involving $H$ and $B$, it helps to observe that
$k_{\Iso1} \lrcorner H = - \rd ( k_{\Iso1} \lrcorner B)$
from which the desired property can be deduced. A key step is to exploit
\begin{align}\label{E:kkkH}
k_{\Iso1} \lrcorner k_{\Iso2} \lrcorner k_{\Iso3} \lrcorner H
    - 3 f_{[\Iso1\Iso2|}{}^{\Iso4} k_{\Iso4} \lrcorner k_{|\Iso3]} \lrcorner B = 0
\end{align}
which follows from the explicit form of $H$ in terms of the $B$ given in \eqref{E:FlattenedGB}.
The expression \eqref{E:dQ} can be interpreted purely as a bosonic equation once we address the first term involving $H$. It is given by
\begin{align}
k_{\fIso1} \lrcorner k_{\fIso2} \lrcorner H
    &=
    i \,\rd x^m \Big(
    \bar \veps_{\fIso1}^1 \gamma_m \veps_{\fIso2}^1
    - \beta_\Lambda \bar\veps_{\fIso1}^2 \gamma_m \veps_{\fIso2}^2
\Big) 
= 
    i\,\rd x^m \times 
    \begin{cases}
    \bar \veps^1_{\fIso1} \gamma_m \veps_{\fIso2}^1
    - \bar \veps^2_{\fIso1} \gamma_m \veps_{\fIso2}^{2}
    & \text{IIB/IIA$^*$} \\[2ex]
    \bar \veps^1_{\fIso1} \gamma_m \veps_{\fIso2}^1
    + \bar \veps^2_{\fIso1} \gamma_m \veps_{\fIso2}^{2}
    & \text{IIA/IIB$^*$} 
\end{cases}~.
\end{align}
Note the crucial relative sign difference with \eqref{E:SUSYIsometries}. The importance of this sign difference was already noted in the context of non-abelian fermionic T-duality in \cite{Astrakhantsev:2021rhj}.

\paragraph{Abelian fermionic T-duality.}
The fermionic T-duality discussed by Berkovits and Maldacena \cite{Berkovits:2008ic} corresponds to a single abelian fermionic isometry, for which the left-hand side of \eqref{E:SUSYIsometries} vanishes. No bosonic isometries are involved and so the vielbein and $B$-field are unchanged. However, the dilaton and Ramond-Ramond complex change as
\begin{align}
\varphi' &= \varphi_0
    + \frac{1}{2} \log \det \cQ_{\fIso1 \fIso1}~, \\
e^{\varphi'} \widehat \cF'{}^{1\halpha \, 2\hbeta}
    &= e^{\varphi} \widehat \cF^{1\halpha \, 2\hgamma} 
    + 32i\, \veps_{\fIso1}^{1\halpha} \,(\cQ^{-1})^{\fIso1 \fIso1} \veps_{\fIso1}^{2\hgamma}~.
\end{align}
Note that there is no Lorentz factor since ${\bm\Lambda}_a{}^b = \delta_a{}^b$. The function $\cQ$ obeys
\begin{align}
-i \,\pa_m \cQ_{\fIso1 \fIso1}
    = 
    \begin{cases}
    \bar \veps^1_{\fIso1} \gamma_m \veps_{\fIso1}^1
    - \bar \veps^2_{\fIso1} \gamma_m \veps_{\fIso1}^{2}
    & \text{IIB/IIA$^*$} \\[2ex]
    \bar \veps^1_{\fIso1} \gamma_m \veps_{\fIso1}^1
    + \bar \veps^2_{\fIso1} \gamma_m \veps_{\fIso1}^{2}
    & \text{IIA/IIB$^*$} 
    \end{cases}
\end{align}
Since there is no duality in a bosonic direction, there are no dual bosonic coordinates for $\cQ$ to depend on.

\paragraph{Non-abelian fermionic T-duality.}
Slightly more generally, one can consider a single \emph{non-abelian} fermionic isometry \cite{Astrakhantsev:2021rhj}, which generates a single bosonic isometry:
\begin{align}
\{k_{\fIso1}, k_{\fIso1}\} = -i k_{\iso1}~, \qquad
[k_{\iso1}, k_{\fIso1}] = 0~, \qquad  f_{\fIso1 \fIso1}{}^{\iso1} = -i~.
\end{align}
Because we must dualize the full set of closed Killing supervectors, this is actually two dualities: a fermionic one $k_{\fIso1}$ and a bosonic one $k_{\iso1}$. 
In our conventions,
\begin{align}
k_{\iso1}
    = 
    \bar \veps_{\fIso1}^1 \gamma^a \veps_{\fIso1}^1
    + \beta_\Lambda \bar \veps_{\fIso1}^2 \gamma^a  \veps_{\fIso1}^{2}
    =
    \begin{cases}
    \bar \veps^1_{\fIso1} \gamma^a \veps_{\fIso1}^1
    + \bar \veps^2_{\fIso1} \gamma^a \veps_{\fIso1}^{2}
    & \text{IIB/IIA$^*$} \\[2ex]
    \bar \veps^1_{\fIso1} \gamma^a \veps_{\fIso1}^1
    - \bar \veps^2_{\fIso1} \gamma^a \veps_{\fIso1}^{2}
    & \text{IIA/IIB$^*$} 
\end{cases}
\end{align}
Now the expression \eqref{E:dQ} becomes
\begin{align}
\rd \cQ_{\fIso1 \fIso1}
    &= 
    k_{\fIso1} \lrcorner k_{\fIso1} \lrcorner H
    -i k_{\iso1} \lrcorner B
    + i\rd \nu_{\iso1}~.
\end{align}
This can perhaps more transparently be written in the following way:
\begin{align}\label{E:dmQ}
\pa_m \cQ_{\fIso1 \fIso1} = i (V_m - k_{\iso1}{}^n B_{nm} )~, \qquad
\pa^{\ciso1} \cQ_{\fIso1 \fIso1} = i \,\pa^{\ciso1} \nu_{\iso1}
    = i \,\tilde e_{\iso1}{}^{\ciso1}
\end{align}
where
\begin{align}\label{E:defVm}
V_{m} &= 
    \bar \veps_{\fIso1}^1 \gamma_m \veps_{\fIso1}^1
    - \beta_\Lambda \bar \veps_{\fIso1}^2 \gamma_m  \veps_{\fIso1}^{2}
    = 
\begin{cases}
    \bar \veps^1_{\fIso1} \gamma_m \veps_{\fIso1}^1
    - \bar \veps^2_{\fIso1} \gamma_m \veps_{\fIso1}^{2}
    & \text{IIB/IIA$^*$} \\[2ex]
    \bar \veps^1_{\fIso1} \gamma_m \veps_{\fIso1}^1
    + \bar \veps^2_{\fIso1} \gamma_m \veps_{\fIso1}^{2}
    & \text{IIA/IIB$^*$} 
\end{cases}
\end{align}
where $\tilde e_{\iso1}{}^{\ciso1}$ is the dual vielbein.
Note that $k_{\iso1} \lrcorner V=0$. This is apparent both from the explicit expressions
in terms of the Killing spinors but also from \eqref{E:kkkH} which collapses to
$k_{\iso1} \lrcorner k_{\fIso1} \lrcorner k_{\fIso1} \lrcorner H = 0$.
The expression for $\pa_m \cQ_{\fIso1 \fIso1}$ in \eqref{E:dmQ} matches the result in \cite{Astrakhantsev:2021rhj}
(with $\cQ_{\fIso1 \fIso1} \rightarrow C$ and $B \rightarrow -B$)
but the expression for $\pa^m \cQ_{\fIso1 \fIso1}$ is different, with
$k_{\iso1}{}^m$ there in place of our $\tilde e_{\iso1}{}^{\ciso1}$. 
(In our approach, $\pa^\um \cQ_{\fIso1 \fIso1}$ vanishes since there are no dual coordinates
$\tilde x_\um$ in the $\sigma$-model.) This is actually a possibility,
because in the case of a single bosonic isometry, one can choose coordinates in the original geometry so that $k_{\iso1}{}^{\ciso1}$ is a constant. Then one simply takes
$\nu_{\iso1} = k_{\iso1}{}^{\ciso1} \tilde x_{\ciso1}$.

Next we address the vielbein and $B$-field. They are 
\begin{align}
e'{}^a &= 
    \rd x^\um \Big(
    e_\um{}^a
    - \EE_{\um \iso1} G^{\iso1 \iso1} k_{\iso1}{}^a
    \Big)
    + \rd \nu_{\iso1} G^{\iso1 \iso1} k_{\iso1}{}^a~, \\
B' &= \frac{1}{2} \rd x^\um \wedge \rd x^\un \,
    \Big(B_{\un \um} + \EE_{\un \iso1} \,G^{\iso1 \iso1} \,\bar \EE_{\um \iso1}\Big)
    + \rd \nu_{\iso1} \wedge \rd x^\um \,G_{\um \iso1} G^{\iso1 \iso1} 
\end{align}
where we have exploited that $(\Adj g)_{\iso1}{}^{\iso1} = 1$.
Finally, the Ramond-Ramond complex is
\begin{align}
e^{\varphi'} \widehat \cF'{}^{1\halpha \, 2\hbeta}
    = 
    \Big( e^{\varphi} \widehat \cF^{1\halpha \, 2\hgamma} 
    + 32i\, \veps_{\fIso1}^{1\halpha} \,(\cQ^{-1})^{\fIso1 \fIso1} \veps_{\fIso1}^{2\hgamma}
    \Big) ({\bm{\slashed\Lambda}}^{-1})_{\hgamma}{}^{\hbeta}~.
\end{align}
Recalling that $G_{\iso1 \iso1} = k_{\iso1}^a k_{\iso1}{}_a$,
the Lorentz transformation governing the T-duality frame is
\begin{align}
{\bm\Lambda}_a{}^b = \delta_a{}^b - 2 \,\frac{k_{\iso1 a} k_{\iso1}{}^b}{k_{\iso1} \cdot k_{\iso1}}
\end{align}
Since this is a single bosonic T-duality, it exchanges the type of supergravity, from type IIB/IIB$^*$ to IIA/IIA$^*$. One finds $\det {\bm\Lambda} = -1$ using the general argument reviewed in section \ref{S:SNATD.worldsheet}. Whether this exchanges the star type (e.g. from IIB to IIA$^*$) depends on whether $\Lambda_0{}^0$ is positive or negative, that is whether the T-duality is spacelike or timelike. We find
\begin{align}
{\bm\Lambda}_0{}^0 
    = \frac{\vec{k}_{\iso1} \cdot \vec{k}_{\iso1} + k_{\iso1}{}^0 k_{\iso1}{}^0}{k_{\iso1} \cdot k_{\iso1}}
\end{align}
which is indeed positive for spacelike $k_{\iso1}$ and negative for timelike $k_{\iso1}$.

\section{Generalized dualities and generalized parallelizable spaces}
\label{S:GPS}

\subsection{Construction of \texorpdfstring{$\cV_\cA{}^\cM$}{V\_A\^M} for constant fluxes}
In the preceding section, we focused on $\sigma$-models depending on two sets of fields: spectator fields $Z^{\uM}$ and fields $Y^{\cIso1}$ that were freely acted upon by some set of isometries. After performing non-abelian T-duality, we arrived at a model with dual fields $\tilde Y_{\cIso1}$. A key point we emphasized was that the dualized sector admitted a double field theory interpretation, with two different generalized vielbeins $\mathring \cV$, \eqref{E:zV.original} and \eqref{E:zV.dual}, depending respectively on $Y$ and $\tilde Y$, so that the generalized fluxes $\mathring \cF$ \eqref{E:NATD.fluxes} were identical and constant.

Let us focus on this last point first, and for simplicity, we dispense with spectator fields. In analogy with the bosonic analogue \cite{Demulder:2018lmj,Hassler:2019wvn,Demulder:2019vvh}, we define a \emph{generalized parallelizable superspace} as a $(D+s)$-dimensional super manifold upon which we can introduce a set of $\g{OSp}(D,D|2s)$-valued generalized frame fields $\cV_\cA{}^\cM$ whose generalized flux tensor $\cF_{\cA \cB \cC}$ \eqref{E:GenFluxTensor} is a constant, $F_{\cA \cB \cC}$. The Bianchi identity for the fluxes reduces to the Jacobi identity
\begin{align}
F_{[\cA \cB}{}^{\cE} F_{\cE \cC \cD]} = 0
\end{align}
for some double Lie group $\mathdsl{D}$. In light of the discussion on non-abelian T-duality, there are two natural questions to pose. First, what conditions on $\mathdsl{D}$ are needed in order to ensure that such a $\cV_\cA{}^\cM$ exists? Second, does this have any relation to an underlying $\sigma$-model in which different realizations of $\cV_\cA{}^\cM$ are dual in some sense?

We will not discuss the second question here, but such a model does exist: it is known as the $\cE$-model \cite{Klimcik:1995dy} and corresponds essentially to the Tseytlin duality symmetric string \cite{Tseytlin:1990nb,Tseytlin:1990va}
with the generalized metric $\cV_\cA{}^\cM$ given below. We refer the reader to the original literature as well as the recent discussion in \cite{Sakatani:2021skx}.

To construct the requisite $\cV_\cA{}^\cM$, it turns out that just three conditions are sufficient:
\begin{enumerate}[label*=\arabic*.)]
  \item A double Lie supergroup $\mathdsl{D}$, generated by $T_\cA = (T_A, T^A)$ with an algebra 
  $$[T_\cA, T_\cB] = -F_{\cA \cB}{}^\cC T_\cC.$$
  \item A non-degenerate, ad-invariant pairing $\Pair{T_\cA}{T_\cB} = \eta_{\cA\cB}$. Conventionally, we choose
$$\eta_{\cA\cB} =
\begin{pmatrix}
0 & \delta_{A}{}^{B} \\
\delta^{A}{}_{B} (-)^{\grad{B}} & 0
\end{pmatrix}~.$$
  \item A maximally isotropic subgroup $H$, generated by $T^A$.
\end{enumerate} \label{IngredientsD}
Different choices of $H$ turn out to correspond to different dual geometries and the supervielbein describes a coset $H \backslash \mathdsl{D}$.

For the case of non-abelian T-duality discussed in the previous section, we would have $T_\cA = (t_{\Iso1}, \tilde t^{\Iso1})$, with commutation relations
\begin{align}
[t_{\Iso1}, t_{\Iso2}] &= - f_{\Iso1 \Iso2}{}^{\Iso3} t_{\Iso3}~, \quad
[t_{\Iso1}, \tilde t^{\Iso2}] = -\tilde t^{\Iso3} f_{\Iso3 \Iso1}{}^{\Iso2}~, \quad
[\tilde t^{\Iso1}, \tilde t^{\Iso2}] = 0~.
\end{align}
The $t_{\Iso1}$ generate the isometry group $G$ and $\tilde t^{\Iso1}$ generate an abelian dual group $\tilde G$. The original $\sigma$-model geometry is produced by choosing $H = \tilde G$ and the dual geometry is produced by $H = G$. This case is known as a Drinfeld double, since the quotient of $\mathdsl{D}$ by either maximally isotropic subgroup $G$ or $\tilde G$ generates the other group, i.e. $G = \tilde G \backslash \mathdsl{D}$ and $\tilde G = G \backslash \mathdsl{D}$. The duality exchanges the roles of $G$ and $\tilde G$. It is also possible for both groups $G$ and $\tilde G$ in a Drinfeld double to be non-abelian. This leads to Poisson-Lie T-duality \cite{Klimcik:1995jn}, and this was historically the first step in generalizing non-abelian T-duality.

The construction of $\cV_\cA{}^\cM$ proceeds as follows. A general group element of $\mathdsl{D}$ is denoted $\mathdsl{g}$, and its left coset $M = H \backslash \mathdsl{D}$ corresponds to a decomposition $\mathdsl{g} = h(\tilde y) \times m(y)$.
The generalized frame field is built from $m$. First, we decompose the Maurer-Cartan form as
\begin{equation}
  \rd m m^{-1} = V^A T_A + A_A T^A (-)^{\grad{A}}
\end{equation}
where $V^A$ and $A_A$ are valued respectively on the coset and the subgroup. Next, we build the two-form $\zB_\WZW$ by integrating
\begin{align}
  \rd \zB_\WZW = \zH_\WZW = -\frac1{12} \Pair{\rd m m^{-1}}{[ \rd m m^{-1}, \rd m m^{-1} ]}~.
\end{align}
This is usually only locally defined. Then the generalized frame field is given by
\begin{align}\label{eqn:genframegroup}
\cV_\cA{}^\cM &= M_\cA{}^\cB 
\begin{pmatrix} 
V_B{}^M & -V_B{}^N \zB_{NM} (-)^m \\
0 & V^B{}_M (-)^m
\end{pmatrix}~, \\
M_\cA{}^\cB &:= (\Adj m)_\cA{}^\cB = \Pair{m T_\cA m^{-1}}{T^{\cB}}~, \\
\zB &= \frac{1}{2} V^A \wedge A_A + \zB_\WZW \,,
\label{eqn:defBforEAI}
\end{align}
We have denoted the two-form by $\zB$ rather than $B$ since contributions from $M_\cA{}^\cB$ typically deform the matrix structure and contribute to the physical $B$-field.

For the case of non-abelian T-duality, choosing $H=G$ leads to 
\begin{align}
\cV_\cA{}^\cM =
\begin{pmatrix}
e_{\Iso1}{}^M & 0 \\
0 & e^{\Iso1}{}_M (-)^{\grad{M}}
\end{pmatrix}
\end{align}
where $e_{\Iso1}{}^M$ are the left-invariant vector fields on $G$, 
see \eqref{E:zV.original}. Alternatively, one can choose $H = G$. To arrange indices
as in \eqref{E:zV.dual}, we take $T^A = t_{\Iso1} (-)^{\grad{\Iso1}}$, $T_A = \tilde t^{\Iso1}$, 
and $m = \exp(\nu_{\Iso1} \tilde t^{\Iso1} (-)^{\grad{\Iso1}}) = \exp(\nu^A T_A)$ 
with $\nu^A = \nu_{\Iso1} (-)^{\grad{\Iso1}}$. The result is
\begin{align}
\cV_\cA{}^\cM =
\begin{pmatrix}
\tilde e_A{}^M &0 \\
-\nu^C f_C{}^{A B} \tilde e_B{}^M \,\,& \tilde e^A{}_M (-)^{\grad{M}}
\end{pmatrix}~, \quad
\tilde e_M{}^A = \pa_M \nu^A~, \quad
\tilde e^A{}_M = \tilde e_M{}^A (-)^{\grad{M} \grad{A}}~.
\end{align}
Swapping indices around, one can show this is just
$\cV_\cA{}^\cM = \cU_{(2)} \cU_{(1)} \cU_{(2)}^{-1}$
where $\cU_{(2)}$ and $\cU_{(1)}$ are the subblocks of \eqref{E:Umatrices} in the isometry directions.

More interesting examples are possible for any real Lie supergroup $G$, provided it admits a non-degenerate Killing form. These can be extended in two distinct ways to a double Lie group $\mathdsl{D}$, either by taking the product group $G \times G$ or its complexification $G^{\mathbb C}$. Both of these cases will be extremely important for the remainder of the paper, and we will describe them in some detail.

\subsection{Example: \texorpdfstring{$\mathdsl{D} = G\times G$}{D=GxG}}
\label{S:GPS.GxG}

We denote a group element of $\mathdsl{D} = G\times G$ with the tuple $(g_1, g_2)\in\mathdsl{D}$ with $g_1, g_2 \in G$.  We use the same convention for the Lie algebra $\mathdsl{d}$ to define the pairing
\begin{align}\label{eqn:pairDGxG}
\Pair{\Xi}{\Xi'} = \frac{1}{2} \pair{\xi_1}{\xi_1'} - \frac{1}{2} \pair{\xi_2}{\xi_2'}~,
\end{align}
for $\Xi = (\xi_1, \xi_2) \in \mathdsl{d}$. In terms of the generators $t_A$ of $G$, we choose the basis of generators on the product group as
\begin{equation}
  T_A = ( t_A, - t_A ) \,, \qquad T^A = (t^A, t^A)\,.
\end{equation}
In the second set, we have raised the indices using the graded inverse $\kappa^{AB}$ (with NW-SE conventions) of the non-degenerate Killing form $\kappa_{AB} = \pair{t_A}{t_B}$.
This choice guarantees that $\Pair{T_\cA}{T_\cB} = \eta_{\cA\cB}$ and that $T^A$ generates the maximally isotropic subgroup $H = G_{\mathrm{diag}}$. This is in fact the only viable choice without imposing additional structure on $G$. The resulting coset $M = H \backslash \mathdsl{D}$ is isomorphic to $G$.
The structure constants $F_{\cA \cB \cC}$, defined by
\begin{align}
[T_\cA, T_\cB] = -F_{\cA \cB}{}^\cC T_\cC = -F_{\cA \cB \cC} \,\eta^{\cC \cD} \,T_\cD (-)^{c}
\end{align}
are given by
\begin{align}\label{eqn:FABCGxG}
  F^{AB}{}_C = f^{AB}{}_C\, \quad \text{and} \quad
  F_{ABC}  = f_{ABC}\,.
\end{align}

A convenient coset representative is
\begin{equation}\label{E:GxG.m1}
  M \ni m = ( g, e )\,, \quad g \in G\,,
\end{equation}
where $e$ is the identity element in $G$. With this convention, it is straightforward to compute all the ingredients for \eqref{eqn:genframegroup}, namely
\begin{align}\label{eqn:GxGgroupMAB}
  M_\cA{}^\cB &= \Pair{m T_\cA m^{-1}}{T^\cB} = \frac12 \begin{pmatrix}
    D_A{}^B + \delta_A{}^B & (D_{AB} - \kappa_{AB}) (-)^b \\
    D^{AB} - \kappa^{AB} & D^A{}_B (-)^b + \delta^A{}_B
  \end{pmatrix}\,,\\
  D_A{}^B &= \pair{g t_A g^{-1}}{t^B}, \\
  V^A &= \Pair{T^A}{\rd m m^{-1}} = \frac12 \pair{t^A}{\dd g g^{-1}} = \frac{1}{2} v^A \,, \\\label{eqn:GxGgroupdB}
  \dd \zB &= -\frac1{24} \pl \dd g g^{-1}, [\dd g g^{-1}, \dd g g^{-1}] \pr 
    = -\frac{1}{24} v^A \wedge v^B \wedge v^C \, f_{CBA}
\end{align}
Above we employ the right-invariant vector field $v^A$ on $G$.
To write down the resulting generalised frame field in a simple form, we also introduce
the left-invariant vector fields on $G$,
\begin{equation}
  e^A = \pair{t^A}{g^{-1} \dd g} = D^A{}_B v^B (-)^b = v^B (D^{-1})_B{}^A
\end{equation}
and the respective inverses $v_A{}^M$ and $e_A{}^M$ with the defining properties
\begin{equation}\label{eqn:dualvectorfields}
  v_A \lrcorner v^B = e_A \lrcorner e^B = \delta_A{}^B \,, \quad
  e_A = D_A{}^B v_B \,,
\end{equation}
to eventually obtain
\begin{equation}\label{eqn:(g,e)genframe}
  \cV_\cA{}^\cM = 
  \begin{pmatrix}
    e_A{}^N + v_A{}^N
    & \quad \frac14 (e_{A N} - v_{AN}) (-)^n \\
    e^{A N} - v^{A N}  
    & \quad \frac14 (e^A{}_N + v^A{}_N) (-)^n
  \end{pmatrix} \times
  \begin{pmatrix}
    \delta_N{}^M
    & -\zB_{NM} (-)^m \\
    0 
    & \delta^N{}_M
  \end{pmatrix}   
\end{equation}
where we use the shorthand $e^A{}_M = e_M{}^A (-)^{ma}$ and $v^A{}_M = v_M{}^A (-)^{ma}$,
with $A$ indices raised and lowered as needed with the Cartan metric.

We can perform the same calculation for a different coset representative,
\begin{equation}\label{E:GxG.m2}
  m' = ( g', g'^{-1} )\,, \quad g' \in G
\end{equation}
which is related to \eqref{E:GxG.m1} by an $H$ transformation,
$m = h m' = (g, e) = (h  g', h  g'^{-1})$ for $g =  g'^2$. Explicitly, we find
\begin{align}
  M_\cA{}^\cB &= \frac{1}{2} \begin{pmatrix}
    (D'+D'^{-1})_A{}^B & \quad (D' - D'^{-1})_{A B} (-)^b \\[1ex]
    (D'- D'^{-1})^{A B} & \quad (D' + D'^{-1})^A{}_B (-)^b 
    \end{pmatrix}\,, \qquad
    D'_A{}^B = \pair{g' t_A g'^{-1}}{t^B}, \\
  V^A &= \frac12 \pair{\dd g' g'^{-1} + g'^{-1} \dd g'}{t^A}
    = \frac{1}{2} (v'^A + e'^A)~, \\
  \dd \zB' &= 
    - \frac16 \pair{\dd g' g'^{-1}}{\dd g' g'^{-1} \dd g' g'^{-1}}
    + \frac14 \pair{g'^{-1} \dd g'}{\dd g' \dd g'^{-1}}
    + \frac14 \pair{\dd g' g'^{-1}}{\dd g'^{-1} \dd g'} \eol
    &= -\frac{1}{24} (v'+e')^A (v'+e')^B (v'+e')^C f_{CBA}
\end{align}
and the generalised frame field arises by plugging these quantities into \eqref{eqn:genframegroup}. The resulting frame is related by a diffeomorphism (and $B$-field gauge transformation) induced by $g =  g'^2$ to the frame in \eqref{eqn:(g,e)genframe}, using
\begin{align}\label{E:GxG.ve.veprime}
v^A = (v'+e')^B D'_B{}^A~, \qquad
e^A = (v'+e')^B (D'^{-1})_B{}^A~,
\end{align}
This can equivalently be written
\begin{align}
v^A t_A &= \rd (g'^2) g'^{-2} = g' (\rd g' g'^{-1} + g'^{-1} \rd g') g'^{-1}~, \eol
e^A t_A &= g'^{-2} \rd (g'^{2}) = g'^{-1} (\rd g' g'^{-1} + g'^{-1} \rd g') g'~.
\end{align}
Explicitly, we find the expression
\begin{align}
\cV_\cA{}^\cM  =
\begin{pmatrix}
    e_A{}^N + v_A{}^N
    & \quad \frac14 (e_{A N} - v_{AN}) (-)^n \\
    e^{A N} - v^{A N}  
    & \quad \frac14 (e^A{}_N + v^A{}_N) (-)^n
  \end{pmatrix} \times
  \begin{pmatrix}
    \delta_N{}^M
    & -\zB'_{NM} (-)^m \\
    0 
    & \delta^N{}_M
  \end{pmatrix}   
\end{align}
where now we interpret $v_M{}^A$ and $e_M{}^A$ as the one-forms that solve \eqref{E:GxG.ve.veprime}.
Naturally this is the same expression as \eqref{eqn:(g,e)genframe}, merely interpreted differently, in a different coordinate system. Note that we still have
\begin{align}
\rd \zB' = -\frac{1}{24} v^A v^B v^C f_{CBA} = -\frac{1}{24} e^A e^B e^C f_{CBA}
\end{align}

Though rewriting it in this way may seem to needlessly complicate matters, it will actually make it easy to see how the generalised frame on $G^\mathbb{C}$, which we construct next, can be related to $G\times G$ by analytic continuation. The key feature of the coset representative \eqref{E:GxG.m2} is that it remains in the same class under the involution $\sigma$ that exchanges the left and right factors: that is, $m'$ just goes to its inverse. This same involution flips the sign of $T_A$,
which negates the first row of $\cV_\cA{}^\cM$. This can be achieved equivalently by exchanging $g'$ with its inverse. This trades $v'^A \leftrightarrows -e'^A$ and $v^A \leftrightarrows -e^A$, and flips the sign of $\zB'$. On the actual matrix elements (keeping in mind that $\rd x'$ flips sign), we find 
$v_M{}^A \leftrightarrows e_M{}^A$. This involution effectively takes
\begin{align}
\cV_A{}^\cM(x') \,\pa_\cM \rightarrow -\,\cV_A{}^\cM(x') \,\pa_\cM ~, \qquad
\cV^A{}^\cM(x') \,\pa_\cM \rightarrow +\cV^A{}^\cM(x') \,\pa_\cM
\end{align}
consistent with the relations between $T_A$ in the two cases, provided we transform
$\pa_{\cM} \rightarrow (-\,\pa_M, \,\pa^M)$.
That is, we flip the sign of $x'$ but not of the dual coordinate. This is sensible, since the dual coordinate parametrizes the diagonal subgroup, which is quotiented out by the coset and undergoes no change.

\subsection{Example: \texorpdfstring{$\mathdsl{D} = G^\mathbb{C}$}{D=G\^C}}
\label{S:GPS.Gc}
Another possibility is to identify $\mathdsl{D}$ with the complexification $G^\mathbb{C}$. While the pairing for $G\times G$ is very simple to define, here we have to work a bit harder. First, let us introduce an involution $\sigma$, which is an isomorphism of the complexified Lie algebra Lie($G^\mathbb{C}$). It has the properties 
\begin{equation}
  \sigma^2 = 1\,, \quad 
  \pair{\sigma X}{\sigma Y} = \pair{X}{Y}^* \,, \quad \text{and} \quad
  \sigma [ X, Y ] =  [\sigma X, \sigma Y]
\end{equation}
with $X$, $Y \in \mathfrak{g}^{\mathbb C}$.
In this case a natural choice for the pairing is
\begin{equation}\label{E:Gc.pairing}
  \Pair{X}{Y} = - \frac{i}{2} \left( \pair{X}{Y} - \pair{\sigma X}{\sigma Y}\right)~, \qquad
  \Pair{X}{Y}^* = \Pair{X}{Y}
\end{equation}
where $X$ and $Y$ are elements of the complexified Lie algebra $\mathfrak{g}^{\mathbb C}$. Here, we in particular make use of the Cartan involution $\sigma$ with the properties
\begin{equation}
  \sigma^2 = 1 \quad \text{and} \quad
  \sigma [ X, Y ] = [\sigma X, \sigma Y] \,.
\end{equation}
It specifies how the real Lie algebra $\mathfrak{g}$ is embedded into $\mathfrak{g}^{\mathbb C}$ by identifying the former's generators $t_A$ with the $+1$ eigenspace of $\sigma$, i.e.
$  \sigma t_A = t_A$. We further assume that $\sigma$ is given by $\sigma X = - S^{-1} X^\dagger S$ where $S$ denotes an optional similarity transformation (for compact $G$, we can set $S=1$). This implies that the structure coefficients are (graded) real, meaning $(f_{AB}{}^C)^* = f_{AB}{}^C (-)^{ab}$. The same holds for the Killing metric $(\kappa_{A B})^* = \kappa_{A B} (-)^{ab}$.

For the generators of $\mathdsl{D}$, we are going to explore two distinct cases. 
The first is obvious:
\begin{equation}\label{eqn:DDgens2}
  T_A = i\, t_A\,, \qquad T^A = t^A~,
\end{equation}
with non-vanishing components of the generalised flux $F_{\cA\cB\cC}$ given by
\begin{equation}
  F^{AB}{}_C = f^{AB}{}_C\,, \qquad F_{ABC} = - f_{ABC}\,.
\end{equation}
For the coset representative, we take a Hermitian element of $G^{\mathbb C}$, so that
$\sigma m = m^{-1}$. Effectively, we can think of $m = \exp(i x^A t_A)$.
The building blocks of the generalized vielbein are then
\begin{align}\label{eqn:MABGC2}
  M_\cA{}^\cB &= \frac12 
  \begin{pmatrix} 
  (D + D^{-1})_A{}^B \quad & i (D - D^{-1})_{AB} (-)^b \\
    -i (D - D^{-1})^{AB}  \quad & (D + D^{-1})^A{}_B (-)^b
  \end{pmatrix}\,, \qquad 
  D_A{}^B = \pair{m t_A m^{-1}}{t^B}\,,\\
  V^A &= \frac1{2 i} \pair{\dd m m^{-1} + m^{-1} \dd m}{t^A} \,,\\\label{eqn:dBGC2}
  \dd \zB &= 
    - \frac1{6 i} \pair{\dd m m^{-1}}{\dd m m^{-1} \dd m m^{-1}}
    + \frac1{4 i} \pair{m^{-1} \dd m}{\dd m \dd m^{-1}}
    + \frac1{4 i} \pair{\dd m m^{-1}}{\dd m^{-1} \dd m}\,.
\end{align}
We introduce the one-form $e'^A$ and its complex conjugate $\bar e'^A$,
\begin{align}
m^{-1} \rd m = i \,e'^A t_A ~, \qquad
\rd m m^{-1} = i \,\bar e'^A t_A ~.
\end{align}
The primes are for later convenience as these will be related to $e'$ and $v'$
in the previous section. For these we recover
\begin{align}
V^A = \frac{1}{2} (e'^A + \bar e'^A)~, \qquad
\rd \zB = \frac{1}{24} (e'+\bar e')^A (e'+\bar e')^B (e'+\bar e')^C f_{CBA}~.
\end{align}
Now the full generalized vielbein can be written
\begin{align}
\cV_\cA{}^\cM  =
\begin{pmatrix}
    e_A{}^N + \bar e_A{}^N
    & \quad \frac{i}{4} (e_{A N} - \bar e_{AN}) (-)^n \\
    -i (e^{A N} - \bar e^{A N})
    & \quad \frac{1}{4} (e^A{}_N + \bar e^A{}_N) (-)^n
  \end{pmatrix} \times
  \begin{pmatrix}
    \delta_N{}^M
    & -\zB_{NM} (-)^m \\
    0 
    & \delta^N{}_M
  \end{pmatrix}   
\end{align}
where we use
\begin{align}
e^A = (e'^B + \bar e'^B) D_B{}^A~, \qquad
\bar e^A = (e'^B + \bar e'^B) (D^{-1})_B{}^A~,
\end{align}
or equivalently,
\begin{align}
i \,e^A t_A &= \rd m^{2} m^{-2} = m (\rd m m^{-1} + m^{-1} \rd m) m^{-1}~, \\
i \,\bar e^A t_A &= m^{-2} \rd m^2 = m^{-1} (\rd m m^{-1} + m^{-1} \rd m) m~.
\end{align}

This case and the one for $G\times G$ with coset representative \eqref{E:GxG.m2} are related by an analytic continuation. There are several ways of seeing it. From the level of the building blocks
\eqref{eqn:MABGC2} -- \eqref{eqn:dBGC2} and the algebra, we can see it by continuing
$T_A \rightarrow i T_A$. To maintain $\eta_{\cA\cB}$, we must substitute 
$\Pair{\cdot}{\cdot} \rightarrow - i \Pair{\cdot}{\cdot}$, too. Consequentially, we obtain
\begin{equation}\label{eqn:anacont1}
  M_A{}^B \rightarrow M_A{}^B\,, \quad
  M_{A B} \rightarrow i M_{A B}\,, \quad
  M^{A B} \rightarrow -i M^{A B}\,, \quad
  M^A{}_B \rightarrow M^A{}_B\,,
\end{equation}
while for the two remaining constituents of the generalised frame field we find
\begin{equation}\label{eqn:anacont2}
  V^A \rightarrow -i V^A \quad \text{and} \quad
  \zB \rightarrow -i \zB\,.
\end{equation}

This is somewhat formal, and we can make it more concrete by observing that both coset representatives $m$ are inverted by their respective involutions, and we use this involution to track how factors of $i$ are inserted. Here, $m = \exp(i x^A t_A)$
and for \eqref{E:GxG.m2} we have $g' = \exp(x'^A t_A)$. We want to analytically continue by taking $x'= i x$. By comparing explicit formulae, we see that $D'(x') = D(x)$ and so
$e_M{}^A(x')$ and $v_M{}^A(x')$ become, respectively, $e_M{}^A(x)$ and $\bar e_M{}^A(x)$.\footnote{The forms pick up factors of $i$ because $\rd x' = i \rd x$.}
The $B$ fields are related as $\zB'_{MN}(x') = -i \zB_{MN}(x)$. Putting this together we see that 
the two generalized vielbeins $\cV_\cA{}^\cM$ turn out to be related by
\begin{align}
\cV'_A{}^\cM(x') \,\pa'_\cM = -i\,\cV_A{}^\cM(x) \,\pa_\cM ~, \qquad
\cV'^A{}^\cM(x') \,\pa'_\cM = \cV^A{}^\cM(x) \,\pa_\cM
\end{align}
consistent with the relations between $T_A$ in the two cases, provided we identify
$\pa'_{\cM} = (-i \,\pa_M,  \pa^M)$.
That is, on the doubled space, we transform $x'= i x$ but leave the dual coordinate unchanged. This makes sense on the coset since the dual coordinate describes a copy of $G$ itself in both cases (being the same isotropic subgroup $H$), and undergoes no analytic continuation.

There is another possibility that will be of interest to us,\footnote{The decomposition \eqref{eqn:DDgens} is actually a Drinfeld double, and one could exchange the roles of $T_A$ and $T^A$. The result is essentially equivalent to taking \eqref{eqn:DDgens2}, up to a similarity transformation and coordinate transformation.}
\begin{equation}\label{eqn:DDgens}
  T_A = t_A\,, \qquad T^A = (R^{AB} + i \,\kappa^{AB}) t_B
\end{equation}
for a matrix $R^{AB}$ obeying certain properties. Requiring $\Pair{T_\cA}{T_\cB} = \eta_{\cA\cB}$ implies that $R^{AB}$ is graded real and antisymmetric. Requiring that $T^A$ generate a maximally isotropic subgroup implies
\begin{equation}\label{E:mCYBE}
  [R X, R Y] - R\left( [R X, Y] + [X, R Y] \right) = [X, Y] \qquad \forall X, Y \in \mathrm{Lie}(G) \,,
\end{equation}
where we employ operator notation for $R$, i.e. $R \cdot \xi = \xi^A R_A{}^B t_B$. From this equation, we learn that $R$ must solve the modified classical Yang-Baxter equation (mCYBE). For the coset representative $m=g$, which is now fixed by the involution $\sigma$,
we again compute all  ingredients required for the generalised frame field,
\begin{alignat}{2}
M_\cA{}^\cB &= 
\begin{pmatrix} 
D_A{}^B & 0 \\
R^{AC} D_C{}^B - D^A{}_C R^{CB} (-)^c  \quad & D^A{}_B (-)^b
\end{pmatrix}\,, &\qquad 
D_A{}^B &= \pair{g t_A g^{-1}}{t^B},  \\
V^A &= \pair{\dd g g^{-1}}{t^A} = v^A~,  &\quad
B &= 0\,.
\end{alignat}
We can streamline the result further by defining
\begin{equation}
  e^A = \pair{g^{-1} \dd g}{t^A} = v^B (D^{-1})_B{}^A~,
\end{equation}
and the corresponding dual vector fields $v_A{}^M$ and $e_A{}^M$ (see \eqref{eqn:dualvectorfields}). With them, we eventually find
\begin{equation}\label{E:Gc.V.Pi}
\cV_\cA{}^\cM = \begin{pmatrix}
    e_A{}^M & 0 \\
    \Pi^{AB} e_B{}^M \quad & e^A{}_M (-)^m
  \end{pmatrix}\,, \quad \text{where} \quad
  \Pi^{AB} = R^{AB} - (R_g)^{A B}
\end{equation}
where
\begin{align}
(R_g)^{AB} := \pair{g t^A g^{-1}}{R ( g \,t^B g^{-1} )}
  = D^A{}_C R^{CD} D^B{}_D (-)^{c + bd}~.
\end{align}
It is interesting to note that $\Pi^{MN} = e_A{}^M \Pi^{AB} e_B{}^N (-)^{am+a}$ defines a Poisson bracket $\{f, g\} = \Pi^{MN} \partial_N f \partial_M g$ which turns $G$ into a Poisson-Lie group. Moreover, we can easily extract the generalised fluxes
\begin{equation}
  F_{AB}{}^C = f_{AB}{}^C\,, \quad \text{and} \quad
    F^{AB}{}_C = 2 R^{[A D} f_D{}^{B]}{}_{C}\,.
\end{equation}
consistent with the structure constants of the generators \eqref{eqn:DDgens}.

It is useful to make a similarity transformation on the generalized vielbein and the generators in this case to give
\begin{equation}\label{E:Gc.simplebasis}
  T_A = t_A\,, \qquad T^A = i \, t^A
\end{equation}
and
\begin{equation}\label{E:genV.Rg}
\cV_\cA{}^\cM = \begin{pmatrix}
    e_A{}^M & 0 \\
    -(R_g)^{A B} e_B{}^M \quad & e^A{}_M (-)^m
  \end{pmatrix}
\end{equation}
with generalized fluxes
\begin{equation}
  F_{AB}{}^C = f_{AB}{}^C\,, \qquad F^{ABC} = - f^{ABC}~.
\end{equation}
Up to the interchange of $T_A \leftrightarrows T^A$, the generalized vielbein \eqref{E:genV.Rg}
and the one constructed from \eqref{eqn:MABGC2}-\eqref{eqn:dBGC2} are Poisson-Lie T-dual to each other.

\subsection{The role of the dilaton}
\label{S:GPS.dilaton}
We have not yet discussed the role of the dilaton on a generalized parallelizable space. Let us address this briefly now. In terms of the generalized dilaton $\Phi$, the dilaton flux is given by \eqref{E:DilatonFlux}, which for the supervielbein \eqref{eqn:genframegroup} becomes
\begin{align}
\cF_\cA = M_\cA{}^B V_B{}^M \Big(
    \pa_M \log \Phi
    - \pa_M \log \det V + A_{M C} F^{C D}{}_D (-)^c
    \Big)
    - M_{\cA B} F^{B D}{}_D (-)^b\,.
\end{align}
In the case of generalized double field theory, we replace 
$\pa_\cM \log\Phi \rightarrow \cX_\cM$ 
and relax the section condition on the free index of $\cX_\cM$. Solving
for $\cX_\cM = (\cX_M, \cX^M)$, we find
\begin{align}
\cX^M &= (M^{-1})^{A \cB} \cF_\cB V_A{}^M + F^{A B}{}_B V_A{}^M , \eol
\cX_M - \zB_{M N} \cX^N &= V_M{}^A \, (M^{-1})_A{}^\cB \cF_\cB
    + \pa_M \log \det V - A_{M C} F^{C D}{}_D (-)^c~.
\label{E:cX.Sol1}
\end{align}
The dilaton is not completely arbitrary since we still require $\cF_\cA$ to obey the usual Bianchi identities. In the context of a generalized parallelizable space, when the fluxes $\cF_{\cA \cB \cC}$ are taken to be constants $F_{\cA \cB \cC}$, the most natural choice is to take the dilatonic fluxes to be constants as well, $\cF_\cA = F_\cA$. The Bianchi identities then imply $F_{\cA \cB}{}^\cC F_\cC = 0$, and the conditions \eqref{E:cX.Sol1} simplify to
\begin{align}
\cX^M &= (F^A + F^{A B}{}_B) V_A{}^M , \eol
\cX_M - \zB_{M N} \cX^N &= V_M{}^A F_A
    + \pa_M \log \det V - A_{M C} F^{C D}{}_D (-)^c~.
\label{E:cX.Sol2}
\end{align}
These can be interpreted as \emph{solutions} for the vector $\cX_\cM$. In order to admit a dilaton solution consistent with the section condition, one must restrict $F^A = -F^{A B}{}_B$.

As a special case, we can consider both $G \times G$ and $G^{\mathbb C}$.
For $G \times G$ using the coset representative \eqref{E:GxG.m1}, we find
\begin{align}\label{E:GxG.cX}
\cX^M = 2 F^A v_A{}^M~, \qquad
\cX_M - \zB_{MN} \cX^N = \frac{1}{2} v_M{}^A F_A + \pa_M \log \det v
\end{align}
with $F^A$ and $F_A$ obeying
\begin{align}
f_{AB}{}^C F_C = F^C f_{C A}{}^B  = 0~.
\end{align}
A dilaton solution requires $F^A = 0$.

For $G^{\mathbb C}$ using the coset representative $g$ in the basis \eqref{eqn:DDgens}, we find
\begin{align}\label{E:Gc.cX}
\cX^M = (F^A + R^{B C} F_{CB}{}^A) v_A{}^M~, \qquad
\cX_M - \zB_{MN} \cX^N = v_M{}^A F_A + \pa_M \log \det v
\end{align}
with $F_A$ and $F^A$ obeying
\begin{align}
f_{AB}{}^C F_C = (F^C - R^{CD} F_D) f_{C A}{}^B = 0~.
\end{align}
If we make the similarity transformation to the simpler basis \eqref{E:Gc.simplebasis} with
$T'^A = i \,\kappa^{AB} t_B$ instead, one replaces $F^A$ with $F'^A + R^{AB} F_B$ in the above formulae. To admit a dilaton solution, we must have the following condition 
\begin{align}
F^A = F'^{A} + R^{AB} F_B = - R^{B C} F_{CB}{}^A~.
\end{align}

\section{Generalized supercosets}
\label{S:GCoset}

\subsection{Review of conventional supercosets}
To motivate the construction of generalized supercosets, we first recall how conventional supercosets are constructed. Let $G$ be a group and $F$ be a subgroup. Denote the generators of $G$ by $t_\hA$, the generators of $F$ by $t_\sa$, and the remaining generators by $t_A$. 
The structure constants are normalized so that $[t_\hA, t_\hB] = - f_{\hA \hB}{}^\hC t_\hC$.
We decompose a generic group element $g$ as $g = m(z) f(y)$ with coset representative $m$. The local coordinates are chosen as $z^\hM = (z^M, y^I)$. 
The Maurer-Cartan form $\dd z^{\hM} \widehat E_{\hM}{}^{\hA} t_{\hA} = g^{-1} \dd g$ decomposes as
\begin{equation}\label{eqn:splittingcoset}
  \widehat E_{\hM}{}^{\hA} =
  \begin{pmatrix}
    \delta_{\ci}{}^{\cj} & 0 \\
    0 & \widetilde{v}_{\si}{}^{\sc}
  \end{pmatrix}   
  \begin{pmatrix} 
  E_{\cj}{}^{\cb} \quad & \Omega_{\cj}{}^{\sb} \\
    0  \quad & \delta_{\sc}{}^{\sb}
  \end{pmatrix}
(\Adj f^{-1})_{\hB}{}^{\hA}
\end{equation}
with
\begin{equation}
  \dd y^{\si} \widetilde{v}_{\si}{}^{\sa}  t_{\sa} = \dd h h^{-1}\,.
\end{equation}
This decomposition shows how the full group can be reconstructed from the coset. In particular, it has three important properties:
\begin{enumerate}[label*=\arabic*.)]
  \item All quantities relevant for the coset are contained in the middle matrix in \eqref{eqn:splittingcoset}. These depend only on the physical coordinates on the coset.
  \item This matrix is in upper triangular form.
  \item It is dressed by an adjoint $f\in F$ action on the right and right-invariant Maurer-Cartan form of the subgroup $F$ on the left. These depend only on the subgroup coordinates.
\end{enumerate}
With the dual vector fields corresponding to \eqref{eqn:splittingcoset},
\begin{equation}\label{eqn:splittingcosetit}
  \widehat E_{\hA}{}^{\hM} = (\Adj f)_{\hA}{}^{\hB} \begin{pmatrix}
    E_{\cb}{}^{\cj} \quad & - E_{\cb}{}^{\ck} \Omega_{\ck}{}^{\sc} \\
    0 \quad & \delta_{\sb}{}^{\sc}
  \end{pmatrix} \begin{pmatrix}
    \delta_{\cj}{}^{\ci} & 0 \\
    0 & \widetilde{v}_{\sc}{}^{\si}
  \end{pmatrix}\,,
\end{equation}
one can compute the anholonomy coefficients
\begin{equation}
  \widehat F_{\hA \hB}{}^{\hC} := 
  -2 \,\widehat{E}_{[\hA}{}^{\hM} \partial_{\hM} \widehat{E}_{\hB]}{}^{\hN} 
    \widehat E_{\hN}{}^{\hC}~.
\end{equation}
With the $y$ coordinate dependence isolated in the first and third factors, one can show that the anholonomy coefficients with a lower index valued in $F$ are completely fixed in terms of the structure constants. Up to an adjoint action of $f$, which we discard in the definition
of $\widehat F_{\hA \hB}{}^{\hC}$, we find
\begin{alignat}{2}\label{E:SuperCosetF1}
\widehat F_{\sa \sb}{}^{\sc} &= f_{\sa\sb}{}^\sc~, &\qquad
\widehat F_{\sa \sb}{}^{C} &= 0~, \eol
\widehat F_{\sa B}{}^{\sc} &= f_{\sa B}{}^\sc~, &\qquad
\widehat F_{\sa B}{}^{C} &= f_{\sa B}{}^C~, 
\end{alignat}
while the remaining two correspond 
to the covariant torsion and curvature tensors
\begin{subequations}\label{E:SuperCosetF2}
\begin{align}
T_{AB}{}^C  = \widehat F_{A B}{}^C &= F_{A B}{}^C
    - 2 \,\Omega_{[A|}{}^{\sd} f_{\sd |B]}{}^C~, \\
R_{AB}{}^\sc  = \widehat F_{A B}{}^\sc &= 2 \,D_{[A} \Omega_{B]}{}^\sc
    + F_{A B}{}^C \Omega_C{}^{\sc}
    + \Omega_A{}^\sa \Omega_B{}^\sb f_{\sb \sa }{}^\sc (-)^{\grad{B} \grad{\sa}}
    - 2\, \Omega_{[A}{}^{\sd} f_{\sd B]}{}^\sc
\end{align}
\end{subequations}
where $F_{A B}{}^C = - 2\,E_{[A}{}^M \pa_M E_{B]}{}^N E_{N}{}^C$.
The results \eqref{E:SuperCosetF1} and the covariance of \eqref{E:SuperCosetF2} follow from
the general fiber bundle structure of \eqref{eqn:splittingcoset} with local symmetry group $F$ acting on the frame bundle. When $E_M{}^A$ and $\Omega_M{}^\sa$ are determined from a larger group $G$, the covariant torsion and curvature tensors are fixed as
\begin{align}
T_{AB}{}^C = f_{AB}{}^C~, \qquad R_{AB}{}^\sc = f_{AB}{}^{\sc}~.
\end{align}

\subsection{Generalized supercoset construction}
Let's apply similar considerations to the case of a double Lie group $\mathdsl{D}$. As before, we presume a maximally isotropic subgroup $H$, consistent with the assumptions made in section \ref{IngredientsD}. We denote the generators of $\mathdsl{D}$ as $T_\hcA = (T_\hA, T^\hA)$ with $T^\hA$ the generators of $H$.  In addition, we presume that $\mathdsl{D}$ possesses \emph{another} isotropic subgroup $F$, with generators $T_\sa$, with respect to which we will construct a generalized coset $H \backslash \mathdsl{D} / F$. 

There is a subtlety here, which we should address at this point. We make no assumptions about how $F$ and $H$ are related. This means we will need two distinct bases for the generators of $\mathdsl{D}$: the original basis $T_\hcA = (T_\hA, T^\hA)$ where $T^\hA$ are the generators of $H$, and a new basis,
\begin{align}
T_{\hcA'} = (T_\sa, T_\ca, T^\ca, T^\sa)
\end{align}
where $T_\sa$ are the generators of $F$. For this new basis, we take the Killing metric to be
\begin{align}\label{eqn:etacoset}
  \eta_{\hcA' \hcB'} = \begin{pmatrix}
    0 & 0 & \delta_{\sa}{}^{\sb} \\
    0 & \eta_{\cA \cB} & 0 \\
    \delta^{\sa}{}_{\sb} (-)^b & 0 & 0
  \end{pmatrix}
\end{align}
with $\eta_{\cA \cB}$ an $\g{OSp}$ metric on the coset. The change of basis matrix between $T_{\hcA'}$ and $T_\hcA$ may in principle be quite complicated.
To avoid a proliferation of indices, we won't explicitly exhibit the prime on $\hcA$, but it should be understood to be in the appropriate basis.

On the generalized frame field in \eqref{eqn:genframegroup}, we aim to impose a similar decomposition inspired by \eqref{eqn:splittingcosetit}. The role of the group $G$ and subgroup $H$ will be played by the left coset $H \backslash \mathdsl{D}$ and the subgroup $F$ respectively:
\begin{equation}\label{eqn:PSform}
  \widehat \cV_{\hcA}{}^{\hcM} = 
  (\Adj f)_{\hcA}{}^{\hcB} 
  \begin{pmatrix}
    \delta_{\sb}{}^{\sc} \quad & 0 \quad & 0 \\
    - \Omega_{\cB}{}^{\sc} \quad & \cV_{\cB}{}^{\cN} \quad & 0 \\
    \rho^{\sb\sc} - \frac12 \Omega^{\sb \cP} \Omega_\cP{}^{\sc} \quad & \Omega^{\sb \cN} \quad & \delta^{\sb}{}_{\sc}
  \end{pmatrix}
  \begin{pmatrix}
    \widetilde{v}_{\sc}{}^{\si} & 0 & 0\\
    0 & \delta_{\cN}{}^{\cM} & 0 \\
    0 & 0 & \widetilde{v}^{\sc}{}_{\si} (-)^i
  \end{pmatrix}~.
\end{equation}
By preserving the $\g{OSp}$ pairing on the generalised tangent space and splitting it into coset and subgroup contributions, we obtain
\begin{equation}
  \eta_{\hcM \hcN} = \begin{pmatrix}
    0 & 0 & \delta_{\si}{}^{\sj} \\
    0 & \eta_{\cM\cN} & 0 \\
    \delta^{\si}{}_{\sj} (-)^{j} & 0 & 0
  \end{pmatrix}~.
\end{equation}
With the tangent space metric \eqref{eqn:etacoset}, this ensures
$\widehat \cV_{\hcA}{}^{\hcM}$ is an $\g{OSp}$ element. In fact, it decomposes into a product of \emph{three} $\g{OSp}$ matrices. The first and the last are naturally comparable to the factors in  \eqref{eqn:splittingcosetit}. For the matrix in the middle, we have imposed a lower triangular form with the diagonal inspired by the geometric coset. Taking $\cV_\cB{}^\cN$ to itself be an $\g{OSp}$ element, the remaining free parameters are $\Omega_{\cM}{}^{\sa}$, with $\Omega_\cB{}^{\sa} = \cV_\cB{}^\cN \Omega_\cN{}^\sa$ and 
$\Omega^{\sb \cN} = \Omega^{\cN \sb} (-)^{\grad{\cN} \grad{\sb}}$
and the graded antisymmetric matrix $\rho^{\sa\sb}$. The former plays obviously a similar role as the connection $\Omega_{\ci}{}^{\sa}$ in the geometric coset, while the latter is a new ingredient required only in generalised geometry. Remarkably, $\rho^{\sa\sb}$ also appears in the work by \Polacek{} and Siegel to construct a natural curvature with manifest T-duality \cite{Polacek:2013nla}. There, the subgroup $F$ is the double Lorentz group and the contracted version $\rho^{\se\sf} F_{\se AB} F_{\sf CD} = r_{ABCD}$ is used. Hence, we call $\rho^{\sa\sb}$ the Pol\'{a}\v{c}ek-Siegel (PS) field. For a deeper discussion on the \Polacek-Siegel formalism in the related context of consistent truncations, we refer the reader to \cite{Butter:2022iza}.

From now on, we will refer to $\widehat \cV_{\hcA}{}^\hcM$ as the \emph{megavielbein} and the enlarged superspace on which it acts the \emph{megaspace}, when we need to distinguish it from the coset supervielbein $\cV_\cA{}^\cM$. Similarly, we use
\begin{align}
\widehat D_{\hcA} = \widehat \cV_{\hcA}{}^\hcM \pa_\hcM~, \qquad
D_\cA = \cV_\cA{}^\cM \pa_\cM
\end{align}
to denote their respective flat derivatives. From \eqref{eqn:PSform} and recalling that $\pa^\si$ vanishes, the flat derivative on the megaspace becomes
\begin{align}
\widehat D_{\hcA} = (\Adj f)_{\hcA}{}^{\hcB}
\begin{pmatrix}
\widetilde{v}_{\sb}{}^{\si} \pa_\si \\[1.5ex]
D_\cB - \Omega_\cB{}^\sc \widetilde{v}_{\sc}{}^{\si} \pa_\si \\[1.5ex]
(\rho^{\sb\sc} - \frac12 \Omega^{\sb \cP} \Omega_\cP{}^{\sc})  \widetilde{v}_{\sc}{}^{\si} \pa_\si
+ \Omega^{\sb \cA} D_\cA
\end{pmatrix}~.
\end{align}

Just as in the conventional supercoset, the middle matrix in \eqref{eqn:PSform} depends only on the coset coordinates. With the $y$ coordinate dependence isolated in the first and third factors, one can show that up to an overall adjoint action of $f$, which we discard,
the generalized fluxes with a lower index valued in $F$ are completely fixed as
\begin{alignat}{4}
\widehat \cF_{\sa \sb \sc} &= 0~, &\qquad
\widehat \cF_{\sa \sb \cC} &= 0~, &\qquad
\widehat \cF_{\sa \sb}{}^\sc &= F_{\sa \sb}{}^{\sc}~, \eol
\widehat \cF_{\sa \cB \cC} &= F_{\sa \cB \cC} &\qquad
\widehat \cF_{\sa \cB}{}^{\sc} &= F_{\sa \cB}{}^{\sc} &\qquad
\widehat \cF_{\sa }{}^{\sb\sc} &= F_{\sa}{}^{\sb \sc}~.
\end{alignat}
The remaining fluxes correspond to generalized curvature tensors.
The torsion tensor is
\begin{align}
\cT_{\cA \cB \cC} &= \widehat \cF_{\cA \cB \cC}
        = - 3 \,\cV_{[\cA}{}^{\cM} \pa_{\cM} \cV_{\cB}{}^{\cN} \cV_{\cN \cC]} 
        - 3 \,\Omega_{[\cA|}{}^{\sd} F_{\sd| \cB \cC]}
        = \cF_{\cA \cB \cC}
        - 3 \,\Omega_{[\cA|}{}^{\sd} F_{\sd| \cB \cC]}
\end{align}
where $\cF$ is the generalized flux of the coset supervielbein. The generalized $\Omega$
curvature is
\begin{align}
\cR_{\cC \cB}{}^\sa = \widehat \cF_{\cC\cB}{}^\sa &=
    2\, D_{[\cC} \Omega_{\cB]}{}^\sa
    + \cF_{\cC \cB}{}^\cD \Omega_\cD{}^\sa
    + \Omega_\cC{}^\sc \Omega_\cB{}^\sb F_{\sb \sc}{}^\sa (-)^{\grad{\cB} \grad{\sc}}
    \eol & \quad
    - 2\, \Omega_{[\cC}{}^{\sd} F_{\sd \cB]}{}^\sa
    - F_{\cC \cB \,\sd} \, 
        \Big(\rho^{\sd \,\sa} + \tfrac{1}{2} \Omega^{\sd \cF} \Omega_\cF{}^\sa\Big) 
            (-)^{\grad{\sd}}~.
\end{align}
Finally, there are two additional curvatures that are not present in the standard supercoset. These are the covariantized gradient of the \Polacek-Siegel field
\begin{align}
\cR_\cC{}^{\sb \sa} = \widehat \cF_{\cC}{}^{\sb\sa} 
    &= - D_\cC \rho^{\sb \sa} + \cdots
\end{align}
and an additional curvature
\begin{align}
\cR^{\sc \sb \sa} = \widehat \cF^{\sc \sb \sa}
    &= 3 \,\rho^{[\sc| \sd} F_\sd{}^{|\sb \sa]}
    - 3 \,\Omega^{[\sc |\cC} D_\cC \rho^{|\sb \sa]}
    + \cdots
\end{align}
where we have shown only the leading terms. The precise forms of these curvatures
will not be relevant for us. The curious reader can find them in \cite{Butter:2021dtu,Butter:2022gbc}.

These generalized torsion and curvature tensors hold for generic $\cV$, $\Omega$, and $\rho$.
When they are determined from a larger doubled Lie group, they are constrained as
\begin{align}
\cT_{\cA \cB \cC} = F_{\cA \cB \cC}~, \qquad
\cR_{\cC \cB}{}^\sa = F_{\cC \cB}{}^\sa~, \qquad
\cR_\cC{}^{\sb \sa} = F_\cC{}^{\sb \sa} ~, \qquad
\cR^{\sc \sb \sa} = F^{\sc \sb \sa}~.
\end{align}

Now let us compute the quantities \eqref{eqn:genframegroup}-\eqref{eqn:defBforEAI} required to construct the generalised frame field for the coset representative $m = n f$, where $f$ is an element of an isotropic subgroup $F$ generated by $T_{\sa}$. We find the adjoint action
\begin{equation}
  M_{\hcA}{}^{\hcB} := (\Adj m)_\hcA{}^{\hcB} =
    \widetilde{M}_{\hcA}{}^{\hcC} \overline{M}_{\hcC}{}^{\hcB}\,, \qquad 
  \widetilde{M}_{\hcA}{}^{\hcB} = (\Adj f)_\hcA{}^\hcB~, \quad
  \overline{M}_{\hcA}{}^{\hcB} = (\Adj n)_\hcA{}^\hcB~,
\end{equation}
the Maurer-Cartan form components
\begin{align}
  V^{\hA} &= \Pair{\dd n n^{-1} + n \dd f f^{-1} n^{-1}}{T^{\hA}} \\
  A_{\hA} &= \Pair{\dd n n^{-1} + n \dd f f^{-1} n^{-1}}{T_{\hA}}~,
\end{align}
and the $B$-field
\begin{align}
  \zB &= \frac12 \left( 
    V^{\hA} \wedge A_{\hA} 
    - \Pair{\dd f f^{-1}}{n^{-1} \dd n} \right) 
    + \overline{\zB}_\WZW\,, \eol
  \dd \overline{\zB}_\WZW &= -\frac1{12} \Pair{\dd n n^{-1}}{[ \dd n n^{-1}, \dd n n^{-1} ]}~.
\end{align}
Above we have split the original $\zB_\WZW$ from \eqref{eqn:defBforEAI} into an exact term
and a term $\overline{\zB}_\WZW$ defined purely on the coset.

A straightforward calculation gives rise to 
\begin{align}\label{eqn:iotahtvB}
  \widetilde{v}_\sa \lrcorner \zB = 
    \Pair{n T_{\sa} n^{-1}}{V^\hB T_\hB}
    = \overline{M}_{\sa \hB} \,V^\hB (-)^{\grad{\hB}}~, \qquad
  \widetilde{v}_\sa \lrcorner V_{\hB} \lrcorner \zB = -\overline{M}_{\sa \hB}
\end{align}
where $\widetilde{v}_{\sa} = \widetilde{v}_\sa{}^\si \pa_\si$ denotes the vector field dual to the one-form $\dd y^{\si} \widetilde{v}_{\si}{}^{\sa} T_{\sa} = \dd f f^{-1}$. From this equation, we immediately obtain
\begin{equation}
  \widetilde{v}_{\sa} \lrcorner \left( -V_\hB \lrcorner \zB \,T^\hB (-)^{\grad{\hB}} + V^\hB T_\hB \right) = n T_{\sa} n^{-1}\,, 
\end{equation}
which proves that 
\begin{equation}
  \widehat\cV_{\hcA \,\si}= \widetilde{M}_{\hcA}{}^{\hcB}
  \begin{pmatrix}
    0 \\
    0 \\
    \widetilde{v}^{\sb}{}_{\si} (-)^i
  \end{pmatrix}
\end{equation}
holds. This verifies the form of the last column in the middle matrix of \eqref{eqn:PSform}. But because $\widehat \cV_{\hcA}{}^{\hcM}$ is an $\g{OSp}$ element, the first row also has the desired form. We can finally read off
\begin{align}
\Omega_{\cB}{}^{\sa} &= - \overline{M}_{\cB}{}^{\hC}  S_{\hC}{}^{\sa}\,, \\
\rho^{\sa\sb} &= \overline{M}^{[\sa| \hC} S_{\hC}{}^{|\sb]}\,, \\
\cV_{\cA}{}^{\cM} &= \overline{M}_{\cA}{}^{\hcB}
    \begin{pmatrix}
    V_{\hB}{}^{\ci} \quad & 
        - V_{\hB}{}^{\cj} \zB_{\cj\ci} 
            - S_{\hB}{}^{\sc} \,\overline{M}_{\sc \hD} \,V^{\hD}{}_{\ci} (-)^{\grad{\ci} + \grad{\hD}}\\
    0 \quad &  V^{\hB}{}_{\ci} (-)^{\grad{\ci}}
  \end{pmatrix}\,,
\end{align}
where we introduced for convenience the quantity
\begin{align}\label{E:G/F.defS}
S_{\hA}{}^{\sb} := V_{\hA}{}^{\si} \widetilde{v}_{\si}{}^{\sb} ~, \qquad
\overline{M}_{\sa}{}^{\hB} S_{\hB}{}^{\sc} = \delta_\sa{}^\sc~.
\end{align}
It is a somewhat involved calculation to show that both
$S_{\hA}{}^{\sb}$ and $V_\hA{}^M$ are $y$-independent, while
$\zB_{NM}$ and $V_M{}^{\hA}$ are $y$-independent by construction.

\subsection{The dilaton on the generalized supercoset}
Now we will equip the \Polacek-Siegel megaspace with a dilaton $\widehat \Phi$. Its generalized flux tensor is
\begin{align}\label{E:DilatonFlux.mega}
\widehat \cF_{\hcA} = \widehat \cV_\hcA{}^\hcM \pa_\hcM \log\widehat\Phi + \pa^\hcM \widehat\cV_{\hcM \hcA}~.
\end{align}
In analogy to the decomposition of the megavielbein \eqref{eqn:PSform}, we expand
\begin{align}\label{E:MegaPhiToPhi}
\log \widehat \Phi = \log \Phi + \log \tilde e
\end{align}
where $\tilde e^\sa T_\sa = f^{-1} \rd f$ is the left-invariant vector field on $F$ and $\Phi$ is chosen to be independent of $y$. The extracted term is responsible for generating the density behavior of $\widehat\Phi$ under $y$ diffeomorphisms. One can now show that
\begin{align}\label{E:PSDilF.Decomp}
\widehat \cF_{\hcA} = (\Adj{f})_\hcA{}^\hcB
\begin{pmatrix}
0 \\
\cT_\cB \\
\cR^\sb
\end{pmatrix}
    + F_{\hcA \sb}{}^{\sb} (-)^{\grad{\sb}}
\end{align}
where
\begin{align}
\cT_\cA &= \cV_\cA{}^\cM \pa_\cM \log\Phi + \pa^\cM \cV_{\cM \cA} - \Omega^{\cB \sc} F_{\sc \cB \cA}~, \\
\cR^\sa &= \Omega^{\sa \cM} \pa_\cM \log\Phi + \pa^\cM \Omega_{\cM}{}^{\sa}
    - \Omega^{\cB \sc} F_{\sc \cB }{}^\sa
    + \rho^{\sb \sc} F_{\sc \sb}{}^\sa
\end{align}
are the dilatonic torsion and curvature respectively.

In the case of generalized DFT, one should replace
$\pa_\hcM \log \widehat\Phi \rightarrow \widehat \cX_\hcM$
in \eqref{E:DilatonFlux.mega}. A natural replacement of the constraint \eqref{E:MegaPhiToPhi}  is
\begin{align}
\widehat \cV_\hcA{}^\hcM \Big(\widehat \cX_\hcM - \pa_\hcM \log \tilde e \Big) = 
(\Adj f)_\hcA{}^\hcB
\begin{pmatrix}
0 \\
\cV_\cB{}^\cM \cX_\cM \\
\Omega^{\sb \cM} \cX_\cM + \cX^\sb
\end{pmatrix}
\end{align}
where $\cX_\cM$ and $\cX^\sa$ transform under coset diffeomorphisms and
$F$-gauge transformations as
\begin{align}
\delta \cX_\cM  = \xi^\cN \pa_\cN \cX_\cM
    + \pa_\cM \pa^\cN \xi_\cN~, \qquad
\delta \cX^\sa = \xi^\cN \pa_\cN \cX^\sa 
    - \cX^\cM \pa_\cM \lambda^\sa - \lambda^\sb \cX^\sc F_{\sc \sb}{}^\sa~.
\end{align}
Now the dilatonic torsion and curvature are
\begin{align}
\label{E:PS.DilT}
\cT_\cA &= \phantom{\cX^\sa + \,} \cV_\cA{}^\cM \cX_\cM + \pa^\cM \cV_{\cM \cA} - \Omega^{\cB \sc} F_{\sc \cB \cA}~, \\
\label{E:PS.DilR}
\cR^\sa &= \cX^\sa + \Omega^{\sa \cM} \cX_\cM + \pa^\cM \Omega_{\cM}{}^{\sa}
    - \Omega^{\cB \sc} F_{\sc \cB }{}^\sa
    + \rho^{\sb \sc} F_{\sc \sb}{}^\sa~.
\end{align}
The dilaton solution corresponds to $\cX_\cM = \pa_\cM \log\Phi$ and $\cX^\sa = 0$
where $\Phi$ is gauge invariant under $F$.

\subsection{Example: \texorpdfstring{$\mathdsl{D} = G \times G$}{D=GxG}}
\label{S:GCoset.GxG}
The examples we will consider are based on the ones presented in the previous section, namely $G \times G$ and $G^{\mathbb C}$. We employ the same real semisimple Lie group $G$ as before, but additionally, we presume the existence of a subgroup $F \subset G$. The most relevant cases are when the coset $G/F$ is a symmetric space, but we will remain rather general here.

When embedded into the double Lie group $\mathdsl{D} = G \times G$, the subgroup $F$ must be isotropic. The pairing \eqref{eqn:pairDGxG} makes this constraint very restrictive and only allows for diagonal subgroups. We denote the generators $T_{\sa} = ( t_{\sa}, t_{\sa} )$ for the generators of $F$. In other words, $F$ here is a subgroup of $H$ itself. The remaining generators are assigned by requiring that the pairing $\eta_{\hcA' \hcB'}$ has to be of the form given in \eqref{eqn:etacoset} and we get
\begin{equation}
  T_{\sa} = ( t_{\sa} ,  t_{\sa} )\,, \quad
  T_{\ca} = ( t_{\ca} ,  t_{\ca} )\,, \quad
  T^{\ca} = ( t^{\ca} ,  -t^{\ca} )\,, \quad
  T^{\sa} = ( t^{\sa} , -t^{\sa} )\,.
\end{equation}
There is a subtle point here: in defining the left coset, $H \backslash \mathdsl{D}$, we arranged the generators as
$T_\hA = (t_\hA, -t_\hA)$ and $T^{\hA} = (t^{\hA}, t^{\hA})$, with the latter defining $H$.
In defining the right coset now, we have swapped the roles of lower and upper indices.\footnote{We \emph{additionally} could swap the roles of $T_A$ and $T^A$ (raising/lowering the indices respectively) to restore the original positioning of the coset indices, but this only works if $\kappa_{\ca \sb}$ vanishes, since we need $\Pair{T_\ca}{T_{\sb}} = 0$.}

Now we can build the components of the generalized vielbein.
As the coset representative, we take
\begin{equation}\label{E:GxG/F.cosetrep}
  m = ( n f, f ) = (n, e) \times (f,f)
\end{equation}
with $f\in F$ and $n$ in the dressing coset $G_{\mathrm{diag}}\backslash (G\times G) / F$.
Because $F \subset H$, some care must be taken in the choice of $n$, because this coset
representative may be rewritten as
\begin{equation}\label{E:GxG/F.cosetrep.2}
  m = (f,f) \times (f^{-1} n f, e)~.
\end{equation}
The factor on the left is an element of $H$, so its only effect is to add an exact term to the $B$-field. For this to be a good coset representative, we must be careful to choose $n$ so that $f^{-1} n f$ is a sufficiently generic element of $G$ -- namely, that it generates invertible left-invariant and right-invariant vielbeins. This is not always possible --- e.g. if $F$ contains an abelian factor that commutes with all elements of $G$.\footnote{It is even problematic for symmetric spaces if we choose $n = \exp (x^\ca t_\ca)$, since then the effect of $f$ is merely to rotate the coordinates $x^\ca$. Then the left and right-invariant vielbeins vanish on the subgroup $F$ since there is no $\rd y$ component.}

In fact, the coset representative \eqref{E:GxG/F.cosetrep.2} is nothing but the coset representative used in \eqref{E:GxG.m1} for the case $g = f^{-1} n f$. This means that the generalized vielbein we will construct must actually be equivalent to the generalized vielbein there \eqref{eqn:(g,e)genframe}, up to an exact shift in the $B$-field, and an overall $\g{OSp}$ transformation acting on the left to swap the roles and index positions of $T_\hA$ and $T^\hA$.

We can begin to see this already when we compute $V_\hM{}^\hA$:
\begin{align}
V^\hA t_\hA = \frac{1}{2} \Big(
    \rd n n^{-1} + n \rd f f^{-1} n^{-1} - \rd f f^{-1}
    \Big) = \frac{1}{2} f \,\rd g g^{-1} \,f^{-1}~.
\end{align}
It is nothing but the adjoint action of $f$ on the right invariant vector field of $g$.
More explicitly, we take
\begin{align}
V_\hM{}^\hA =
\begin{pmatrix}
\delta_\ci{}^\cj & 0 \\
0 & \tilde v_\si{}^\sb
\end{pmatrix}
\times
\begin{pmatrix}
V_\cj{}^\ca & V_\cj{}^\sa \\
\frac{1}{2} D_\sb{}^\ca & \frac{1}{2} (D-1)_\sb{}^\sa
\end{pmatrix}
\end{align}
where we use $D_\hB{}^\hA t_\hA := n t_\hB n^{-1}$ and $\tilde v = \rd f f^{-1}$.
Its inverse we denote
\begin{align}
V_\hA{}^\hM =
\begin{pmatrix}
V_\ca{}^\cj & S_\ca{}^\sb \\
V_\sa{}^\cj & S_\sa{}^\sb
\end{pmatrix} \times
\begin{pmatrix}
\delta_\cj{}^\ci & 0 \\
0 & \tilde v_\sb{}^\si
\end{pmatrix}
\end{align}
where $S_\hA{}^\sb$ was defined in \eqref{E:G/F.defS}.
Importantly, we will need the two conditions
\begin{align}
\frac{1}{2} (D-1)_\sa{}^\hB S_\hB{}^\sc &= \delta_\sa{}^\sc~, \\
\frac{1}{2} (D-1)_\sa{}^\hB V_\hB{}^\ci &= 0 \quad \implies \quad D_\sa{}^\sb V_\hB{}^\ci = V_\sa{}^\ci~.
\end{align}

We need to compute $\overline{M}_{\hcA'}{}^\hcB$. Here one needs to keep in mind that the $\hcA'$ index is in the $F$-adapted basis, whereas the $\hcB$ index is in the $H$-adapted basis. This leads to
\begin{alignat}{2}
\overline{M}{}_\sa{}^\hB 
    &= \frac{1}{2} (D - 1)_\sa{}^\hB
& \qquad
\overline{M}{}_{\sa \hB}
    &= \frac{1}{2} (D \kappa + \kappa)_{\sa \hB} (-)^{\grad{\hB}} ~, \eol
\overline{M}{}_\ca{}^\hB 
    &= \frac{1}{2} (D - 1)_\ca{}^\hB
& \qquad
\overline{M}{}_{\ca \hB}
    &= \frac{1}{2} (D \kappa + \kappa)_{\ca \hB} (-)^{\grad{\hB}} ~, \eol
\overline{M}{}^{\ca\hB }
    &= \frac{1}{2} (\kappa D + \kappa)^{\ca\hB}
& \qquad
\overline{M}{}^{\ca}{}_{\hB}
    &= \frac{1}{2} (\kappa D \kappa - 1)^\ca{}_\hB (-)^{\grad{\hB}} ~, \eol
\overline{M}{}^{\sa\hB }
    &= \frac{1}{2} (\kappa D + \kappa)^{\sa\hB}
& \qquad
\overline{M}{}^{\sa}{}_{\hB}
    &= \frac{1}{2} (\kappa D \kappa - 1)^\sa{}_\hB(-)^{\grad{\hB}} ~.
\end{alignat}
The vector pieces of the generalized vielbein are
\begin{align}
\cV_\ca{}^\ci &= \overline{M}_\ca{}^\hB V_\hB{}^\ci
    = \frac{1}{2} (D-1)_\ca{}^\hB V_\hB{}^\ci~, \\
\cV^\ca{}^\ci &= \overline{M}^{\ca\hB} V_\hB{}^\ci
    = \frac{1}{2} (\kappa D +\kappa)^{\ca \hB} V_\hB{}^\ci
\end{align}
Because $\frac{1}{2} (D-1)_\sa{}^\hB V_\hB{}^\ci = 0$
we can rewrite the first term as
\begin{align}
\kappa^{\ca \cb} \cV_\cb{}^\ci = 
    \frac{1}{2} (\kappa D-\kappa)^{\ca \hB} \,V_\hB{}^\ci~.
\end{align}

At this point, we denote the coset part of the inverse Killing metric $\kappa^{\ca \cb}$,
which we presume to be invertible with graded inverse $\eta_{\ca \cb}$,
\begin{align}
\kappa^{\ca \cb} \eta_{\cb \cc} = \delta_\cc{}^\ca (-)^{\grad{\ca}}
\end{align}
Note that $\eta_{\ca \cb}$ does not equal $\kappa_{\ca \cb}$ unless $\kappa_{\ca \sb}$ vanishes. Now on the coset, we introduce the vector fields
\begin{align}
\kappa^{\ca \cb} e_\cb{}^\ci := \frac{1}{2} \,(\kappa D)^{\ca \hB} V_\hB{}^\ci~, \qquad
\kappa^{\ca \cb} v_\cb{}^\ci := \frac{1}{2} \,\kappa^{\ca \hB} V_\hB{}^\ci~.
\end{align}
We presume these are invertible. Then we find
\begin{align}
\cV_\ca{}^\ci = e_\ca{}^\ci - v_\ca{}^\ci~ \qquad 
\cV^{\ca \ci} = \kappa^{\ca \cb} (e_\cb{}^\ci + v_\cb{}^\ci)~.
\end{align}
At this point, we can exploit a fact more familiar from $\g{O}(D,D)$ elements that can be extended to $\g{OSp}$ elements when we have a metric $\kappa^{\ca\cb}$ with inverse $\eta_{\ca\cb}$. In general, we may write
\begin{align}\label{E:GxG.cosetV}
\cV_\cA{}^\cM =
\begin{pmatrix}
e_\ca{}^\cj - v_\ca{}^\cj & \frac{1}{4} \eta_{\ca \cb} (e^\cb{}_\cj + v^\cb{}_\cj) (-)^{n+b} \\
\kappa^{\ca \cb} (e_\cb{}^\cj + v_\cb{}^\cj) & \frac{1}{4} (e_\cj{}^\ca - v_\cj{}^\ca) (-)^n 
\end{pmatrix} \times
\begin{pmatrix}
\delta_\cj{}^\ci & -\widetilde \zB_{\cj \ci} (-)^m \\
0 & \delta^\cj{}_\ci
\end{pmatrix} 
\end{align}
for some graded antisymmetric $\widetilde \zB$. We have already identified $e_\ca{}^\ci$
and $v_\ca{}^\ci$.  In our case, the two vielbeins are the pure coset parts of the left and right invariant $G \times G$ vielbeins $e_\hA{}^\hM$ and $v_\hA{}^\hM$ for $g = f^{-1} n f$, but dressed with an additional adjoint action of $f$.
Using the explicit form of the generalized vielbein, one can confirm it falls into the above form for
\begin{align}
\widetilde \zB
    &= 
    \frac{1}{8} \Big((\kappa S)^{\ca \sc} D_{\sc}{}^\cb 
    - \frac{1}{2} (\kappa S)^{\ca\sc} (\kappa S)^{\cb\sd} (D\kappa)_{\sd \sc}
    (-)^{cb}
    \Big) \,\,v^\cd \eta_{\cd \cb} \wedge v^{\cc} \eta_{\cc \ca}
    + \overline{\zB}_{\WZW}~,
\end{align}
or equivalently,
\begin{align}
\widetilde  \zB
    &= 
    \frac{1}{8} \Big((\kappa D S)^{\ca \sc} (D^{-1})_{\sc}{}^\cb 
    - \frac{1}{2} (\kappa D S)^{\ca\sc} (\kappa D S)^{\cb\sd} (D^{-1}\kappa)_{\sd \sc}
    (-)^{cb}
    \Big) \,\,e^\cd \eta_{\cd \cb} \wedge e^{\cc} \eta_{\cc \ca}
    \eol & \quad
    + \overline{\zB}_{\WZW}~,
\end{align}
In these expressions, the suppressed indices between $\kappa$ and other objects run over both the coset and subgroup indices, i.e. $(\kappa S)^{\ca \sc} = \kappa^{\ca \hB} S_\hB{}^{\sc}$.
The pure WZW term on the coset is
\begin{align}
\rd \overline{\zB}_{\WZW}
    = -\frac{1}{24} \pair{\rd n n^{-1}}{[\rd n n^{-1}, \rd n n^{-1}]}~.
\end{align}
For reference we give the translation between $\rd n n^{-1}$ and $n^{-1} \rd n$
and the 1-forms $e^A$ and $v^A$ introduced on the coset:
\begin{align}
\rd n n^{-1}
    &= v^\ca \Big(t^\cb - \frac{1}{2} (\kappa S)^{\cb \sc} (D-1)_{\sc}{}^{\hD} t_{\hD} \Big)
    \eta_{\cb \ca} \eol
n^{-1} \rd n &= e^\ca \Big(t^\cb - \frac{1}{2} (\kappa D S)^{\cb \sc} (1-D^{-1})_{\sc}{}^{\hD} t_{\hD} \Big)
    \eta_{\cb \ca}~.
\end{align}
The two vielbeins are related by a graded version of a Lorentz transformation,
\begin{align}\label{E:Lambda.GxG}
\Lambda_{\ca}{}^\cb
    := e_\ca{}^\ci v_\ci{}^\cb~, \qquad 
    \eta^{\ca \cc} \Lambda_{\cc}{}^{\cd} \eta_{\cd \cb}
    = \Lambda^{\ca}{}_{\cb} = (\Lambda^{-1})_{\cb}{}^{\ca} (-)^{ba}
\end{align}
where explicitly
\begin{align}
\Lambda^{A}{}_B = (\kappa D \kappa)^\ca{}_\cb - (\kappa D \kappa)^\ca{}_\sc \Big((D \kappa - \kappa)^{-1}\Big){}^{\sc \sd} (D\kappa)_{\sd \cb}~.
\end{align}
The remainder of the megavielbein is characterized by $\Omega$ and $\rho$:
\begin{align}
\Omega_{\ca}{}^{\sb} &= -\frac{1}{2} (D-1)_\ca{}^{\hC} S_\hC{}^{\sb}~, \qquad
\Omega^{\ca \sb} = -\frac{1}{2} (\kappa D+\kappa )^{\ca\hC} S_\hC{}^{\sb}~,\\
\rho^{\sa \sb} &= \frac{1}{2} (\kappa D + \kappa)^{[\sa |\hC} S_\hC{}^{|\sb]}~.
\end{align}

It will actually be useful for us to consider a slightly different coset representative,
which will be relevant for analytic continuation:
\begin{equation}
  m = (n', n'^{-1}) \times (f,f)
\end{equation}
The coset element $(n', n'^{-1})$ goes to its inverse under the involution $\sigma$
that exchanges left and right group factors. Thankfully, we do not however
need to perform any new computation. Similar to the generalized group manifold case,
this coset representative is related to the previous one merely by an $H$-action on
the left (which is just an exact shift in the $B$-field) and a coordinate transformation,
taking $n = n'^2$, exploiting the identification
\begin{align}
  (n', n'^{-1}) \times (f,f) = (n'^{-1}, n'^{-1}) \times (n'^2, e) \times (f,f)~.
\end{align}
Of course, it is related to the two $G \times G$ generalized group manifold cases as well.

With these facts in mind, and using what we have learned in the previous cases, we can simply describe the result here in a manner that will be useful for analytic continuation. Let $g = f^{-1} n f$ be a generic element of $G$, and similarly for $g' = f^{-1} n' f$ with $n = n'^2$ (so $g = g'^2$) . Define on the full group the modified left and right invariant forms
\begin{align}
\widehat v^\hA t_\hA &= f \rd g g^{-1} f^{-1} = \rd n n^{-1} + n \rd f f^{-1} n^{-1} - \rd f f^{-1}
~, \\
\widehat e^\hA t_\hA &= f g^{-1} \rd g f^{-1} = n^{-1} \rd n + \rd f f^{-1} - n^{-1} \rd f f^{-1} n~. \end{align}
In terms of $n'$, these can be written
\begin{align}
\widehat v^\hA t_\hA &= n' \Big(\rd n' n'^{-1} + n'^{-1} \rd n'
    + n' \rd f f^{-1} n'^{-1} - n'^{-1} \rd f f^{-1} n' \Big) n'^{-1}
~, \\
\widehat e^\hA t_\hA &= n'^{-1} \Big(\rd n' n'^{-1} + n'^{-1} \rd n'
    + n' \rd f f^{-1} n'^{-1} - n'^{-1} \rd f f^{-1} n' \Big) n'
\end{align}
Then define two vielbeins on the coset by
\begin{align}
\kappa^{\ca \cb} e_\cb{}^\ci := \kappa^{\ca \hB} \widehat e_\hB{}^{\ci}~, \qquad
\kappa^{\ca \cb} v_\cb{}^\ci := \kappa^{\ca \hB} \widehat v_\hB{}^{\ci}~,
\end{align}
and additional fields
\begin{align}
S_{\hA}{}^{\sb} := D'_{\hA}{}^{\hC} \widehat v_{\hC}{}^{\si} \tilde v_{\si}{}^{\sb}
    = (D'^{-1})_{\hA}{}^{\hC} \widehat e_{\hC}{}^{\si} \tilde v_{\si}{}^{\sb}~.
\end{align}
These equations imply that
\begin{align}
\rd n'^2 n'^{-2}
    &= v^\ca \Big(t^\cb - \frac{1}{2} (\kappa D'^{-1} S)^{\cb \sc} (D'^2-1)_{\sc}{}^{\hD} t_{\hD} \Big)
    \eta_{\cb \ca} \\
n'^{-2} \rd n'^2 &= e^\ca \Big(t^\cb - \frac{1}{2} (\kappa D' S)^{\cb \sc} (1-D'^{-2})_{\sc}{}^{\hD} t_{\hD} \Big)
    \eta_{\cb \ca}~.
\end{align}
Then the generalized supervielbein on the large space is given by \eqref{eqn:PSform}.
The connection $\Omega$ and \Polacek-Siegel field are
\begin{align}
\Omega_{\ca}{}^{\sb} &= -\frac{1}{2} (D'-D'^{-1})_\ca{}^{\hC} S_\hC{}^{\sb}~, \\
\Omega^{\ca \sb} &= -\frac{1}{2} (\kappa D'+\kappa D'^{-1})^{\ca\hC} S_\hC{}^{\sb}~,\\
\rho^{\sa \sb} &= \frac{1}{2} (\kappa D' + \kappa D'^{-1})^{[\sa |\hC} S_\hC{}^{|\sb]}~.
\end{align}
and $\widetilde \zB$ is given by
\begin{align}
\widetilde \zB
    &= 
    \frac{1}{8} \Big((\kappa D' S)^{\ca \sc} (D'^{-2})_{\sc}{}^\cb 
    - \frac{1}{2} (\kappa D' S)^{\ca\sc} (\kappa D' S)^{\cb\sd} (D'^2\kappa)_{\sd \sc}
    (-)^{cb}
    \Big) \,\,e^\cd \eta_{\cd \cb} \wedge e^{\cc} \eta_{\cc \ca}
    \eol & \quad
    + \overline{\zB}_{\WZW}
\end{align}
with
\begin{align}
\rd \overline{\zB}_{\WZW}
    &= -\frac{1}{24} \pair{\rd n'^2 n'^{-2}}{[\rd n'^2 n^{-2}, \rd n'^2 n'^{-2}]}~.
\end{align}

\subsection{Example: \texorpdfstring{$\mathdsl{D} = G^{\mathbb{C}}$}{D=G\^C}}
\label{S:GCoset.Gc}
Next, we take the complexified group $\mathdsl{D} = G^{\mathbb C}$ discussed in section \ref{S:GPS.Gc}. The subgroup $F \subset G$ is again an isotropic subgroup using the pairing
\eqref{E:Gc.pairing}. The basis \eqref{eqn:DDgens} already introduced for the $G^{\mathbb C}$ case is perfectly suitable here: we merely split the generators up so that
\begin{equation}
  T_{\sa} = t_{\sa}\,, \quad
  T_{\ca} = t_{\ca}\,, \quad
  T^{\ca} = (R^{\ca \hB} + i \kappa^{\ca \hB}) t_\hB\,, \quad
  T^{\sa} = (R^{\sa \hB} + i \kappa^{\sa \hB}) t_\hB~.
\end{equation}
Again, we do not need to impose that $\kappa_{\sa \cb}$ vanishes, although this will certainly be the case most of interest.  A natural coset representative lies in $G$ itself,
\begin{equation}\label{eqn:decompm}
  m = n f = g \in G~.
\end{equation}
Introducing the usual left invariant vector fields suitable for $G/F$,
\begin{align}\label{E:ndn}
n^{-1} \rd n = e^\ca t_\ca + \omega^\sa t_\sa
\end{align}
we easily find
\begin{align}
v_\hM{}^{\hA} = 
\begin{pmatrix}
e_\ci{}^\cb & \omega_\ci{}^\sb \\
0 & \tilde v_\si{}^{\sb}
\end{pmatrix} D_\hB{}^{\hA}~, \qquad
v_\hA{}^{\hM} = 
(D^{-1})_\hA{}^{\hB}
\begin{pmatrix}
e_\cb{}^\ci & -\omega_\ca{}^\sb \tilde v_\cb{}^\si \\
0 & \tilde v_\sb{}^{\si}
\end{pmatrix} ~, 
\end{align}
where $D_{\hA}{}^{\hB} t_{\hB} = n t_{\hA} n^{-1}$. It follows that
$S_\hA{}^\sb = (D^{-1})_\hA{}^\sb - (D^{-1})_\hA{}^\cb \omega_\cb{}^\sb$.
Computing $\overline{M}_{\hcA}{}^{\hcB}$, one finds
\begin{align}
\overline{M}_{\hcA}{}^{\hcB} =
\begin{pmatrix}
D_\hA{}^\hB & 0 \\
(R D - D R)^{\hA \hB} & D^\hA{}_\hB (-)^b
\end{pmatrix}~.
\end{align}
This leads to a generalized vielbein on the coset of
\begin{align}
\cV_\cA{}^\cM =
\begin{pmatrix}
e_\ca{}^\ci & 0 \\
\overline{\Pi}^{\ca \cb} e_\cb{}^\ci \quad & e^\ca{}_\ci (-)^m
\end{pmatrix}~.
\end{align}
The connection $\Omega$ and \Polacek-Siegel field are
\begin{align}
\Omega_\ca{}^\sb = \omega_\ca{}^\sb~, \qquad 
\Omega^{\ca \sb} = -\overline{\Pi}^{\ca \sb} + \overline{\Pi}^{\ca \cc} \omega_\cc{}^{\sb}~, \qquad
\rho^{\sa \sb} = \overline{\Pi}^{\sa \sb} - \overline{\Pi}^{[\sa |\cc} \omega_{\cc}{}^{\sb]}~.
\end{align}
The matrices $\overline{\Pi}^{\hA \hB}$ appearing above are given by
\begin{align}
\overline{\Pi}^{\hA \hB} = R^{\hA \hB} - D^{\hA \hC} R_\hC{}^{\hD} (D^{-1})_{\hD}{}^{\hB}~.
\end{align}
Note that $\overline{\Pi}^{\hA \hB}$ resembles the matrix $\Pi^{\hA \hB}$ given in \eqref{E:Gc.V.Pi}, except we have restricted the group element used to construct $D_\hA{}^\hB$ from $g$ to $n$. Of course, this is no accident: the megavielbein on the generalized coset is nothing but the generalized vielbein on the full space (up to a $B$-field gauge transformation). It is an instructive exercise to check these formulae emerge directly by comparing with the expression \eqref{E:Gc.V.Pi} and extracting $(\Adj f)_{\hcA}{}^{\hcB}$.

Just as on the generalized parallelizable space, we can make a similarity transformation to the basis
\begin{equation}
  T_{\sa} = t_{\sa}\,, \quad
  T_{\ca} = t_{\ca}\,, \quad
  T^{\ca} = i \,\kappa^{\ca \hB} t_\hB\,, \quad
  T^{\sa} = i \,\kappa^{\sa \hB} t_\hB~.
\end{equation}
This remains in \Polacek-Siegel form, except now the various constituents of the megavielbein are given by
\begin{align}\label{E:genV.Rg.coset}
\cV_\cA{}^\cM &=
\begin{pmatrix}
e_\ca{}^\ci & 0 \\
- (R_n)^{\ca \cb} e_\cb{}^\ci \quad & e^\ca{}_\ci (-)^m
\end{pmatrix}~, \qquad
(R_n)_\hA{}^{\hB} := D_\hA{}^\hC R_\hC{}^{\hD} (D^{-1})_{\hD}{}^{\hB}~, \eol
\Omega_\ca{}^\sb &= \omega_\ca{}^\sb~, \qquad 
\Omega^{\ca \sb} = -(R_n)^{\ca \cc} \omega_\cc{}^{\sb}~, \quad
\rho^{\sa \sb} = -(R_n)^{\sa \sb} + (R_n)^{[\sa \cc} \omega_{\cc}{}^{\sb]}~.
\end{align}
This case can also be easily compared with the corresponding megavielbein in \eqref{E:genV.Rg}
after extracting a factor of $(\Adj f)_{\hcA}{}^{\hcB}$.

Another interesting case is to choose
\begin{equation}
  T_{\sa} = t_{\sa}\,, \quad
  T_{\ca} = t_{\ca}\,, \quad
  T^{\ca} = i \,t^{\ca}\,, \quad
  T^{\sa} = i \,t^{\sa}
\end{equation}
We use the same decomposition for $m$ as given in \eqref{eqn:decompm} with $n$ being hermitian and $f$ unitary. For the elements of $\overline{M}_{\hcA'}{}^{\hcB}$ we find
\begin{alignat}{2}
\overline{M}{}_\sa{}^\hB 
    &= \frac{1}{2i} (D - D^{-1})_\sa{}^\hB
& \qquad
\overline{M}{}_{\sa \hB}
    &= \frac{1}{2} (D \kappa + D^{-1} \kappa)_{\sa \hB} (-)^{\grad{\hB}} ~, \eol
\overline{M}{}_\ca{}^\hB 
    &= \frac{1}{2i} (D - D^{-1})_\ca{}^\hB
& \qquad
\overline{M}{}_{\ca \hB}
    &= \frac{1}{2} (D \kappa + D^{-1} \kappa )_{\ca \hB} (-)^{\grad{\hB}} ~, \eol
\overline{M}{}^{\ca\hB }
    &= \frac{1}{2} (\kappa D + \kappa D^{-1})^{\ca\hB}
& \qquad
\overline{M}{}^{\ca}{}_{\hB}
    &= -\frac{1}{2i} (\kappa D - \kappa D^{-1})^\ca{}_\hB (-)^{\grad{\hB}} ~, \eol
\overline{M}{}^{\sa\hB }
    &= \frac{1}{2} (\kappa D + \kappa D^{-1})^{\sa\hB}
& \qquad
\overline{M}{}^{\sa}{}_{\hB}
    &= -\frac{1}{2i} (\kappa D - \kappa D^{-1})^\sa{}_\hB (-)^{\grad{\hB}} ~.
\end{alignat}
The computation is very nearly identical to the $G \times G$ coset.
We find for $V^\hA$ and $A^\hA$
\begin{align}
V^\hA = \frac{1}{2i} (\rd n n^{-1} + n^{-1} \rd n)
    + \frac{1}{2i} \tilde v^\sb (D - D^{-1})_\sb{}^\hA t_\hA~, \\
A^\hA = \frac{1}{2} (\rd n n^{-1} - n^{-1} \rd n)
    + \frac{1}{2} \tilde v^\sb (D + D^{-1})_\sb{}^\hA t_\hA~~.
\end{align}
The vector pieces of the generalized vielbein are
\begin{alignat}{3}
\kappa^{\ca \cb} \cV_\cb{}^\ci &= \kappa^{\ca \cb}
    \overline{M}_\ca{}^\hB V_\hB{}^\ci
    &\,
    &= -\frac{i}{2} (\kappa D- \kappa D^{-1})^{\ca\hB} V_\hB{}^\ci
    &\,
    &=: -i \,\kappa^{\ca \cb} (e_\cb{}^\ci - \bar e_\cb{}^\ci)
    ~, \\
\cV^\ca{}^\ci &= \phantom{\kappa^{\sa \sb}} \overline{M}^{\ca\hB} V_\hB{}^\ci
    &\,
    &= \phantom{+} \frac{1}{2} (\kappa D +\kappa D^{-1})^{\ca \hB} V_\hB{}^\ci
    &\,
    &=: \phantom{-i\,}  \kappa^{\ca \cb} (e_\cb{}^\ci + \bar e_\cb{}^\ci)
\end{alignat}
where we again exploit the vanishing of $(D - D^{-1})_{\sa}{}^{\hB} V_\hB{}^\ci$ in the first line. These expressions define the doublet of coset supervielbeins $e$ and $\bar e$.
These alternatively can be understood as $\kappa^{\ca \hB} \widehat e_\hB{}^{\ci}$ where $\widehat e_\hA{}^{\hM}$ is the inverse of
\begin{align}
i \,\widehat e^\hA t_\hA &= n^{-1} \Big(\rd n n^{-1} + n^{-1} \rd n
    + n \rd f f^{-1} n^{-1} - n^{-1} \rd f f^{-1} n \Big) n
\end{align}
and similarly for its complex conjugate,
\begin{align}
i \,\widehat {\bar e}^\hA t_\hA &= n \Big(\rd n n^{-1} + n^{-1} \rd n
    + n \rd f f^{-1} n^{-1} - n^{-1} \rd f f^{-1} n \Big) n^{-1}
\end{align}
Inspired by $G \times G$ case, these can also be written
\begin{align}
i \,\widehat e^\hA t_\hA = f m^{-1} \rd m f^{-1}~, \qquad
i \,\widehat {\bar e}^\hA t_\hA = f \rd m m^{-1} f^{-1}~, \qquad m = f^{-1} n^2 f
\end{align}
The generalized supervielbein on the coset is
\begin{align}
\cV_\cA{}^\cM =
\begin{pmatrix}
-i \,(e_\ca{}^\cj - \bar e_\ca{}^\cj) &
    \frac{1}{4} \eta_{\ca \cb} (e^\cb{}_{\cj} + \bar e^\cb{}_{\cj}) (-)^{b+n}\\[2ex]
\kappa^{\ca \cb} \,(e_\cb{}^{\cj} + \bar e_\cb{}^{\cj}) &
    \frac{i}{4} (e^\ca{}_{\cj} - \bar e^\ca{}_{\cj}) (-)^n
\end{pmatrix} \times
\begin{pmatrix}
\delta_\cj{}^\ci & -\widetilde \zB_{\cj \ci} (-)^m \\
0 & \delta^\cj{}_\ci
\end{pmatrix}
\end{align}
where
\begin{align}
\widetilde \zB &= 
    -\frac{1}{8}\Big(
    i (\kappa D S)^{\ca \sc} (D^{-2})_{\sc}{}^{\cb} \, 
    - \tfrac{1}{2} (\kappa D S)^{\ca \sc} (\kappa D S)^{\cb \sd}
    (D^2 \kappa)_{\sd \sc} (-)^{bc}
    \Big)
    e^\cd \eta_{\cd \cb} \wedge
    e^\cc \eta_{\cc \ca}
    \eol & \quad
    + \overline{\zB}_{\WZW}~,
\end{align}
with
\begin{align}
\rd \overline{\zB}_{\WZW} &= -\frac{1}{24} \pair{\rd n^2 n^{-2}}{[\rd n^2 n^{-2}, \rd n^2 n^{-2}]}~.
\end{align}
The $\Omega$ connection and \Polacek-Siegel field are
\begin{align}
\Omega_{\ca}{}^{\sb} &= -\frac{1}{2i} (D-D^{-1})_\ca{}^{\hC} S_\hC{}^{\sb}~, \\
\Omega^{\ca \sb} &= -\frac{1}{2} (\kappa D+\kappa D^{-1})^{\ca\hC} S_\hC{}^{\sb}~,\\
\rho^{\sa \sb} &= \frac{1}{2} (\kappa D + \kappa D^{-1})^{[\sa |\hC} S_\hC{}^{|\sb]}~.
\end{align}

\section{Generalized supercosets for supergravity backgrounds}
\label{S:SUGRA}

\subsection{Supergravity backgrounds in double field theory}
In order for the generalized supervielbein to describe a valid background of supersymmetric DFT, the generalized flux tensor must obey a certain set of constraints \cite{Butter:2022gbc} (for earlier work, see \cite{Hatsuda:2014qqa,Hatsuda:2014aza} and \cite{Cederwall:2016ukd}).
At dimension -1/2, all flux tensors vanish
\begin{align}\label{E:FluxConstraint1}
\cF_{\alpha \beta \gamma} = 
\cF_{\alpha \beta \bgamma} = 
\cF_{\alpha \bbeta \bgamma} = 
\cF_{\balpha \bbeta \bgamma} = 0
\end{align}
while at dimension 0,
\begin{align}\label{E:FluxConstraint2}
\cF_{\alpha \beta \rc} = -i \sqrt{2}\, (\gamma_\rc)_{\alpha \beta}~, \quad
\cF_{\balpha \bbeta \rbc} = -i \sqrt{2}\, (\bar\gamma_\rbc)_{\balpha \bbeta}~, \quad
\cF_{\alpha \bbeta \rc} = \cF_{\alpha \bbeta \rbc} =
\cF_{\alpha \beta \rbc} = \cF_{\balpha \bbeta \rc} = 0~.
\end{align}
We refer to these as $\kappa$-symmetric constraints, in analogy to their supergravity analogues \cite{Wulff:2016tju}. In addition, one imposes \emph{conventional constraints} at dimension 1/2
\begin{align}\label{E:FluxConstraint3}
\cF_{\alpha \beta}{}^{\beta} = \tfrac{1}{4} \cF_{\beta \rb \rc} (\gamma^{\rb\rc})_{\alpha}{}^\beta~, \quad
    \cF_{\balpha \bbeta}{}^{\bbeta} = \tfrac{1}{4} \cF_{\bbeta \rbb \rbc} (\gamma^{\rbb\rbc})_{\balpha}{}^\bbeta~, \quad
    \cF_{\alpha \rb \rbc} (\gamma^{\rb})^{\alpha \beta} = 
    \cF_{\balpha \rbb \rc} (\gamma^{\rbb})^{\balpha \bbeta} = 0~,
\end{align}
which amount only to redefinitions of the physical dilatini and gravitini. A final conventional constraint at dimension 1 redefines the Ramond-Ramond bispinor,
\begin{align}\label{E:FluxConstraint4}
(\gamma^{\rc})^{\alpha\beta} \cF_{\rc \beta}{}^\balpha = 
    - (\gamma^{\rbc})^{\balpha\bbeta} \cF_{\rbc \bbeta}{}^\alpha~.
\end{align}

As argued in \cite{Butter:2022gbc} (and in analogy with \cite{Wulff:2016tju}), these constraints alone lead to a generalized double field theory (which is related to \emph{modified DFT} \cite{Sakatani:2016fvh}), the DFT analogue of generalized type II supergravity, where one does not presume a dilaton to exist, see section \ref{S:SNATD.dilaton}.  We will return to the question of conventional supergravity (i.e. where a dilaton exists) in section \ref{S:SUGRA.dilaton}.

Now we can pose the question whether the generalized vielbeins we have constructed in previous sections, namely for the double Lie groups $G^{\mathbb C}$ and $G \times G$, satisfy these constraints so that they describe supergravity backgrounds. If we presume that the group $G$ should have 32 supercharges (to accommodate the full range of $\alpha$ and $\balpha$ indices we seek), ten corresponding translation generators $P_a$, and a subgroup $F$ corresponding to any Lorentz and/or $R$-symmetry groups, we are essentially restricting our attention to maximally supersymmetric type II backgrounds. These were analyzed long ago \cite{Figueroa-OFarrill:2002ecq}, with only the $\g{AdS}_5 \times \g{S}^5$ background of IIB and its Penrose limit (an Hpp wave) \cite{Blau:2001ne} relevant to us here.\footnote{There is also the IIB$^*$ background $dS_5 \times H^5$ \cite{Hull:1998vg} and its Penrose limit \cite{Singh:2004uz}, but we won't consider these.}

The supergroup $G$ of isometries for $\g{AdS}_5 \times \g{S}^5$ is $\g{PSU}(2,2|4)$ (see e.g.
\cite{Arutyunov:2009ga}). Only some of the details of this algebra are important to us, so we will treat it in rather general language. It consists of generators $t_\hA = \{ t_a, t_\alpha, t_\balpha,  t_{\iso1}\}$. The generators $t_{\iso1}$ span a (bosonic) subgroup $F = \g{SO}(4,1) \times \g{SO}(5)$. The generators $t_A = \{t_a, t_\alpha, t_\balpha\}$ comprise spatial translations and supersymmetries, and the supercoset $G / F$ is a superspace whose bosonic body is $\g{AdS}_5 \times \g{S}^5$. The superalgebra admits a $\mathbb Z_4$ grading under which $t_{\iso1}$, $t_\alpha$, $t_a$, and $t_\balpha$ carry charge 0, 1, 2, and 3. The non-vanishing (anti)commutators are
\begin{alignat}{4}
[t_{\iso1}, t_\beta] &= -f_{\iso1 \beta}{}^{\gamma} t_{\gamma}~, &\quad
[t_{\iso1}, t_\bbeta] &= -f_{\iso1 \bbeta}{}^{\bgamma} t_{\bgamma}~, &\quad
[t_{\iso1}, t_b] &= -f_{\iso1 b}{}^{c} t_{c}~, \quad
[t_{\iso1}, t_{\iso2}]= -f_{\iso1 \iso2}{}^{\iso3}~, \eol{}
\{t_\alpha, t_\beta\} &= - f_{\alpha \beta}{}^c t_c~, &\quad
\{t_\balpha, t_\bbeta\} &= - f_{\balpha \bbeta}{}^c t_c~, &\quad
\{t_\alpha, t_\bbeta\} &= - f_{\alpha \bbeta}{}^{\iso1} t_{\iso1}~, \eol{}
[t_{a}, t_\beta] &= -f_{a \beta}{}^{\bgamma} t_{\bgamma}~, &\quad
[t_{a}, t_\bbeta] &= -f_{a \bbeta}{}^{\gamma} t_{\gamma}~, &\quad
[t_a, t_b] &= -f_{a b}{}^{\iso1} t_{\iso1}~.
\end{alignat}
We normalize the generators so that the SUSY algebra is conventional with
\begin{align}
f_{\alpha \beta}{}^c = -i\, (\gamma^c)_{\alpha \beta}~, \qquad
f_{\balpha \bbeta}{}^c = -i\, (\gamma^c)_{\balpha \bbeta}~.
\end{align}
Then the structure constants $f_{AB}{}^C$ may be interpreted as the torsion tensor $T_{AB}{}^C$ of the undeformed $\g{AdS}_5 \times \g{S}^5$ background. The algebra admits a non-degenerate Cartan metric $\kappa_{\hA \hB}$ with nonzero pieces $\kappa_{ab} = \eta_{ab}$, 
$\kappa_{\alpha \bbeta} = -\kappa_{\bbeta \alpha}$, and $\kappa_{\iso1\iso2}$. 
The (graded) inverse component $\kappa^{\alpha \bbeta}$ is proportional to the Ramond-Ramond bispinor of the undeformed $\g{AdS}_5 \times \g{S}^5$ background, i.e.
$\kappa^{\alpha \bbeta} \propto \widehat F_{a_1 a_2 a_3 a_4 a_5} (\gamma^{a_1 a_2 a_3 a_4 a_5})^{\alpha\bbeta}$, since it appears in the constant torsion 
\begin{align}
T_{a \beta}{}^{\bgamma} = f_{a \beta}{}^{\bgamma} 
    = -i \,\kappa^{\bgamma \gamma} (\gamma_a)_{\gamma \beta}~, \qquad
T_{a \bbeta}{}^{\gamma} = f_{a \bbeta}{}^{\gamma} 
    = -i \,\kappa^{\gamma \bgamma} (\gamma_a)_{\bgamma \bbeta}~.
\end{align}
A crucial feature of $\kappa^{\alpha \bbeta}$ is that due to the 10D gamma matrix identity
$\gamma_a \gamma_{b_1 b_2 b_3 b_4 b_5} \gamma^a = 0$, one finds
$T_{a \beta}{}^{\bgamma} (\gamma^{a})_{\bgamma \bbeta} = T_{a \bbeta}{}^{\gamma} (\gamma^{a})_{\gamma \beta} = 0$.

\subsection{The \texorpdfstring{$\eta$}{eta}-deformation}
\label{S:SUGRA.eta}
In the context of supercoset sigma models, the $\eta$ deformation is a specific deformation that preserves the classical integrability of the original model. It depends on the existence of an $R$-matrix obeying the modified classical Yang-Baxter equation \eqref{E:mCYBE}; such models are known as (inhomogeneous) Yang-Baxter $\sigma$-models \cite{Klimcik:2002zj,Klimcik:2008eq}. For the case of the $\g{AdS}_5 \times \g{S}^5$ superstring, the Lagrangian is given by \cite{Delduc:2013qra, Delduc:2014kha}
\begin{align}\label{E:sigmaLag.eta}
\cL &= -\frac{(1-\eta^2)}{4 t} (\sqrt{-h} \,h^{ij} - \veps^{ij}) \STr \Big(
    g^{-1} \pa_i g \,\mathbf d \,\cO_-^{-1} \,g^{-1}\pa_j g
    \Big) \eol
    &= -\frac{(1-\eta^2)}{4 t} (\sqrt{-h} \,h^{ij} - \veps^{ij}) \,
    {\widehat e}_i{}^\hA
    {\widehat e}_j{}^\hB (\cO_-^{-1})_{\hB}{}^\hC \mathbf d_{\hC}{}^{\hD} \kappa_{\hD \hA}~.
\end{align}
The group element $g$ is an element of $\g{PSU}(2,2|4)$. The factor $1/t$ can be interpreted as the string tension $T$. The Lie algebra operator $\mathbf d$ is defined in terms of 
$\mathbb Z_4$ graded projectors as
$\mathbf d = P^{(1)} + \frac{2}{1-\eta^2} P^{(2)} - P^{(3)}$. As a diagonal matrix, 
$\mathbf d_\hA{}^\hB$ and its transverse are given by\footnote{Ref. \cite{Borsato:2016ose} relates operators to matrices as
$\cO \cdot \xi^\hA t_\hA = \xi^\hA t_\hB \cO^\hB{}_\hA$,
while we use $\cO \cdot \xi^\hA t_\hA = \xi^\hA \cO_\hA{}^\hB t_\hB$. This amounts to
replacing $\cO^\hB{}_\hA \rightarrow \cO_\hA{}^\hB (-)^{b+ba}$.}
\begin{alignat}{2}
\mathbf d_\alpha{}^\beta &= -(\mathbf d^T)_\alpha{}^\beta = \delta_\alpha{}^\beta~, &\qquad
\mathbf d_\balpha{}^\bbeta &= -(\mathbf d^T)_\balpha{}^\bbeta = -\delta_\balpha{}^\bbeta~,\eol
\mathbf d_a{}^b &= (\mathbf d^T)_a{}^b = \frac{2}{1-\eta^2} \delta_a{}^b &\qquad
\mathbf d_{\iso1}{}^{\iso2} &= (\mathbf d^T)_{\iso1}{}^{\iso2} = 0~.
\end{alignat}
The operator $\cO_-$ and a related operator $\cO_+$ are given in matrix form by
\begin{align}\label{E:cO.defs}
(\cO_-)_\hA{}^\hB = \delta_\hA{}^\hB - \eta \,\mathbf d_{\hA}{}^{\hC} (R_g)_\hC{}^{\hB}~,
\qquad
(\cO_+)_\hA{}^\hB = \delta_\hA{}^\hB + \eta \,(\mathbf d^T)_{\hA}{}^{\hC} (R_g)_\hC{}^{\hB}~. \end{align}
The Lagrangian \eqref{E:sigmaLag.eta} can be rewritten in Green-Schwarz form as
\begin{align}\label{E:sigmaLag.GSform}
\cL = -\frac{T}{2} \sqrt{-h} h^{i j} \STr(A_{-i}^{(2)} A_{-j}^{(2)})
    + \frac{T}{2} \veps^{ij} \STr(A_{-i} \widehat B A_{-j})
\end{align}
where $A_- = \cO_-^{-1} (g^{-1} \rd g)$ and
\begin{align}
T = \frac{1}{t}~, \qquad
\widehat B = \frac{1-\eta^2}{2} \Big(P^{(1)} - P^{(3)} + \eta\, \mathbf d^T R_g \mathbf d\Big)~.
\end{align}
It is straightforward to show that if one decomposes $g = n f$ for $f \in \g{SO}(1,4) \times \g{SO}(5)$, the $f$ factor drops out, so this is indeed describing the supercoset.

In the seminal work \cite{Borsato:2016ose}, Borsato and Wulff analyzed the supergeometry of the $\eta$-model, establishing that its $\kappa$-symmetry was of the GS form and deriving a condition on the $R$-matrix (dubbed a unimodularity condition) for the background to be a supergravity solution. Our goal in this section is to analyze the $\eta$-deformed model purely on group theoretic grounds and show how the relevant structures of the $\sigma$-model emerge purely from the doubled supergeometry.

The starting point is the complexification $G^{\mathbb C}$ of the group $G = \g{PSU}(2,2|4)$. As we have already discussed in section \ref{S:GPS.Gc}, the complexified group involves the addition of generators $\tilde t_\hA = i \,t_\hA$, obeying
\begin{align}
[t_{\hA}, \tilde t_{\hB}] = - f_{\hA \hB}{}^{\hC} \tilde t_{\hC}~, \qquad
[\tilde t_{\hA}, \tilde t_{\hB}] = + f_{\hA \hB}{}^{\hC} t_{\hC}~,
\end{align}
with Killing form built from imaginary part of the Killing form on $G$, so that
\begin{align}
\Pair{t_{\hA}}{t_{\hB}} = \Pair{\tilde t_{\hA}}{\tilde t_{\hB}} = 0~, \qquad
\Pair{t_{\hA}}{\tilde t_{\hB}} = \kappa_{\hA\hB}~.
\end{align}

We want to find a new basis for this supergroup, for which the structure constants can be interpreted as generalized flux tensors for a supergravity background. Denote the generators of this new basis $T_{\hcA} = (T_{\iso1}, T_\cA, T^{\iso1})$ with pairing
\begin{align}
\Pair{T_\hcA}{T_{\hcB}} = \eta_{\hcA \hcB} =
\begin{pmatrix}
0 & 0 & \delta_{\iso1}{}^{\iso2} \\
0 & \eta_{\cA \cB} & 0 \\
\delta^{\iso1}{}_{\iso2} & 0 & 0
\end{pmatrix}~.
\end{align}
The generators 
$T_\cA = (T_{\alpha}, \,T_{\balpha}, \,T_{\ra}, \,T_{\rba}, \,T^{\alpha}, \,T^\balpha)$
will parametrize the generalized supercoset with pairing $\eta_{\cA \cB}$ given by \eqref{E:eta.cAcB}.  A few basic assumptions will help us choose these generators:
\begin{itemize}
\item The only group invariant is presumed to be the Killing superform. This suggests that the
new basis of generators $T_\hcA$ should be very simply written in terms of the old basis,
\begin{align}
T_\hA = a_{(\hA)} \,t_\hA + b_{(\hA)} \,\tilde t_\hA~, \qquad
T^\hA = c_{(\hA)} \,\kappa^{\hA \hB} t_\hB + d_{(\hA)} \,\kappa^{\hA \hB} \tilde t_\hB~, 
\end{align}
where $a$, $b$, $c$, and $d$ correspond to numerical constants and
no summation on the parenthetical indices is assumed. This implies that
the flux tensors will all be proportional to the original structure constants,
$\cF_{\hA \hB \hC} \propto f_{\hA \hB \hC}$.

\item $T_{\iso1} = t_{\iso1}$, in order to preserve the coset interpretation, with the Lorentz generator acting on all other generators in the expected way.

\item The structure constants must obey the supergravity
constraints. This means that all the dimension -1/2 components vanish,
$\cF_{\alpha \beta \gamma} = \cF_{\alpha \beta \bgamma} = \cF_{\alpha \bbeta \bgamma} = 
\cF_{\balpha \bbeta \bgamma} = 0$.
This is automatic because there is no corresponding structure constant in the original algebra (since the structure constants are bosonic quantities). The dimension 0 components should also be constrained to obey
\begin{align}\label{E:DimZeroFlux}
\cF_{\alpha \beta \rc} = \sqrt{2} \,f_{\alpha \beta c}~, \qquad
\cF_{\balpha \bbeta \rbc} = -\sqrt{2}  f_{\balpha \bbeta c} ~, \qquad
\cF_{\alpha \bbeta \rc} = \cF_{\alpha \bbeta \rbc} = 0~.
\end{align}
Additional constraints apply at dimension 1/2; however, these are fermionic  and must vanish since the fluxes correspond to structure constants of a supergroup (just as for dimension -1/2). Finally, at dimension 1, we will also require \eqref{E:FluxConstraint4}.

\end{itemize}

The most general possibility for
$T_\alpha$ and $T_\balpha$ is
\begin{subequations}
\begin{align}\label{E:Ta.eta.1}
T_\alpha = a_1 \Big(t_\alpha + \eta \,\tilde t_\alpha\Big)~, \qquad
T_\balpha = a_2 \Big(t_\balpha - \eta \,\tilde t_\balpha\Big)~.
\end{align}
We choose an arbitrary parameter $\eta$ and normalization $a_1$ to define $T_\alpha$.
The fact that $-\eta$ appears in $T_\balpha$ is a direct consequence of 
$\Pair{T_\alpha}{T_\bbeta} = 0$. From the basic dimension zero flux constraint \eqref{E:DimZeroFlux},
we can deduce $T_\ra$ from $\{T_\alpha, T_\beta\}$ and similarly for $T_\rba$:
\begin{align}
T_\ra = \frac{(a_1)^2}{\sqrt 2} \Big( (1-\eta^2) t_a + 2 \eta \,\tilde t_a\Big)~, \qquad
T_\rba = \frac{(a_2)^2}{\sqrt 2} \Big( (1-\eta^2) t_a - 2 \eta \,\tilde t_a\Big)~.
\end{align}
The dimension zero flux also fixes $T^\alpha$ using $[T_\alpha, T_\rb]$ (and similarly for $T^\balpha$) as
\begin{align}
T^\alpha &= \frac{(a_1)^3}{2} \Big(
    (1-3\eta^2) t^\alpha + \eta (3-\eta^2) \tilde t^\alpha
    \Big)~, \quad
T^\balpha = \frac{(a_2)^3}{2} \Big(
    -(1-3\eta^2) t^\balpha + \eta (3-\eta^2) \tilde t^\balpha
    \Big)~.
\end{align}
The Lorentz generator and its dual can only be
\begin{align}
T_{\iso1} = t_{\iso1}~, \qquad T^{\iso1} = \tilde t^{\iso1}
\end{align}
\end{subequations}
in order to satisfy $\Pair{T_{\iso1}}{T^{\iso2}} = \delta_{\iso1}{}^{\iso2}$ and
$\Pair{T^{\iso1}}{T^{\iso2}} = 0$. From $\Pair{T_\ra}{T_\rb} = \eta_{\ra \rb} = \eta_{ab}$ and
$\Pair{T_\rba}{T_\rbb} = \eta_{\ol{\ra \rb}} = -\eta_{ab}$, we find the normalizations
\begin{align}\label{E:ai.norm}
(a_1)^4 = (a_2)^4 = \frac{1}{2\eta (1-\eta^2)}~.
\end{align}
This fixes the range of $\eta$ as $0 < \eta < 1$ or $\eta < -1$. 
We fix the phases of $a_1$ and $a_2$ by choosing them to be positive real numbers.
We summarize the full set of structure constants in Appendix \ref{A:FluxTensors}.

There are two equivalent paths to the supervielbein, depending on whether we want to view it as the supervielbein for the generalized parallelizable space (section \ref{S:GPS.Gc}) or for the generalized coset (section \ref{S:GCoset.Gc}). While the most direct path is the latter, it will be more instructive to use the former construction to generate the megavielbein directly, since this is closer in spirit to the results of \cite{Borsato:2016ose}. Recall that for $G^{\mathbb C}$, we gave a simple form for the generalized supervielbein in the basis $t_\hA$ and $\tilde t^\hA = i\, t^\hA$ in \eqref{E:genV.Rg} (promoting unhatted indices to hatted ones). The construction involved the left-invariant vector fields $\widehat e^\hA t_\hA = g^{-1} \rd g$ and the $R$-matrix $R^{\hA \hB}$ obeying the mCYBE \eqref{E:mCYBE}. Then one simply can apply the dictionary derived above for relating $t_\hA$ and $\tilde t^\hA$ to the generators $T_\hcA$ we actually want. This gives a simple similarity transformation which can be applied to give the generalized supervielbein.

Actually, in order to match normalizations, we need to rescale the generalized supervielbein with a dimensionful parameter (this is related to rescaling the worldsheet tension):
\begin{align}\label{E:RescaleV.eta}
\widehat \cV'_{\hcA}{}^{\hcM} = 
    \cW_{\hcA}{}^{\hcB} \widehat \cV_{\hcB}{}^{\hcN} \cU_{\hcN}{}^{\hcM}~.
\end{align}
The $\cW$ factor rescales the flat indices, with nonzero entries
\begin{gather}
\cW_{\iso1}{}^{\iso2} = \delta_{\iso1}{}^{\iso2}~, \quad
\cW_{\alpha}{}^{\beta} = v^{1/2} \delta_{\alpha}{}^{\beta}~, \quad
\cW_{\balpha}{}^{\bbeta} = v^{1/2} \delta_{\balpha}{}^{\bbeta}~, \quad
\cW_{\ra}{}^{\rb} = v \,\delta_{\ra}{}^{\rb}~, \quad
\cW_{\rba}{}^{\rbb} = v \,\delta_{\rba}{}^{\rbb}~, \eol
\cW^{\alpha}{}_{\beta} = v^{3/2} \delta^{\alpha}{}_{\beta}~, \quad
\cW^{\balpha}{}_{\bbeta} = v^{3/2} \delta^{\balpha}{}_{\bbeta}~, \quad
\cW^{\iso1}{}_{\iso2} = v^2 \delta^{\iso1}{}_{\iso2}~, 
\label{E:RescaleV.flat}
\end{gather}
The parameter $v$ carries mass dimension, and the choices above reflect the engineering dimensions of $\widehat D_\hcA$. The $\cU$ factor rescales the dual derivative $\pa^\hM$,
\begin{align}
\cU_{\hcN}{}^{\hcM} =
\begin{pmatrix}
\delta_\hN{}^{\hM} & 0 \\
0 & v^{-2} \delta^{\hN}{}_{\hM}
\end{pmatrix}~.
\end{align}
The choice of $v^{-2}$ here is needed to ensure that $\widehat \cV'$ remains an $\g{OSp}$ element with unchanged $\eta_{\hcA \hcB}$ and $\eta_{\hcM \hcN}$. We drop the prime from now on. After this redefinition, the fluxes are unchanged except for an overall rescaling by $v$ consistent with their engineering dimension. To match conventions in \cite{Borsato:2016ose}, we will choose
\begin{align}
v = \sqrt{\frac{2\eta}{1-\eta^2}}~.
\end{align}
The generalized supervielbein can be read off the covariant derivatives. Using the matrices $(\cO_\pm)_\hA{}^\hB$ introduced earlier they are
\begin{subequations}
\label{E:nabla.eta}
\begin{align}
\label{E:nabla.eta.r}
\widehat D_{\iso1} &= {\widehat e}_{\iso1}{}^{\hM} \pa_\hM~, \\[2ex]
\label{E:nabla.eta.alpha}
\widehat D_\alpha
    &= \frac{1}{\sqrt{1-\eta^2}} \Big(
        (\cO_\pm)_\alpha{}^\hB {\widehat e}_\hB{}^\hM \pa_\hM
        + \frac{1}{2} (1-\eta^2) {\widehat e}_\hM{}^{\bbeta} \kappa_{\bbeta \alpha}\, \pa^\hM
        \Big) \\
\label{E:nabla.eta.balpha}
\widehat D_\balpha
    &= \frac{1}{\sqrt{1-\eta^2}} \Big(
        (\cO_\pm)_\balpha{}^\hB {\widehat e}_\hB{}^\hM \pa_\hM
        - \frac{1}{2} (1-\eta^2) {\widehat e}_\hM{}^{\beta} \kappa_{\beta \balpha}\, \pa^\hM
        \Big)~,\\[2ex]
\label{E:nabla.eta.ra}
\widehat D_\ra &= \frac{1}{\sqrt 2} \Big(
    (\cO_-)_a{}^\hB {\widehat e}_\hB{}^\hM \pa_\hM
    + {\widehat e}_\hM{}^{b} \eta_{b a} (-)^m \pa^{\hM} 
\Big)~, \\
\label{E:nabla.eta.rba}
\widehat D_\rba &= \frac{1}{\sqrt 2} \Big(
    (\cO_+)_a{}^\hB {\widehat e}_\hB{}^\hM \pa_\hM
    - {\widehat e}_\hM{}^{b} \eta_{b a} (-)^m \pa^{\hM} 
\Big)~, \\[2ex]
\label{E:nabla.eta.talpha}
\widehat D^\alpha &= \frac{1}{2 \sqrt{1-\eta^2}} \Big(
    + 4 \kappa^{\alpha \bbeta} {\widehat e}_{\bbeta}{}^\hM \pa_\hM
    - \frac{3-\eta^2}{1-\eta^2} (\cO_\pm)^{\alpha \hB} {\widehat e}_{\hB}{}^\hM \pa_\hM
    + \frac{1}{2} (3-\eta^2) {\widehat e}_\hM{}^\alpha \pa^\hM
\Big)~, \\
\label{E:nabla.eta.tbalpha}
\widehat D^\balpha &= \frac{1}{2 \sqrt{1-\eta^2}}  \Big(
    -4 \kappa^{\balpha \beta} {\widehat e}_{\beta}{}^\hM \pa_\hM
    + \frac{3-\eta^2}{1-\eta^2} (\cO_\pm)^{\balpha \hB} {\widehat e}_{\hB}{}^\hM \pa_\hM
    + \frac{1}{2} (3-\eta^2) {\widehat e}_\hM{}^\balpha \pa^\hM
\Big)~, \\[2ex]
\label{E:nabla.eta.tr}
\widehat D^{\iso1} &= -\frac{2\eta^2}{1-\eta^2} (R_g)^{\iso1 \hB} {\widehat e}_\hB{}^{\hM} \pa_\hM
    + {\widehat e}_\hM{}^{\iso1} (-)^m \pa^\hM~.
\end{align}
\end{subequations}
It is worth emphasizing here that $(\cO_+)_{\alpha}{}^\hB = (\cO_-)_{\alpha}{}^\hB$
and similarly for $\balpha$; this is apparent from the operators themselves, but it is a \emph{requirement} from the underlying structure of supersymmetric DFT, see the second line of \eqref{E:cV1}.

The supervielbein implicit in \eqref{E:nabla.eta} is not immediately written in \Polacek-Siegel form. In particular, it has dependence on the subgroup coordinates $y$. However, it is easy enough to put it into that form. Decomposing the group element as $g = n \times f$, the $G$ vielbeins ${\widehat e}_\hA{}^\hM$ employed above can be rewritten as
\begin{align}\label{E:Inverse.e.coset}
{\widehat e}_\hA{}^\hM =
(\Adj f)_\hA{}^\hB \,{\overline e}_{\hB}{}^\hM~, \qquad
{\overline e}_\hA{}^\hM
= \begin{pmatrix}
e_A{}^M & -\omega_A{}^{\iso1} \tilde v_{\iso1}{}^{\si} \\
0 & \tilde v_{\iso1}{}^{\si}
\end{pmatrix}
\end{align}
with $e$ and $\omega$ defined as in \eqref{E:ndn}.
Conjugation by $\Adj f$ leaves the diagonal matrices $\mathbf d$ and $\mathbf d^T$ invariant and replaces $R_g$ with $R_n$. This leaves an overall $\Adj f$ on the very outside of the megavielbein as in \eqref{eqn:PSform}. The fields on the coset simply correspond to replacing $g$ with $n$ in the operators $\cO_\pm$ and dropping the $\Adj f$ factor in \eqref{E:Inverse.e.coset}. We denote $\overline{\cO}_\pm$ as the operators \eqref{E:cO.defs} with $g$ replaced by $n$. The result coincides with applying the  similarity transformation for $T_\cA$ to the coset supervielbein \eqref{E:genV.Rg.coset} directly.

As discussed in section \ref{S:SDFT.OSp}, one can read off from these the components of the physical supervielbein. First, one identifies\footnote{The fact that the index sum is over $B$ and not $\hB$ comes from the upper triangular structure of $e_\hA{}^\hM$ in \eqref{E:Inverse.e.coset}. One could equivalently write
$\cE_\alpha{}^M = \frac{1}{\sqrt{1-\eta^2}} (\Adj f^{-1})_\alpha{}^\beta
    (\cO_-)_\beta{}^\hC e_\hC{}^M$
with the full $\cO_-$ and $e_\hA{}^\hM$ depending on $y$.}
\begin{subequations}
\begin{alignat}{2}
\cE_\alpha{}^M &= \frac{1}{\sqrt{1-\eta^2}} (\overline\cO_-)_\alpha{}^B e_B{}^M~, &\qquad
\bar \cE_\alpha{}^M &= \frac{1}{\sqrt{1-\eta^2}} (\overline\cO_+)_\alpha{}^B e_B{}^M~, \\
\cE_\balpha{}^M &= \frac{1}{\sqrt{1-\eta^2}} (\overline\cO_-)_\balpha{}^B e_B{}^M~, &\qquad
\bar \cE_\balpha{}^M &= \frac{1}{\sqrt{1-\eta^2}} (\overline\cO_+)_\balpha{}^B e_B{}^M~, \\
\cE_\ra{}^M &= (\overline\cO_-)_a{}^B e_B{}^M~, &\qquad
\bar \cE_\rba{}^M &= (\overline\cO_+)_a{}^B e_B{}^M~.
\end{alignat}
\end{subequations}
The fact that it is $e_B{}^M$ rather than $\overline{e}_\hB{}^M$ appearing here is a consequence of the triangular form of \eqref{E:Inverse.e.coset}. Their inverses are
\begin{subequations}
\begin{alignat}{2}
\cE_M{}^\alpha &= \sqrt{1-\eta^2}\, e_M{}^B (\overline\cO_-^{-1})_B{}^\alpha~, &\qquad
\bar \cE_M{}^\alpha &= \sqrt{1-\eta^2}\, e_M{}^B (\overline\cO_+^{-1})_B{}^\alpha~, \\
\cE_M{}^\balpha &= \sqrt{1-\eta^2}\, e_M{}^B (\overline\cO_-^{-1})_B{}^\balpha~, &\qquad
\bar \cE_M{}^\balpha &= \sqrt{1-\eta^2}\, e_M{}^B (\overline\cO_+^{-1})_B{}^\balpha~, \\
\cE_M{}^\ra &= e_M{}^B (\overline\cO_-^{-1})_B{}^a~, &\qquad
\bar \cE_M{}^\rba &= e_M{}^B (\overline\cO_+^{-1})_B{}^a~.
\end{alignat}
\end{subequations}
It is crucial that $(\overline\cO_\pm^{-1})_{\iso2}{}^A = 0$ for the inverses to have such a simple structure.

The $\g{OSp}$ structure \emph{requires} that $\cE_M{}^\ra$ and $\cE_M{}^\rba$ be related by a Lorentz transformation,
\begin{align}
\Lambda_\ra{}^\rbb = (\overline{\cO}_-)_a{}^\hC (\overline{\cO}_+^{-1})_\hC{}^b~.
\end{align}
That this matrix is a Lorentz transformation was observed in \cite{Borsato:2016ose}. There the operator $M = \cO_-^{-1} \cO_+$ was introduced; its matrix form is
\begin{align}
M_\hA{}^\hB =
\begin{pmatrix}
(\Lambda^{-1})_a{}^b & M_a{}^\beta & M_a{}^{\bbeta} & M_a{}^{\iso2} \\
0 & \delta_\alpha{}^\beta & 0 & 0 \\
0 & 0 & \delta_\balpha{}^\bbeta  & 0 \\
0 & 0 & 0 & \delta_{\iso1}{}^{\iso2}
\end{pmatrix}~.
\end{align}
It is not hard to show that $\det \Lambda^{-1} = \sdet M = \sdet \overline \cO_+ / \sdet \overline \cO_- = 1$, with the last equality following from $\sdet \overline\cO_+^T = \sdet \overline\cO_-$. This guarantees that we are dealing with an $\g{SO}(1,9)$ transformation, so the duality frame must be IIB or IIB$^*$. Actually, it is clear that  $\Lambda_\ra{}^\rbb \in \g{SO}^+(1,9)$ for $\eta$ sufficiently small, since it is continuously deformable to the identity; this property should hold so long as we restrict to the $\eta$ locus where $\cO_\pm$ is invertible. Then the vielbein and gravitino one-forms can be read off from \eqref{E:cE-E.dictionary}
\begin{subequations}
\begin{align}
\label{E:eta.deformation.grav}
E_M{}^a &= e_M{}^B (\overline\cO_-^{-1})_B{}^a~, \\
E_M{}^{1\alpha} &= \sqrt{1-\eta^2} \,e_M{}^B (\overline\cO_+^{-1})_B{}^\alpha~,  \\
E_M{}^{2\alpha} &= \sqrt{1-\eta^2} \,e_M{}^B (\overline\cO_-^{-1})_B{}^\bbeta (\Lambda^{-1})_{\bbeta}{}^\alpha~.
\end{align}
\end{subequations}
Since $\Lambda_\ra{}^\rbb \in \g{SO}^+(1,9)$, the second gravitino is of the same chirality as the first, so we have written the above in terms of 16-component Weyl spinors.

These superficially differ from the corresponding formulae in \cite{Borsato:2016ose} in a few ways. The first is that the expressions in \cite{Borsato:2016ose} are defined on the full group manifold rather than the physical coset. This means the expressions above have the indices $M$ and $B$ replaced with $\hM$ and $\hB$ and the operator $\overline\cO_\pm$ replaced with $\cO_\pm$. As we have discussed, an overall $\Adj{f}$ action (a Lorentz transformation) accounts for the change in the operators, and $(\overline\cO_\pm^{-1})_{\iso2}{}^A = 0$ allows for the restriction of the indices to the coset. The second issue also involves a Lorentz transformation: the $\Lambda$ factor is moved off the second gravitino and onto the first gravitino and vielbein (modifying $\overline\cO_-^{-1}$ to $\overline\cO_+^{-1}$ for the latter).

We similarly can read off the dilatini directly using \eqref{E:dilatini.defs}:
\begin{align}
\chi_{1\alpha} &= \frac{i}{2} \cE_\ra{}^M \bar \cE_{M}{}^\beta (\gamma^a)_{\beta \alpha}
    = \frac{i}{2} \sqrt{1-\eta^2}\, (\overline\cO_- \overline\cO_+^{-1})_a{}^\beta (\gamma^a)_{\beta \alpha}~, \\
\chi_{2\alpha} &= \frac{i}{2} \Lambda_{\alpha}{}^{\bbeta}
    \bar \cE_\rba{}^M \cE_{M}{}^\bgamma (\gamma^\rba)_{\bgamma \bbeta}
    = \frac{i}{2} \sqrt{1-\eta^2}\, (\overline\cO_+ \overline\cO_-^{-1})_a{}^\bgamma 
        (\gamma^a)_{\bgamma \bbeta} \Lambda_{\alpha}{}^\bbeta~.
\end{align}
These agree with \cite{Borsato:2016ose} although the intermediate expressions differ.
The Ramond-Ramond bispinor can be read off from either $\widehat D^\alpha$ or $\widehat D^\balpha$ using
\begin{align}
S^{\alpha \bbeta} &= -\cV^{\alpha M} \cE_M{}^{\bbeta}
    = \phantom{+}\frac{1}{2} \Big(
    \frac{3-\eta^2}{1+\eta^2} \kappa^{\alpha \bbeta}
    - 4 \,(\overline\cO_-^{-1})^{\alpha \bbeta}
    \Big) \eol
    &= - \cV^{\bbeta M} \bar \cE_M{}^\alpha
    = - \frac{1}{2} \Big(
    \frac{3-\eta^2}{1+\eta^2} \kappa^{\bbeta \alpha}
    - 4 \,(\overline\cO_+^{-1})^{\bbeta \alpha}
    \Big)
\end{align}
and applying \eqref{E:RRtoF}. 

To recover the original $\sigma$-model is straightforward. It should be of Green-Schwarz form \eqref{E:sigmaLag.GSform}, since we have imposed the Green-Schwarz constraints. The symmetric term matches the vielbein \eqref{E:eta.deformation.grav}. The antisymmetric term is recovered by working out the $B$-field by comparing \eqref{E:nabla.eta} with \eqref{E:cV2}. The result is
\begin{align}
B &= -e^D (\overline\cO_-^{-1})_D{}^B\,
    \wedge e^C (\overline\cO_-^{-1})_C{}^A\,
    \widehat B_{A B}~, \eol
\widehat B_A{}^B &= \frac{1-\eta^2}{2} \Big(\delta_\alpha{}^\beta- \delta_\balpha{}^\bbeta
    + \eta \,(\mathbf d^T R_n \mathbf d)_A{}^B
    \Big)~,
\end{align}
in agreement with \eqref{E:sigmaLag.GSform}. Note that the supergeometry does not determine the overall normalization $T$ of the Lagrangian.

\subsection{The \texorpdfstring{$\lambda$}{lambda}-deformation}
\label{S:SUGRA.lambda}

The $\lambda$-deformation \cite{Sfetsos:2013wia,Hollowood:2014rla} (see also \cite{Tseytlin:1993hm}) was extended to $\g{AdS}_5 \times \g{S}^5$ in \cite{Hollowood:2014qma}. Strictly speaking, this is not a deformation of the $\g{AdS}_5 \times \g{S}^5$ superstring but rather a deformation of its non-abelian T-dual. The Lagrangian can be written\footnote{The normalization in \cite{Borsato:2016ose} differs from \cite{Hollowood:2014qma} by a factor of $1/4$. We follow the normalization of \cite{Hollowood:2014qma}.}
\begin{align}\label{E:sigmaLag.lambda}
\cL = -\frac{k}{8\pi} (\sqrt{-h} h^{ij} - \veps^{ij}) \STr \Big(
    g^{-1} \pa_i g \, (1 + \widehat \zB - 2 \, \cO_-^{-1}) \,g^{-1} \pa_j g 
    \Big)~.
\end{align}
As with the $\eta$-deformation, the group element $g$ lies in $\g{PSU}(2,2|4)$. The constant $k$ is the level of the WZW model, and the antisymmetric operator $\widehat \zB$ generates the WZW term.
The Lie algebra operators $\cO_\pm$ are given by
\begin{alignat}{2}
\cO_- &= 1 - \Adj g^{-1} \Omega~, 
&\qquad
\Omega &= P^{(0)} + \lambda^{-1} P^{(1)} + \lambda^{-2} P^{(2)} + \lambda\, P^{(3)}~, \eol
\cO_+ &= \Adj g^{-1} - \Omega^T~,
&\qquad
\Omega^T &= P^{(0)} + \lambda \,P^{(1)} + \lambda^{-2} P^{(2)} + \lambda^{-1} P^{(3)}~.
\end{alignat}
Just as for the $\eta$ deformation, the Lagrangian \eqref{E:sigmaLag.lambda} can be put into GS form \eqref{E:sigmaLag.GSform} with
\begin{align}
T = \frac{k}{4\pi} (\lambda^{-4} - 1)~, \qquad
\widehat B = (\lambda^{-4} - 1)^{-1} \Big(
    \cO_-^T \widehat \zB \cO_-
    + \Omega^T \Adj g
    - \Adj g^{-1} \Omega
    \Big)~.
\end{align}
The string tension is positive for $k>0$ and $|\lambda|<1$ or $k<0$ and $|\lambda|>1$. These two parameter regions are related by taking $g \rightarrow g^{-1}$.

Just as for the $\eta$-deformation, we want to recover the supergeometry of this Green-Schwarz $\sigma$-model purely from the algebra. The underlying group structure of the $\lambda$ deformation is $\mathdsl{D} = G \times G$ with generators
\begin{align}
t_\hA^{(L)} = (t_\hA,0)~, \qquad
t_\hA^{(R)} = (0,t_\hA)~.
\end{align}
In terms of these, we can build $T_\hcA$ that satisfy the supergravity constraints, under the same simplifying assumptions as for the $\eta$-deformation:
\begin{subequations}
\label{E:Ta.lambda}
\begin{alignat}{2}
\label{E:Ta.lambda.1}
T_\alpha &= b_1 \Big( t_\alpha^{(L)} + \lambda^{-1} t_\alpha^{(R)} \Big)~, 
&\qquad
T_\balpha &= b_2 \Big( \lambda^{-1} t_\alpha^{(L)} + t_\alpha^{(R)} \Big)~, \\
\label{E:Ta.lambda.2}
T_\ra &= \frac{(b_1)^2}{\sqrt 2} \Big( t_a^{(L)} + \lambda^{-2} t_a^{(R)}\Big)~, 
&\qquad
T_\rba &= \frac{(b_2)^2}{\sqrt 2} \Big( \lambda^{-2} t_a^{(L)} + t_a^{(R)}\Big)~, \\
\label{E:Ta.lambda.3}
T^\alpha &= \frac{(b_1)^3}{2} \kappa^{\alpha \bbeta} \Big(
    t^{(L)}_{\bbeta} + \lambda^{-3} t^{(R)}_\bbeta \Big)~,
&\qquad
T^\balpha &= -\frac{(b_2)^3}{2} \kappa^{\balpha \beta}
    \Big(\lambda^{-3} t^{(L)}_\beta + t^{(R)}_\beta \Big)~, \\
\label{E:Ta.lambda.4}
T_{\iso1} &= t^{(L)}_{\iso1} + t^{(R)}_{\iso1}~,
&\quad
T^{\iso1} &= \kappa^{\iso1 \iso2} (t^{(L)}_{\iso2} - t^{(R)}_{\iso2})~.
\end{alignat}
\end{subequations}
The choices for $T_\alpha$ and $T_\balpha$ are the most general expressions subject to the condition $\Pair{T_\alpha}{T_{\bbeta}} = 0$. The expressions for $T_\ra$, $T_\rba$, $T^\alpha$, and $T^\balpha$ follow from requiring the canonical choice of the dimension zero flux tensor. The choice of $T_{\iso1}$ is obvious, and $T^{\iso1}$ is dictated by orthonormality. Requiring $\Pair{T_\ra}{T_\rb} = -\Pair{T_\rba}{T_\rbb} = \eta_{ab}$ fixes the normalizations $b_1$ and $b_2$ as
\begin{align}\label{E:bi.norm}
(b_1)^4 = (b_2)^4 = \frac{4}{1-\lambda^{-4}}~.
\end{align}
We find here $|\lambda|>1$. This comes about for several related reasons -- the choice of $\lambda^{-1}$ rather than $\lambda$ in \eqref{E:Ta.lambda}, the sign choice of Killing metric for the left and right sectors, etc. The reason we keep this choice is that it better matches the explicit expressions in \cite{Borsato:2016ose} \emph{provided} we keep our coset representative \eqref{E:GxG.m1} for $G \times G$. Replacing $g$ with $g^{-1}$ (or equivalently taking $m = (e, g)$) and flipping $\lambda^{-1}$ to $\lambda$ would give the same expressions as \cite{Borsato:2016ose}, but now with $|\lambda|<1$, as in the $\sigma$-model.

Now we apply the generalized parallelizable space construction for $G\times G$ in section \ref{S:GPS.GxG}, using the coset representative \eqref{E:GxG.m1}.
As with the $\eta$-deformation, one can introduce a dimensionful parameter $v$ when defining the generalized supervielbein. We employ the same redefinitions \eqref{E:RescaleV.eta} as for the $\eta$-deformation, but now subject to the normalization
\begin{align}
v^2 = (b_1)^{-4} = (b_2)^{-4}= \frac{1}{4} (1- \lambda^{-4})~.
\end{align}
For convenience, we isolate the phases of $b_i$ by $\hat b_i = b_i / |b_i|$, so that $b_i = v^{-1/2} \hat b_i$. 

The expressions for $\widehat D_{\hcA}$ are a bit more cumbersome than for the $\eta$-deformation:
\begin{subequations}
\label{E:nabla.lambda}
\begin{align}
\widehat D_{\iso1} &= 
    (1 - \Adj{g}^{-1})_{\iso1}{}^{\hB} {\widehat e}_\hB{}^\hM \cD_\hM
    + \frac{1}{4} v^{-2} {\widehat e}_{\hM}{}^{\hB} (1 + \Adj g)_{\hB\iso1} \pa^\hM \,(-)^m \\[2ex]
\widehat D_\alpha &= 
\hat b_1\, \Big[(\cO_-)_\alpha{}^\hB {\widehat e}_\hB{}^{\hM} \cD_\hM
+ \frac{1}{4} v^{-2} {\widehat e}_{\hM}{}^\hB  
    (1+\lambda^{-1} \Adj g)_{\hB \alpha} \pa^\hM\, \Big]~, \\
\widehat D_\balpha &= 
\hat b_2 \Big[-(\cO_+)_\balpha{}^\hB {\widehat e}_\hB{}^{\hM} \cD_\hM
+ \frac{1}{4} v^{-2} {\widehat e}_{\hM}{}^\hB  (\lambda^{-1} + \Adj g)_{\hB \alpha} \pa^\hM \Big]~, \\[2ex]
\widehat D_\ra &= 
\frac{(\hat b_1)^2}{\sqrt 2} \Big[
    (\cO_-)_a{}^\hB {\widehat e}_\hB{}^{\hM} \cD_\hM
    + \frac{1}{4} v^{-2} {\widehat e}_{\hM}{}^{\hB} (1+\lambda^{-2} \Adj g)_{\hB a} \pa^\hM\, (-)^m
    \Big]~, \\
\widehat D_\rba &= 
\frac{(\hat b_2)^2}{\sqrt 2} \Big[
    - (\cO_+)_a{}^\hB {\widehat e}_\hB{}^{\hM} \cD_\hM
    + \frac{1}{4} v^{-2} {\widehat e}_\hM{}^{\hB} (\lambda^{-2} + \Adj g)_{\hB a} \pa^\hM\, (-)^m
    \Big]~,\\[2ex]
\widehat D^\alpha &=
    \frac{1}{2} (\hat b_1)^3 \Big[
    (1-\lambda^{-4} + \cO_-)^{\alpha \hB} {\widehat e}_\hB{}^{\hM} \cD_\hM
    + \frac{1}{4} v^{-2} {\widehat e}_\hM{}^{\hB} (1+\lambda^{-3} \Adj g)_\hB{}^\alpha \pa^\hM
    \Big]~, \\
\widehat D^\balpha &=
    \frac{1}{2} (\hat b_2)^3 \Big[
    (\lambda -\lambda^{-3} + \cO_+)^{\balpha \hB} {\widehat e}_\hB{}^{\hM} \cD_\hM
    - \frac{1}{4} v^{-2} {\widehat e}_\hM{}^{\hB} (\lambda^{-3} + \Adj g)_\hB{}^\balpha \pa^\hM
    \Big]~, \\[2ex]
\widehat D^{\iso1} &= v^2 \Big[
    (1+ \Adj{g}^{-1})^{\iso1 \hB} {\widehat e}_\hB{}^\hM \cD_\hM
    + \frac{1}{4} v^{-2} {\widehat e}_{\hM}{}^{\hB} (1 - \Adj g)_B{}^{\iso1} \pa^\hM \,(-)^m\Big]
\end{align}
\end{subequations}
The construction involves the left-invariant vector fields ${\widehat e}^\hA t_\hA = g^{-1} \rd g$ and the intrinsic WZW $B$-field (see \eqref{eqn:GxGgroupdB}) appearing in $\cD_\hM = \pa_\hM - \zB_{\hM \hN} \pa^{\hN} (-)^n$.  Again, we emphasize that $(\cO_+)_{\alpha}{}^\hB$
and $(\cO_-)_{\alpha}{}^\hB$ are related, consistent with the underlying structure of supersymmetric DFT \eqref{E:cV1}, although here the relation is slightly more complicated:
\begin{align}
(\cO_+)_{\alpha}{}^\hB = -\lambda \,(\cO_-)_{\alpha}{}^\hB~, 
\qquad
(\cO_+)_{\balpha}{}^\hB = -\lambda^{-1} (\cO_-)_{\balpha}{}^\hB~.
\end{align}

As with the $\eta$ deformation, we have first identified the supervielbein on the full generalized parallelizable space. Following the discussion in section \ref{S:GCoset.GxG}, we can pass to the generalized coset by taking $g = f^{-1} n f$. However, we cannot directly apply many of the formulae from that section because of the non-trivial similarity transformation applied to the generators $T_\hcA$ \eqref{E:Ta.lambda}. This is in contrast to the $\eta$-deformation construction, where the triangular structure of the coset supervielbein \eqref{E:genV.Rg.coset} simplified matters. In this instance, it will be easier to proceed from scratch.

The intrinsic WZW $B$-field becomes, for $g = f^{-1} n f$, 
\begin{align}
\zB &= \frac{1}{4} \pair{\rd n n^{-1} + n^{-1} \rd n + n \rd f f^{-1} n^{-1}}{\rd f f^{-1}}
    + \overline{\zB}_\WZW~, \eol
    \qquad
\rd \overline{\zB}_\WZW &= -\frac{1}{24} \pair{\rd n n^{-1}}{[\rd n n^{-1}, \rd n n^{-1}]}~.
\end{align}
The WZW part lives purely on the coset, while the other term has at least one leg in the subgroup $F$. The upshot, while far from obvious from this perspective, is that we recover the \Polacek-Siegel form with
\begin{align}
\widehat D_{\iso1} = (\Adj{f})_{\iso1}{}^{\iso2} \tilde v_{\iso2}{}^{\si} \pa_{\si}~.
\end{align}
We will not show this explicitly for the other terms, although it is a worthwhile exercise.

From the explicit form of the covariant derivatives, we can read off
\begin{subequations}
\label{E:chiralEinv.lambda}
\begin{alignat}{2}
\cE_\alpha{}^M &= \hat b_1 \,(\overline\cO_-)_\alpha{}^\hB {\overline e}_\hB{}^M~, 
&\qquad
\bar \cE_\alpha{}^M &= - \hat b_1 \, \lambda^{-1} \,(\overline\cO_+)_\alpha{}^\hB {\overline e}_\hB{}^M~, \\
\cE_\balpha{}^M &= \hat b_2 \, \lambda^{-1} \,(\overline\cO_-)_\balpha{}^\hB {\overline e}_\hB{}^M~, &\qquad
\bar \cE_\balpha{}^M &= - \hat b_2 \,(\overline\cO_+)_\balpha{}^\hB {\overline e}_\hB{}^M~, 
\\
\cE_\ra{}^M &= (\hat b_1)^2 \,(\overline\cO_-)_a{}^\hB {\overline e}_\hB{}^M~,
&\qquad
\bar \cE_\rba{}^M &= - (\hat b_2)^2 \,(\overline\cO_+)_a{}^\hB {\overline e}_\hB{}^M~.
\end{alignat}
\end{subequations}
The bars on $\cO_\pm$ again signify the restriction to the coset, and by ${\overline e}_\hA{}^\hM$ we mean extracting the $\Adj f$ action from ${\widehat e}_{\hA}{}^\hM$, i.e.
$\widehat e_{\hA}{}^{\hM} = (\Adj f)_\hA{}^{\hB} \overline{e}_{\hB}{}^\hM$.
This quantity is not so simple as in the previous section: its inverse can be written
\begin{align}
\overline{e}^\hA t_\hA = n^{-1} \rd n  + \rd f f^{-1} - n^{-1} \rd f f^{-1} n~, \qquad
\overline{e}_\hM{}^{\hA} =
\begin{pmatrix}
\overline{e}_M{}^A & \overline{e}_M{}^{\iso1}  \\
\tilde v_{\si}{}^{\iso2} (\cO_-)_{\iso2}{}^{A} \quad &
\tilde v_{\si}{}^{\iso2} (\cO_-)_{\iso2}{}^{\iso1}
\end{pmatrix}~.
\end{align}
The inverses of \eqref{E:chiralEinv.lambda} are
\begin{subequations}
\label{E:chiralE.lambda}
\begin{alignat}{2}
\cE_M{}^\alpha &= \frac{1}{\hat b_1 } \,\overline{e}_M{}^\hB (\overline\cO_-^{-1})_\hB{}^\alpha~, &\qquad
\bar \cE_M{}^\alpha &= - \frac{\lambda}{\hat b_1} \, \overline{e}_M{}^\hB (\overline\cO_+^{-1})_\hB{}^\alpha~, \\
\cE_M{}^\balpha &= \frac{\lambda}{\hat b_2} \, \overline{e}_M{}^\hB (\overline\cO_-^{-1})_\hB{}^\balpha~, 
&\qquad
\bar \cE_M{}^\balpha &= - \frac{1}{\hat b_2 } \, \overline{e}_M{}^\hB (\overline\cO_+^{-1})_\hB{}^\balpha~, \\
\cE_M{}^\ra &= \frac{1}{(\hat b_1)^2 } \,\overline{e}_M{}^\hB (\overline\cO_-^{-1})_\hB{}^a~, 
&\qquad
\bar \cE_M{}^\rba &= - \frac{1}{(\hat b_2)^{2}} \,\overline{e}_M{}^\hB (\overline\cO_+^{-1})_\hB{}^a~.
\end{alignat}
\end{subequations}
Here we have exploited $(\overline\cO_+)_{\iso1}{}^\hB = -(\overline\cO_-)_{\iso1}{}^\hB$ and the structure of the $\overline{e}_\hM{}^{\hA}$.

The Lorentz transformation that connects $\cE_M{}^\ra$ to $\bar \cE_M{}^{\rba}$ is
\begin{align}\label{E:Lambda.lambda}
\Lambda_\ra{}^\rbb = - \frac{(\hat b_1)^2}{(\hat b_2)^2} \times 
(\overline \cO_-)_a{}^\hC (\overline \cO_+^{-1})_\hC{}^b
    = - (\overline \cO_-)_a{}^\hC (\overline \cO_+^{-1})_\hC{}^b
\end{align}
for $b_1$ and $b_2$ both real. The matrix 
$M_\hA{}^\hB = (\overline{\cO}_+)_\hA{}^\hC (\overline{\cO}_-^{-1})_\hC{}^\hB$ is
\begin{align}
M_\hA{}^\hB =
\begin{pmatrix}
-(\Lambda^{-1})_a{}^b & M_a{}^\beta & M_a{}^{\bbeta} & M_a{}^{\iso2} \\
0 & -\lambda \,\delta_\alpha{}^\beta & 0 & 0 \\
0 & 0 & - \lambda^{-1} \delta_\balpha{}^\bbeta  & 0 \\
0 & 0 & 0 & -\delta_{\iso1}{}^{\iso2}
\end{pmatrix}~.
\end{align}
Again, it is not hard to show
$\det \Lambda^{-1} = \sdet M = \sdet \overline\cO_+ / \sdet \overline\cO_- = 1$, which follows from $\sdet(\Adj g) = 1$. This guarantees a IIB or IIB$^*$ duality frame.

The supervielbein is
\begin{subequations}
\begin{align}
E_M{}^\ra &= \overline{e}_M{}^\hB (\overline\cO_-^{-1})_\hB{}^a~, \\
E_M{}^{1\alpha} &= -\frac{\lambda}{\hat b_1} \, \overline{e}_M{}^\hB (\overline\cO_+^{-1})_\hB{}^\alpha~, \\
E_M{}^{2\alpha} &= \frac{\lambda}{\hat b_2} \, \overline{e}_M{}^\hB (\overline\cO_-^{-1})_\hB{}^\bbeta
    (\Lambda^{-1})_{\bbeta}{}^\alpha~,
\end{align}
\end{subequations}
where we are free to use 16-component spinors because the duality frame is IIB/IIB$^*$. Following similar steps as before, we find the dilatini
\begin{subequations}
\begin{align}
\chi_{1\alpha} &= \frac{i}{2} \cE_\ra{}^M \bar \cE_{M}{}^\beta (\gamma^a)_{\beta \alpha}
    = -\frac{i}{2} \hat b_1 \,\lambda \, 
    (\overline\cO_-)_a{}^\hC (\overline\cO_+^{-1})_\hC{}^\beta (\gamma^a)_{\beta \alpha}~, \\
\chi_{2\alpha} &= \frac{i}{2} \Lambda_{\alpha}{}^{\bbeta}
    \bar \cE_\rba{}^M \cE_{M}{}^\bgamma (\gamma^\rba)_{\bgamma \bbeta}
    = -\frac{i}{2} \hat b_2 \,\lambda\, 
    (\overline\cO_+)_a{}^\hC (\overline\cO_-^{-1})_\hC{}^\bgamma 
        (\gamma^a)_{\bgamma \bbeta} \,\Lambda_{\alpha}{}^\bbeta~.
\end{align}
\end{subequations}
and two equivalent expressions for the Ramond-Ramond bispinor 
\begin{align}
S^{1\alpha \,2\beta} &= -\cV^{\alpha M} \cE_M{}^{\bbeta} (\Lambda^{-1})_{\bbeta}{}^\beta
    = -\frac{1}{2} \frac{(\hat b_1)^3}{\hat b_2} \, \Big(
    \lambda  (1-\lambda^{-4}) (\overline\cO_-^{-1})^{\alpha \bbeta}
    + \lambda^{-3} \kappa^{\alpha \bbeta}
    \Big) (\Lambda^{-1})_{\bbeta}{}^\beta \eol
    &= - \cV^{\bbeta M} \bar \cE_M{}^\alpha (\Lambda^{-1})_{\bbeta}{}^\beta
    = \phantom{+}\frac{1}{2} \frac{(\hat b_2)^3}{\hat b_1} \Big(
    \lambda^2 (1-\lambda^{-4}) (\overline\cO_+^{-1})^{\bbeta \alpha}
    + \lambda \,\kappa^{\bbeta \alpha}
    \Big) (\Lambda^{-1})_{\bbeta}{}^\beta~.
\end{align}

Again, we can directly recover the Green-Schwarz $\sigma$-model \eqref{E:sigmaLag.GSform}. The vielbein $E^a$ matches the desired expression and the $B$-field is given by
\begin{align}
B &= \overline{\zB}_\WZW
    - {\overline e}^\hD (\overline\cO_-^{-1})_\hD{}^\hB\,
    \wedge {\overline e}^\hC (\overline\cO_-^{-1})_\hC{}^\hA\,
    \widehat B_{A B}~, \eol
\widehat B_A{}^B &= 
    \frac{1}{1-\lambda^{-4}} \Big(
    \Adj n^{-1}\,\Omega - \Omega^T \,\Adj n 
    \Big){}_A{}^B~.
\end{align}
An overall factor involving the tension must be separately specified. Here it is
$T = \frac{|k|}{4\pi} (1-\lambda^{-4})$ with the understanding that $k$ should be
taken to be negative and $|\lambda|>1$.

To recover the results of \cite{Borsato:2016ose}, we should choose $\hat b_1 = -1$ and $\hat b_2 = -i$. The latter choice is not technically allowed since $b_i$ should be real to ensure the Majorana condition holds. However, one can interpret this as arising from writing IIB$^*$ results in IIB conventions: this introduces factors of $i$ for objects carrying $\balpha$ indices (see e.g. footnote 20 of \cite{Borsato:2016ose} or section 5 of \cite{Butter:2022gbc}). Now the sign in \eqref{E:Lambda.lambda} is eliminated, so that 
$\Lambda_\ra{}^\rbb = + (\overline \cO_-)_a{}^\hC (\overline \cO_+^{-1})_\hC{}^b$. Presuming this to lie in $\g{SO}^+(1,9)$, we recover the results of \cite{Borsato:2016ose} up to an overall Lorentz transformation.
However, it is by no means obvious that this is fixed in $\g{SO}^+(1,9)$ (or $\g{SO}^-(1,9)$). Actually, one can show by randomly sampling elements of $\g{SU}(2,2) \times \g{SU}(4)$ that $\Lambda_\ra{}^\rbb$ can lie in either connected part.  Moreover, 
$(\overline\cO_+)_a{}^\hC (\overline\cO_-^{-1})_\hC{}^b$ turns out to be \emph{independent} of $\lambda$ and determined entirely by the group element $g$; it in fact matches the Lorentz transformation on the coset $G / F$ determined using $\Adj g$ as in \eqref{E:Lambda.GxG}, in remarkable contrast to the $\eta$-deformation. This surprising condition follows because the element defined in \eqref{E:Lambda.GxG} appears always to be idempotent.\footnote{We could find no proof of this last point, but it seems to hold for all random matrices we sampled.} \emph{This seems to imply that the $\lambda$ deformation is not purely fixed in either a IIB or IIB$^*$ duality frame, but that this depends on the specific group element $g$.}

This is unexpected because one might very naturally expect a IIB$^*$ duality frame since the $\lambda$-model can be understood as a deformation of the non-abelian T-dual of the $\g{AdS}_5 \times \g{S}^5$ superstring, as argued in \cite{Borsato:2016ose}. Certainly it is possible to find IIB backgrounds for very specific cases involving $\g{AdS}_n \times \g{S}^n$ factors (see e.g. \cite{Sfetsos:2014cea,Hoare:2015gda,Borsato:2016zcf}). It would be good to understand this point better, and whether some other factor forbids these choices of group element or invalidates the naive duality argument.\footnote{We thank Riccardo Borsato and Linus Wulff for discussions about this point and for pointing out references \cite{Sfetsos:2014cea,Hoare:2015gda,Borsato:2016zcf} to us.}

\subsection{Analytic continuation and PL T-duality}
Let us briefly comment about how the $\eta$ and $\lambda$ models are related \cite{Hoare:2015gda, Sfetsos:2015nya,Klimcik:2015gba}. As discussed in section \ref{S:GPS.Gc}, there exist coset representatives for $G \times G$ and $G^{\mathbb C}$ that are straightforwardly connected by analytic continuation, and so the same holds for their generalized supervielbeins. For $G^{\mathbb C}$, this corresponds to a different choice of isotropic subgroup  \eqref{E:Gc.simplebasis} than the one \eqref{eqn:DDgens} relevant for the $\eta$ deformation; in other words, the $\eta$ deformation should be the Poisson-Lie dual of the analytic continuation of the $\lambda$ deformation.

Of course, the generalized supervielbeins built in sections \ref{S:GPS.GxG} and \ref{S:GPS.Gc} carry no reference to $\lambda$ or $\eta$. These parameters arose from a similarity transformation to recover the physical supervielbeins with the correct supergravity flux constraints. To understand the connection, we need only compare \eqref{E:Ta.eta.1} to \eqref{E:Ta.lambda.1}. Since the generators on 
$G^{\mathbb C}$  map to generators on $G \times G$  as
$t_\hA \rightarrow (t_\hA, t_\hA) $
and
$\tilde t_\hA \rightarrow i \,(t_\hA, -t_\hA)$, it must be that
\begin{align}\label{E:eta.lambda.connection}
\eta \rightarrow i\,\frac{1-\lambda}{1+\lambda}~, \qquad
a_i \rightarrow \frac{1+\lambda}{2\lambda}\, b_i~.
\end{align}
This is consistent with the normalizations \eqref{E:ai.norm} and \eqref{E:bi.norm} up to a factor of $i$, coming from the analytic continuation of the Killing form on $\mathdsl{D}$. 

Finally, it is worth mentioning that the $\eta$ and $\lambda$ $\sigma$-models \eqref{E:sigmaLag.eta} and \eqref{E:sigmaLag.lambda} each involve one additional parameter corresponding with an overall normalization: these are $1/t$ and $\frac{k}{\pi}$. These parameters are related to the deformation parameter of the quantum group $U_q(\mathfrak{psu}(2,2|4))$ governing the deformed models as
\begin{align}
q = 
\begin{cases}
e^{- \varkappa t} & \text{$\eta$-deformation} \\
e^{i \pi / k} & \text{$\lambda$-deformation} 
\end{cases}
\end{align}
for $\varkappa = \frac{2\eta}{1-\eta^2}$. The analytic continuation from $t$ to $k/\pi$ can be checked at the classical level by comparing the respective Hamiltonians. For these models, we find $\cH = \frac{1}{2 T} \Pi_\ra \eta^{\ra\rb} \Pi_\rb + \frac{1}{2T} \Pi_\rba \eta^{\rba\rbb} \Pi_\rbb$, where $\Pi_\cA = \cV_\cA{}^\cM (p_M, T \pa_\sigma x^M)$. Undoing the rescaling the supervielbein replaces $T$ by $T / v^2$. This leads to canonical Poisson brackets
\begin{align}
\{\Pi_\cA(\sigma), \Pi_\cB(\sigma')\} = 
    T v^{-2} \,\eta_{\cA \cB} \,\pa_\sigma \delta(\sigma -\sigma') 
    + F_{\cA \cB}{}^{\cC} \Pi_\cC \,\delta(\sigma -\sigma')~.
\end{align}
The normalization of the Schwinger term is
\begin{align}
T v^{-2} =
\begin{cases}
\dfrac{1}{\varkappa t} & \text{$\eta$-deformation} \\[2.5ex]
\dfrac{|k|}{\pi} & \text{$\lambda$-deformation}
\end{cases}
\end{align}
and captures how the parameters must change, with a factor of $i$ coming from analytically continuing the Killing form.

\subsection{Results for the dilaton}
\label{S:SUGRA.dilaton}

We have not yet addressed the question of whether these supergravity backgrounds admit a dilaton. It was shown in \cite{Borsato:2016ose} that the $\lambda$-deformation always admits a dilaton while the $\eta$-deformation admits a dilaton only when a certain unimodularity condition on the $R$-matrix is satisfied. We can now see how these conditions arise naturally within double field theory.

As discussed in section \ref{S:SNATD.DFT}, one can replace $\pa_\cM \log\Phi$ in the dilatonic flux tensor by a vector $\cX_\cM$ \eqref{E:GenDilatonFlux} and impose the same constraints on this flux as in super DFT \cite{Butter:2022gbc}. This implies no additional constraints on the supergeometry: the vector $\cX_\cM$ is the DFT analogue of $X_M$ and $K^M$ in generalized supergravity. The constraints in question amount to fixing
\begin{align}\label{E:DilFluxConstraint}
\cF_\alpha = -\cF_{\alpha \beta}{}^\beta~, \qquad
\cF_\balpha = -\cF_{\balpha \bbeta}{}^\bbeta~.
\end{align}
From these expressions, one can compute $\cX_\alpha$. The question is whether that can be written as $D_\alpha$ of some superfield.

Rather than compute this directly for the models in question, we will follow a less direct but more rewarding route, and address the full set of dilatonic fluxes in one fell swoop. The crucial point is that the covariant dilatonic torsions
\begin{align}
\cT_\cA = \cV_\cA{}^\cM \cX_\cM + \pa_\cM \cV_\cA{}^\cM + \Omega^{\cB}{}_{\cB \cA}~.
\end{align}
all vanish when the constraint \eqref{E:DilFluxConstraint} and the Bianchi identities are imposed \cite{Butter:2022gbc}. These differ from the fluxes $\cF_\cA$ by the $\Omega$ connection of type II DFT, which is composed of not only the double Lorentz connection but also connections associated with the additional parameters given in Table \ref{T:lambdaConstraints}.

What exactly are these $\Omega$? Recall that the \Polacek-Siegel framework furnished us a Lorentz spin connection
\begin{align}
\Omega_{\cM \ra}{}^\rb = 
\Omega_{\cM \rba}{}^\rbb = -\Omega_\cM{}^{\iso1} F_{\iso1 a}{}^b
\end{align}
where $\Omega_\cM{}^{\iso1}$ was a piece of the megavielbein. Is this the right one? That question is easy enough to answer. At dimension 1/2, choosing the DFT torsion tensors $\cT_{\alpha \rb \rc}$ and $\cT_{\balpha \rb \rc}$ to vanish fixed the $\alpha$ component of $\Omega$. Indeed, we can check that (similarly for the barred versions)
\begin{align}
\cT_{\alpha \rb \rc} = \widehat \cF_{\alpha \rb \rc} = 0~, \qquad
\cT_{\balpha \rb \rc} = \widehat \cF_{\balpha \rb \rc} = 0
\end{align}
where $\widehat \cF_{\alpha \rb \rc}$ is the flux for the megavielbein (which vanishes for both cases of interest). The other dimension 1/2 torsion tensors $\cT_{\alpha\beta}{}^\gamma$,
$\cT_{\alpha\bbeta}{}^\gamma$, $\cT_{\ol{\alpha\beta}}{}^\gamma$, and their barred versions similarly match the corresponding generalized flux tensors (all also vanishing). At dimension 1, we find
\begin{align}
\cT_{\ra \rb \rc} = \widehat \cF_{\ra \rb \rc} = 0~, \qquad
\cT_{\ra \rb \rbc} = \widehat \cF_{\ra \rb \rbc} = 0~, \qquad
\cT_{\ra \ol{\rb \rc}} = \widehat \cF_{\ra \ol{\rb \rc}} = 0~, \qquad
\cT_{\ol{\ra\rb\rc}} = \widehat \cF_{\ol{\ra\rb\rc}} = 0
\end{align}
implying that $\Omega_{[\ra\rb\rc]}$ and $\Omega_{\rba \rb \rc}$ and their barred versions are chosen properly. At dimension 1 we also have
\begin{align}
\cT_{\balpha \rb}{}^\gamma
    = \widehat \cF_{\balpha \rb}{}^\gamma + \Omega_{\balpha \rb}{}^\gamma~, \qquad
\cT_{\alpha \rb}{}^\gamma
    = \widehat \cF_{\alpha \rb}{}^\gamma + \Omega_{\alpha \rb}{}^\gamma~.
\end{align}
Both of these should vanish. Since
$\widehat \cF_{\balpha \rb}{}^\gamma \propto \kappa^{\gamma \bgamma} (\gamma_{b})_{\bgamma \balpha}$ is $\gamma$-traceless, using the properties of $\kappa^{\alpha \balpha}$, there is no obstruction to choosing $\Omega_{\balpha \rb}{}^\gamma = -\widehat \cF_{\balpha \rb}{}^\gamma$ so that first torsion vanishes.
The second vanishes since $\widehat \cF_{\alpha \rb}{}^\gamma=0$ and so we can
choose $\Omega_{\alpha \rb}{}^\gamma = 0$. 
At dimension 3/2, we have
\begin{alignat}{3}
\cT_{\ra \rb}{}^\gamma
    &= \widehat \cF_{\ra \rb}{}^\gamma + \Omega^{\gamma}{}_{\ra \rb} + 2 \,\Omega_{[\ra, \rb]}{}^\gamma~, 
&\quad
\cT_{\rba \rb}{}^\gamma
    &= \widehat \cF_{\ra \rbb}{}^\gamma + \Omega_{\rba, \rb}{}^\gamma~, 
&\quad
\cT_{\ol{\ra \rb}}{}^\gamma
    &= \widehat \cF_{\ol{\ra \rb}}{}^\gamma + \Omega^{\gamma}{}_{\ol{\ra \rb}}~, \eol
\cT_{\alpha}{}^{\beta \gamma}
    &= \widehat \cF_{\alpha}{}^{\beta \gamma}
    + \Omega_\alpha{}^{\beta \gamma}~, 
&\qquad
\cT_{\alpha}{}^{\beta \bgamma}
    &= \widehat \cF_{\alpha}{}^{\beta \bgamma}~, 
&\qquad
\cT_{\alpha}{}^{\ol{\beta \gamma}}
    &= \widehat \cF_{\alpha}{}^{\ol{\beta \gamma}}~.
\end{alignat}
All the generalized flux tensors vanish on the right, and so we are free to choose
all the corresponding $\Omega$'s to vanish.\footnote{Strictly speaking, we can only fix $\Omega$ up to the residual shift symmetries discussed in \cite{Butter:2022gbc}.}

What does this mean for $\cT_\cA$? From the conditions derived on the non-Lorentz $\Omega$, we find
\begin{align}
\cT_\alpha = \cF_\alpha - \Omega_{\beta \alpha}{}^\beta~,
\qquad
\cT_\ra = \cF_\ra - \Omega_\rb{}_{\ra}{}^\rb + \Omega_\beta{}_\ra{}^\beta~,
\qquad
\cT^\alpha = \cF^\alpha 
    + \Omega^\beta{}_{\beta}{}^\alpha~.
\end{align}
Each of these can be interpreted as pieces of the dilaton flux tensor on the \Polacek-Siegel megaspace \eqref{E:PS.DilT}. We know for a supergravity solution, all of these
must vanish. Moreover, from the dilatonic Bianchi identity, we also know that the dilatonic $\g{SO}(4,1) \times \g{SO}(5)$ curvature $\cR_{a b} = - \cR^{\iso1} f_{\iso1 a b}$ vanishes. The upshot is from \eqref{E:PSDilF.Decomp} we can impose the strictest possible condition on the \Polacek-Siegel dilatonic flux,
\begin{align}
\widehat \cF_\hcA = F_{\hcA \iso1}{}^{\iso1}  = 0
\end{align}
with the vanishing of the second term following from the properties of $\g{PSU}(2,2|4)$.

This means that for both the $\eta$ and $\lambda$ deformations, the generalized dilatonic torsion in the \Polacek-Siegel framework must be taken to vanish, $\widehat \cF_\hcA = 0$. The results in section \ref{S:GPS.dilaton} apply for $F_\hA = F^\hA = 0$. 
For $G \times G$, we have from \eqref{E:GxG.cX}
\begin{align}
\cX^\hM = 0~, \qquad \cX_\hM = \pa_\hM \log \sdet \hat v_{\hN}{}^{\hB}
\end{align}
where $\hat v_{\hN}{}^{\hB}$ is the right-invariant vielbein for the group $G$.
This solution admits a dilaton solution with
\begin{align}
\log \widehat \Phi = \log \sdet \hat v_\hM{}^\hA + \text{constant}~.
\end{align}
To derive the supergravity dilaton requires two steps. First, we pass from the \Polacek-Siegel framework to DFT on the coset. This involves defining 
$\log \Phi = \log\widehat \Phi - \log \sdet \tilde e_{\si}{}^{\iso1}$.
Then we translate from the DFT dilaton to the supergravity dilaton, using
$\Phi = e^{-2 \varphi} \times \sdet E_M{}^A$. From \eqref{E:sdetE.same},
we can replace $\sdet E_M{}^A$ with $\sdet \cE_M{}^A$ or $\sdet \bar \cE_M{}^A$,
discarding any overall sign difference as an irrelevant constant factor. Combining these factors gives
\begin{align}
e^{-2 \varphi} &= \sdet \hat v_\hM{}^\hA \times
\sdet \tilde e_{\iso1}{}^\si \times \sdet \cE_A{}^M \times 
\text{constant}~.
\end{align}
For the $\lambda$ deformation, this amounts to
\begin{align}\label{E:SugraDilatonOpm}
e^{-2 \varphi} = \sdet \overline{\cO}_\pm \times \text{constant}~.
\end{align}
To see this, one first exploits
\begin{align}
\begin{pmatrix}
\delta_{\iso1}{}^{\iso2} & 0 & 0 & 0 \\
0 & \hat b_1 \delta_\alpha{}^\beta & 0 & 0 \\
0 & 0 & \hat b_2 \lambda^{-1} \delta_\balpha{}^\bbeta & 0 \\
0 & 0 & 0 & (\hat b_1)^2 \delta_a{}^b
\end{pmatrix}
\times
(\overline\cO_-)_\hB{}^\hC \overline{e}_\hC{}^\hM
= 
\begin{pmatrix}
\tilde v_{\iso1}{}^{\si} & 0  \\
{\bullet} & \cE_A{}^M
\end{pmatrix}
\end{align}
where the $\bullet$ denotes an irrelevant quantity. From this, we can immediately see
\begin{align}
\sdet \overline\cO_- =
    \sdet \tilde v_{\iso1}{}^{\si} \times \sdet \cE_A{}^M \times
    \sdet \overline{e}_\hM{}^{\hA} \times \text{constant}~.
\end{align}
But $\tilde v_{\iso1}{}^{\si}$ and $\overline{e}_\hM{}^\hA$ differ from 
$\tilde e_{\iso1}{}^{\si}$ and $\hat v_\hM{}^\hA$ only by factors of $(\Adj f)_{\iso1}{}^{\iso2}$
and $(\Adj{f^{-1} n})_\hA{}^\hB$, respectively, and the superdeterminants of these are just $\pm1$. A similar line of argument establishes that
$\sdet \overline{\cO}_-$ is proportional to $\sdet \overline{\cO}_+$, and these are also proportional to the full operators $\overline{\cO}_\pm$. This recovers the result of
\cite{Borsato:2016ose}.

For $G^{\mathbb C}$, we first observe from \eqref{E:Gc.cX} that
\begin{align}
\cX^\hM = R^{\hB \hC} f_{\hC\hB}{}^\hA \,\hat v_\hA{}^\hM~.
\end{align}
Therefore, the existence of a dilaton solution requires the unimodularity condition for the $R$-matrix, $R^{\hB \hC} f_{\hC\hB}{}^\hA = 0$. Provided this holds, we recover the same conditions, and an identical line of reasoning leads to \eqref{E:SugraDilatonOpm} for the corresponding operators $\cO_\pm$. This again is in full agreement with \cite{Borsato:2016ose}.

\section{Discussion}

In this paper we have discussed how to employ superspace double field theory, involving a generalized supervielbein, an element of $\g{OSp}(D,D|2s)$, to describe generalized dualities. We confirmed our initial expectation that all algebraic structures relevant for dualities of the bosonic string carry over to generalized supergeometry naturally. When the generalized flux tensor is constant, the space is generalized parallelizable (or a generalized coset thereof), and one can construct the generalized supervielbein explicitly in terms of the group theoretic data.

A considerable advantage is that the generalized supervielbein unifies all fields of type II supergravity, except for the dilaton, in one object. To appreciate this fact, recall the salient features of established generalized geometries for type II strings:
\begin{itemize}
    \item In $\g{O}(D,D)$ generalized geometry, the metric and $B$-field are unified by the generalized frame, while the Ramond-Ramond sector can be captured either with an $\g{O}(D,D)$ Majorana-Weyl spinor \cite{Hohm:2011zr,Hohm:2011dv}
    or an $\g{O}(D-1,1) \times \g{O}(1,D-1)$ bispinor \cite{Jeon:2012kd,Jeon:2012hp}
    (see \cite{Butter:2022sfh} for the relation between them).
    The Ramond-Ramond sector and the generalized frame are \emph{a priori}
    independent objects, related only by the field equations.
    \item Exceptional generalized geometry improves the situation by incorporating the Ramond-Ramond sector into the generalized frame. However, this requires the transition from a T-duality covariant  description to a U-duality covariant one. Consequentially, strings are no longer the fundamental objects. They are replaced by membranes, which come with their own challenges. When the full ten-dimensional spacetime needs a unified treatment, like for the $\eta$ and $\lambda$-deformations of the $\g{AdS}_5\times\g{S}^5$ superstring, one has to deal with the infinite dimensional duality group $E_{11(11)}$ \cite{West:2003fc,West:2010ev,Rocen:2010bk} which is not completely understood yet
    (see \cite{Bossard:2017wxl,Bossard:2019ksx,Bossard:2021ebg} for recent progress).
\end{itemize}
Additionally, neither approach directly incorporates fermionic dualities. All these problems are resolved by generalized supergeometry making it the ideal framework to analyze integrable deformations of superstrings. Therefore, one main focus of our efforts was to explain the $\eta$ and $\lambda$ deformations within superspace double field theory. While their $\sigma$-model actions are fairly complicated, their explanation within super-DFT is rather straightforward, in terms of the double Lie groups $G \times G$ and $G^{\mathbb C}$, with a single parameter ($\eta$ and $\lambda$, respectively) describing how the supergravity frame is embedded in the doubled space.

\begin{table}[t]
\centering
\renewcommand{\arraystretch}{1.5}
\begin{tabular}{r|l}
\toprule
    dim. & constraint \\\hline
    $-\tfrac12$ & $\cF_{\alpha \beta \gamma} = 
        \cF_{\alpha \beta \bgamma} = 
        \cF_{\alpha \bbeta \bgamma} = 
        \cF_{\balpha \bbeta \bgamma} = 0$ \\
    $0$ & $\cF_{\alpha \beta \rc} = -i \sqrt{2}\, (\gamma_\rc)_{\alpha \beta}~, \quad
\cF_{\balpha \bbeta \rbc} = -i \sqrt{2}\, (\bar\gamma_\rbc)_{\balpha \bbeta}~, \quad
\cF_{\alpha \bbeta \rc} = \cF_{\alpha \bbeta \rbc} =
\cF_{\alpha \beta \rbc} = \cF_{\balpha \bbeta \rc} = 0$ \\
    \hline
    $\tfrac12$ & 
    $\cF_{\alpha \beta}{}^{\beta} = \tfrac{1}{4} \cF_{\beta \rb \rc} (\gamma^{\rb\rc})_{\alpha}{}^\beta~, \quad
    \cF_{\balpha \bbeta}{}^{\bbeta} = \tfrac{1}{4} \cF_{\bbeta \rbb \rbc} (\gamma^{\rbb\rbc})_{\balpha}{}^\bbeta~, \quad
    \cF_{\alpha \rb \rbc} (\gamma^{\rb})^{\alpha \beta} = 
    \cF_{\balpha \rbb \rc} (\gamma^{\rbb})^{\balpha \bbeta} = 0
    $
    \\
    $ 1 $ &
    $(\gamma^{\rc})^{\alpha\beta} \cF_{\rc \beta}{}^\balpha =- 
    (\gamma^{\rbc})^{\balpha\bbeta} \cF_{\rbc \bbeta}{}^\alpha$ \\
\bottomrule
\end{tabular}
\captionsetup{width=\textwidth}
\caption{Flux constraints in supersymmetric DFT.
The ones at dimension $\leq 0$ are necessary for $\kappa$-symmetry.
The higher dimension constraints are conventional, amounting to redefinitions of the dilatini and Ramond-Ramond bispinor to absorb unphysical fields.}
\label{T:FluxConstrSummary}
\end{table}


A major novelty compared to the purely bosonic approach is the necessity of additional torsion constraints, which restrict the generalized fluxes beyond their Bianchi identities. They fix the form of their dimension $-\tfrac{1}{2}$ and dimension 0 components as in Table \ref{T:FluxConstrSummary}: these imply similar constraints in generalized type II supergravity \cite{Wulff:2016tju}. From the worldsheet perspective, these are required for the underlying Green-Schwarz superstring to possess $\kappa$-symmetry. Consequentially, the target space supergeometry satisfies the field equations of generalized supergravity \cite{Arutyunov:2015mqj, Wulff:2016tju}. Moreover, they put the theory on-shell; otherwise, supersymmetry transformations would not close into an algebra. 

As one can see from Table \ref{T:FluxConstrSummary}, these flux constraints are not covariant under $\g{OSp}(D,D|2s)$ transformations. Rather, they break the duality group to the local symmetry group $\g{H}_L \times \g{H}_R$, which plays the same role as the double Lorentz group in bosonic DFT. In the latter, the generalized metric is responsible for the breaking. Due to the absence of a generalized supermetric in the supersymmetric extension, the flux constraints take over this function, too. This is analogous to the situation in conventional supergravity: there the torsion constraints are essential and there is no Riemannian supermetric.

There are several additional avenues one could explore at this point. One issue we avoided discussing was the $\sigma$-model interpretation of generalized dualities. These are described in terms of the $\cE$-model \cite{Klimcik:1995dy, Klimcik:1996nq, Klimcik:2015gba}
and its dressing coset extension \cite{Klimcik:1996np, Klimcik:2019kkf}. These models can be straightforwardly built for supergroups, but a subtlety involves finding the right constraints to ensure that the $\sigma$-model is of Green-Schwarz form. This would undoubtedly be related to a duality-symmetric formulation of the GS superstring using the language of super-DFT \cite{Butter:2022gbc}.

Another avenue to explore is the potential connection with integrability. The $\eta$ and $\lambda$ deformations were initially constructed as integrable deformations of the $\g{AdS}_5 \times \g{S}^5$ superstring, and a key role is played by the $\mathbb{Z}_4$ grading in the supergroup. It is already known that there are connections between the structure of $\cE$-models and integrability \cite{Lacroix:2020flf,Lacroix:2021iit}. It would be interesting to explore the connection for the case of super $\cE$-models.

Generalized dualities have proven to be useful solution generating techniques. Examples include non-abelian T-duals of backgrounds like $\g{AdS}_5 \times \g{S}^5$ and $\g{AdS}_3 \times \g{S}^3 \times \g{T}^4$ \cite{Sfetsos:2010uq} which are relevant for the AdS/CFT correspondence. In this context, an important question is how much supersymmetry of the original background is preserved by T-duality. In our framework the amount of supersymmetry is fixed by the number of fermionic generalized Killing vectors. Therefore, one should study how they transform under duality transformations. Perhaps one could construct a systematic treatment within super-DFT. One could then revisit known examples and try to exhaust all possible dualities to find new solutions.

Finally, we should add that very significant work on U-duality extensions of Poisson-Lie T-duality and its generalizations has appeared recently \cite{Sakatani:2019zrs, Malek:2019xrf, Blair:2020ndg, Musaev:2020nrt, Musaev:2020bwm,Sakatani:2022auu,Hassler:2022egz}. These would undoubtedly have natural descriptions in supersymmetric extensions of U-dual formulations, of the type explored e.g. in \cite{Butter:2018bkl,Bossard:2019ksx,Cederwall:2021ejp}.

\acknowledgments

We would like to thank Riccardo Borsato, Sybille Driesen, Gabriel Larios, Gr\'{e}goire Josse, Edvard Musaev, Yuho Sakatani, and Linus Wulff for helpful discussions, and Evgeny Ivanov, Martin Wolf, Pietro Grassi, Peter West, and Ali Eghbali for helpful comments. FH wants to thank the organizers of the workshop ``Supergravity, Strings and Branes'' at Bogazici University, Turkey for giving him the opportunity to present this work. The work of FH is supported by the SONATA BIS grant 2021/42/E/ST2/00304 from the National Science Centre (NCN), Poland. CNP is supported in part by DOE grant DE-FG02-13ER42020.

\appendix

\section{Supergroup conventions}
\label{A:Supergroups}

\subsection{Lie superalgebras and supergroups}
\label{A:Supergroups.Lie}
We summarize here our conventions for supergroups and superalgebras. A Lie superalgebra $\mathfrak{g}$ is spanned by elements $\xi = \xi^A t_A$ obeying
\begin{align}\label{E:Fabc.Appendix}
[\xi_1, \xi_2] = \xi_1^B \xi_2^C f_{CB}{}^A t_A = - [\xi_2, \xi_1]~.
\end{align}
The elements $\xi^A = (\xi^a, \xi^\alpha)$ are graded, with $\xi^a$ bosonic (commuting) and $\xi^\alpha$ fermionic (anticommuting), so that the structure constants are graded antisymmetric,
\begin{align}
f_{AB}{}^C = - f_{BA}{}^C (-)^{ab}
\end{align}
and are themselves commuting quantities, so that precisely zero or two of $A$,$B$, and $C$ may be fermionic.

Above we use the notation $(-)^{\grad{A} \grad{B}}$ as shorthand for the mathematically cleaner but bulkier $(-1)^{\epsilon(A) \epsilon(B)}$ where $\epsilon(A)$ is $0$ for bosonic $A$ and $1$ for fermionic $A$. Essentially, this gives $-1$ if both $A$ and $B$ are fermonic and $+1$ otherwise. This shorthand follows the classic text \cite{Wess:1992cp}.

When $\mathfrak{g}$ admits a Killing supermetric $\kappa_{AB}$, we introduce the pairing
\begin{align}\label{E:Pairing.Appendix}
\pair{\xi_1}{\xi_2} = \pair{\xi_2}{\xi_1} = \xi_1^A \xi_2^B \kappa_{BA} ~, \qquad 
\kappa_{AB} = \kappa_{BA} (-)^{ab}
\end{align}
and use $\kappa$ to raise and lower indices using NW-SE conventions, so that
\begin{align}
\xi_A = \xi^B \kappa_{BA}~, \qquad
\xi^A = \kappa^{A B} \xi_B~, \qquad \kappa^{AB} \kappa_{B C} = \delta_C{}^A (-)^{c a}~.
\end{align}
The structure constants with three lowered indices,
$f_{ABC} = f_{AB}{}^D \kappa_{DC}$, are totally (graded) antisymmetric.

Both the algebra and the pairing can expressed purely in terms of the generators $t_A$, but it depends on whether the generators $t_A$ are treated as commuting quantities, $\xi^\alpha t_\alpha = t_\alpha \xi^\alpha$ or as formal graded objects themselves, $\xi^\alpha t_\alpha = - t_\alpha \xi^\alpha$. The first situation applies when the superalgebra $\mathfrak{g}$ is embedded in a supermatrix algebra $\mathfrak{gl}(m|n)$; in this case, the \emph{generators} themselves are matrices of (commuting) complex numbers, and \eqref{E:Fabc.Appendix} and
\eqref{E:Pairing.Appendix} imply
\begin{align}
[t_A, t_B] := t_A t_B - t_B t_A (-)^{ab} = - f_{AB}{}^C t_C (-)^{ab}~, \qquad
\pair{t_A}{t_B} = \kappa_{AB} (-)^{ab}~.
\end{align}
The second situation, where the $t_A$ are themselves graded, leads to the more conventional expressions
\begin{align}
[t_A, t_B] := t_A t_B - t_B t_A (-)^{ab} = - f_{AB}{}^C t_C~, \qquad
\pair{t_A}{t_B} = \kappa_{AB}
\end{align}
where gradings arise primarily because of index ordering and the direction of contraction. We will employ the latter conventions when explicit indices are exhibited. The sign convention for $f_{AB}{}^C$ is a bit unconventional; this is to ensure the torsion tensors for supergroup manifolds have a plus sign, i.e. $T_{AB}{}^C = + f_{AB}{}^C$.

\subsection{The orthosymplectic group \texorpdfstring{$\g{OSp}(D,D|2s)$}{OSp(D,D|2s)}}
\label{A:Supergroups.OSp}
An element of $\g{OSp}(D,D|2s)$ is described by a graded supermatrix 
$\cU_\cM{}^\cN \in \g{GL}(2D|2s)$ satisfying the condition
\begin{align}
(\cU^{-1})_\cM{}^\cN = \eta^{\cN \cP} \cU_\cP{}^\cQ \eta_{\cQ \cM} (-)^{\grad{\cM} \grad{\cN}}
\end{align}
for a graded symmetric matrix $\eta_{\cM \cN}$ with graded inverse $\eta^{\cM \cN}$,
\begin{align}
\eta^{\cM \cP} \eta_{\cP \cN} = -\delta_\cN{}^\cM (-)^{\grad{\cM} \grad{\cN}}~.
\end{align}
It can be naturally described in terms of its $\g{GL}(D|s)$ subgroup where a generalized
vector $V_\cM$ decomposes as a one-form and vector $V_\cM = (V_M, V^M)$. In this basis,
$\eta$ is given by 
\begin{align}
\eta^{\cM \cN} =
\begin{pmatrix}
0 & \delta^M{}_N  \\
\delta_M{}^N (-)^{\grad{M} \grad{N} } & 0
\end{pmatrix}~, \qquad
\eta_{\cM \cN} =
\begin{pmatrix}
0 & \delta_M{}^N  \\
\delta^M{}_N (-)^{\grad{M} \grad{N}} & 0
\end{pmatrix}~.
\end{align}
Because of the grading present in $\eta$, it matters whether an index is raised or lowered. We conventionally identify elements of a matrix $\cU_\cM{}^\cN$ as if they were elements of $\cU_{\cM \cN}$, i.e.
\begin{align}
\cU_{\cM \cN} =
\begin{pmatrix}
U_{MN} & U_M{}^N \\
U^{M}{}_N & U^{MN}
\end{pmatrix}
\quad \implies \quad
\cU_{\cM}{}^\cN =
\begin{pmatrix}
U_M{}^N & U_{MN} (-)^{\grad{N}} \\
U^{M N} & U^M{}_N (-)^{\grad{N}}
\end{pmatrix}~.
\end{align}
This ensures that multiple contractions $(\cU_1)_\cM{}^\cN (\cU_2)_\cN{}^\cP$ follow the usual $\g{GL}(D|s)$ grading conventions, i.e. NW-SE contractions ${}^M {}_M$ are natural while SW-NE contractions ${}_M{}^M$ are accompanied by a grading $(-)^{\grad{M}}$. It also gives a natural expression for the inverse,
\begin{align}
(\cU^{-1})_{\cM}{}^\cN =
(-)^{\grad{N} \grad{M}} \begin{pmatrix}
U^N{}_M & U_{NM} (-)^{\grad{N}} \\
U^{NM} & U_N{}^M (-)^{\grad{N}}
\end{pmatrix}~.
\end{align}

\section{Democratic Type II supergravity conventions}
\label{A:DemoTypeII}
We summarize here our conventions for democratic type II supergravity and how they arise from DFT. Conventions for 10D gamma matrices and spinors can be found in \cite{Butter:2022gbc}. The inspiration for such a ``democratic'' approach to type II was inspired by Wulff, see the appendices of \cite{Wulff:2013kga}.

The supervielbein emerging from DFT consists of two copies of the vielbein super one-form $E_M{}^\ra$ and $E_M{}^\rba$, as well as two gravitino super one-forms, $E_M{}^\alpha$ and $E_M{}^\balpha$. The two vielbeins are related by a Lorentz transformation that determines the duality frame relative to IIB. That is, $\Lambda_\ra{}^{\rbb}$ is an element of $\g{O}^{(\alpha, \beta)}(1,9)$, where $\alpha_\Lambda=-1$ or $\beta_\Lambda=-1$ if $\Lambda$ involves a temporal or spatial orientation reversal, and $+1$ otherwise, see Table \ref{T:TypeIISugras}.

\begin{table}[t]
\begin{center}\begin{tabular}{ccccc}\toprule
Type & $\Lambda_\ra{}^{\rbb} $ \\ \midrule
IIB & $\g{O}^{(+,+)}$ \\
IIA${}^*$ & $\g{O}^{(-,+)}$ \\
IIA & $\g{O}^{(+,-)}$  \\
IIB${}^*$ & $\g{O}^{(-,-)}$ \\ \midrule
\end{tabular}
\end{center}
\caption{Classification of type II duality frame.}
\label{T:TypeIISugras}
\end{table}

We may think of $\Lambda_\ra{}^\rbb$ as a similarity transformation to convert barred vector indices to unbarred ones. In order to convert barred spinors to unbarred ones, we introduce the spinorial matrix $\slashed{\Lambda} = (\Lambda_\halpha{}^{\hbbeta})$, which obeys
\begin{align}
\slashed{\Lambda} \bar\gamma^{\rba} \slashed{\Lambda}^{-1} 
    = \gamma_* \gamma^\rb \Lambda_\rb{}^\rba~, \qquad
\slashed{\Lambda} \bar\gamma_* \slashed{\Lambda}^{-1} = \alpha_\Lambda \beta_\Lambda \gamma_*~,\qquad
\slashed{\Lambda} \bar C^{-1} \slashed{\Lambda}^{T} = \alpha_\Lambda C^{-1}~.
\end{align}
The last condition implies that 
$\slashed{\Lambda}^{-1} = \alpha_\Lambda \bar C^{-1} \slashed{\Lambda}^T C$.

The left Lorentz group is conventionally chosen to be the supergravity Lorentz group.
This identifies the supergravity vielbein as $E_M{}^\ra$. The barred gravitino and dilatino must be converted to the left Lorentz group with $\slashed{\Lambda}$. To do this, we rewrite
gravitini one-forms as 32-component Majorana spinors, with raised indices, $E_M{}^{i\halpha}$ for $i=1,2$. The dilatini have lower indices, $\chi_{i\halpha}$:
\begin{alignat}{2}
\label{E:Gravitini.Dirac}
E_M{}^{1 \halpha} &=
\begin{pmatrix}
E_M{}^{\alpha} \,& 0
\end{pmatrix}~, 
&\qquad
E_M{}^{2 \halpha} &=
\begin{pmatrix}
E_M{}^{\bbeta} \,& 0
\end{pmatrix}
(\Lambda^{-1})_\hbbeta{}^{\halpha}~, \\
\chi_{1\halpha} &=
\begin{pmatrix}
\chi_{\alpha} \\
0
\end{pmatrix}~, 
&\qquad
\chi_{2\halpha} &=
\Lambda_\halpha{}^{\hbbeta}
\begin{pmatrix}
\chi_{\bbeta} \\
0
\end{pmatrix}~,
\end{alignat}
The supercharges $Q_{i \halpha}$ obey analogous formulae as the dilatini and satisfy the SUSY algebra
\begin{align}
\{Q_{1 \halpha}, Q_{1 \hbeta}\} = i\,(P_L \gamma^a C^{-1})_{\halpha \hbeta} P_a~, \qquad
\{Q_{2 \halpha}, Q_{2 \hbeta}\} = \frac{i}{2} \,\alpha_\Lambda (\tilde P_{L}\gamma^a C^{-1})_{\halpha \hbeta} P_a
\end{align}
where we use the chiral projector $P_L = \frac{1}{2} (1+\gamma_*)$. The second SUSY involves a projector $\tilde P_L = \frac{1}{2} (1+\alpha_\Lambda \beta_\Lambda \gamma_*)$, which is
$P_L$ for IIB/IIB$^*$ and $P_R$ for IIA/IIA$^*$.

For type IIB/IIB$^*$ duality frames, $\alpha_\Lambda = \beta_\Lambda$, and 
\begin{gather}
\Lambda_\halpha{}^{\hbeta} =
\begin{pmatrix}
\Lambda_\alpha{}^\bbeta & 0 \\
0 & \Lambda^\alpha{}_\bbeta
\end{pmatrix}~, \qquad
(\Lambda^{-1})_\hbbeta{}^{\halpha} =
\begin{pmatrix}
(\Lambda^{-1})_\bbeta{}^{\alpha} & 0 \\
0 & (\Lambda^{-1})^{\bbeta}{}_\alpha
\end{pmatrix} =
\alpha_\Lambda \begin{pmatrix}
\Lambda^{\alpha}{}_\bbeta & 0 \\
0 & \Lambda_\alpha{}^\bbeta
\end{pmatrix}~, \\[2ex]
\Lambda_\alpha{}^\bgamma \Lambda_{\beta}{}^{\bdelta} (\gamma^\rba)_{\bgamma \bdelta}
    = \alpha_\Lambda\, (\gamma^\rb)_{\alpha \beta} \Lambda_\rb{}^\rba~, \qquad
\Lambda^\alpha{}_\bgamma \Lambda^{\beta}{}_{\bdelta} (\gamma^\rba)^{\bgamma \bdelta}
    = -\alpha_\Lambda\, (\gamma^\rb)^{\alpha \beta} \Lambda_\rb{}^\rba~.
\end{gather}
For type IIA/IIA$^*$ duality frames, $\alpha_\Lambda = -\beta_\Lambda$, and 
\begin{gather}
\Lambda_\halpha{}^{\hbeta} =
\begin{pmatrix}
0 & \Lambda_{\alpha \bbeta} \\
\Lambda^{\alpha \bbeta} & 0
\end{pmatrix}~, \qquad
(\Lambda^{-1})_\hbbeta{}^{\halpha} =
\begin{pmatrix}
0 & (\Lambda^{-1})^{\bbeta \alpha} \\
(\Lambda^{-1})_{\bbeta \alpha} & 0
\end{pmatrix} =
\alpha_\Lambda \begin{pmatrix}
0 & \Lambda^{\alpha \bbeta} \\
\Lambda_{\alpha \bbeta} & 0
\end{pmatrix}~, \\[2ex]
\Lambda^{\alpha \bgamma} \Lambda^{\beta \bdelta} (\gamma^\rba)_{\bgamma \bdelta}
    = -\alpha_\Lambda\, (\gamma^\rb)^{\alpha \beta} \Lambda_\rb{}^\rba~, \qquad
\Lambda_{\alpha\bgamma} \Lambda_{\beta \bdelta} (\gamma^\rba)^{\bgamma \bdelta}
    = \alpha_\Lambda\, (\gamma^\rb)_{\alpha \beta} \Lambda_\rb{}^\rba~.
\end{gather}

Democratic type II superspace is described by a supervielbein 
$E_M{}^A = (E_M{}^a, E_M{}^{i\halpha})$, a Kalb-Ramond super two-form $B_{MN}$, a scalar dilaton $e^{-2 \varphi}$, and a set of Ramond-Ramond super $(p-1)$-forms $\widehat \cC_{M_1 \cdots M_{p-1}}$ with $p$ even for IIA/IIA$^*$ and $p$ odd for IIB/IIB$^*$. The supervielbein is subject to local $\g{SO}^+(9,1)$ Lorentz transformations, gauged by a spin connection $\Omega_{M A}{}^B \in \mathfrak{so}(9,1)$. The Kalb-Ramond two-form and Ramond-Ramond $p$-forms transform as
\begin{align}
\delta B = \rd \tilde \xi~, \qquad
\delta \widehat \cC_{p-1} = \rd \widehat \lambda_{p-2} + \widehat \lambda_{p-4} \wedge H~.
\end{align}
The torsion tensors $T^A$ and field strengths $H$ and $\widehat\cF_p$ are given by
\begin{alignat}{2}
T^A &= \rd E^A + E^B \wedge \Omega_B{}^A 
    &\,
    &= \frac{1}{2} E^B E^C T_{CB}{}^A~, \\
H &= \rd B 
    &\,
    &= \frac{1}{3!} E^A E^B E^C H_{CBA}~, \\
\widehat \cF_p &= \rd \widehat \cC_{p-1} + \widehat \cC_{p-3} \wedge H
    &\,
    &=\frac{1}{p!} E^{A_1} \cdots E^{A_p} \widehat \cF_{A_p \cdots A_1}~.
\end{alignat}
The complex of $p$-form field strengths is encoded in the supercovariant Ramond-Ramond bispinor
\begin{align}\label{E:RRtoF}
S^{1\halpha \,2\hbeta} =
\begin{pmatrix}
S^{\alpha \bgamma} & 0 \\
0 & 0
\end{pmatrix} (\Lambda^{-1})_{\hbgamma}{}^{\hbeta}
= \frac{e^\varphi}{32i} 
    \begin{cases}
    \sum_p
        \frac{1}{p!} \widehat \cF_{a_1 \cdots a_p} 
        (C P_R \gamma^{a_1 \cdots a_p})^{\halpha \hbeta}
        & \text{IIB/IIB$^*$ ($p$ odd)} \\[2ex]
    \sum_p
        \frac{1}{p!} \widehat \cF_{a_1 \cdots a_p} 
        (C P_R \gamma^{a_1 \cdots a_p})^{\halpha \hbeta}
        & \text{IIA/IIA$^*$ ($p$ even)}
    \end{cases}~.
\end{align}
This $S$ differs from \cite{Wulff:2013kga, Wulff:2016tju} by a factor of $-16i$. An
extra factor of two comes from employing the democratic formulation with both field strengths and their duals.

Employing 32-component Majorana spinors can be inconvenient when exhibiting the various torsion tensors. This was addressed in \cite{Butter:2022gbc} by introducing tilde spinors for the second copy of the gravitini and dilatini
\begin{align}
E_M{}^{2\halpha} =
\begin{pmatrix}
E_M{}^{\talpha} \\
0
\end{pmatrix}~, \qquad
\chi_{2 \halpha} =
\begin{pmatrix}
\chi_\talpha \\
0
\end{pmatrix}~,
\end{align}
so that 16-component Majorana-Weyl notation can be used throughout. Effectively, tilde spinors are just barred spinors of DFT, reinterpreted as either same chirality or opposite chirality as 
unbarred spinors, depending on the duality frame, i.e.
$E_M{}^\talpha$ is $E_M{}^{2\alpha} \delta_\alpha{}^\talpha$ or 
$E_M{}^2{}_\alpha \delta^{\alpha \talpha}$. We do not employ tilde spinors in the main body of this paper, but they are convenient for describing the superspace curvatures without sprinkling chiral projectors everywhere. First, we introduce tilde $\gamma$ matrices as
\begin{align}
(\tilde\gamma^c)_{\talpha \tbeta} = 
\begin{cases}
\phantom{+} (\gamma^c)_{\alpha \beta} & \text{IIB} \\[1ex]
- (\gamma^c)_{\alpha \beta} & \text{IIB${}^*$} \\[1ex]
- (\gamma^c)^{\alpha \beta} & \text{IIA} \\[1ex]
\phantom{+} (\gamma^c)^{\alpha \beta} & \text{IIA${}^*$}
\end{cases}~, \qquad
(\tilde\gamma^c)^{\talpha \tbeta} = 
\begin{cases}
\phantom{+} (\gamma^c)^{\alpha \beta} & \text{IIB} \\[1ex]
- (\gamma^c)^{\alpha \beta} & \text{IIB${}^*$} \\[1ex]
- (\gamma^c)_{\alpha \beta} & \text{IIA} \\[1ex]
\phantom{+} (\gamma^c)_{\alpha \beta} & \text{IIA${}^*$}
\end{cases}~.
\end{align}
In terms of these, the non-vanishing torsion tensors are given through dimension 1 by
\begin{subequations}
\begin{alignat}{3}
T_{\alpha \beta}{}^c &= -i (\gamma^c)_{\alpha \beta}~, &\qquad
T_{\talpha \tbeta}{}^c &= -i (\gamma^c)_{\talpha \tbeta}~, \\
T_{\gamma \beta}{}^\alpha &= 
    2 \,\chi_{(\gamma} \delta_{\beta)}{}^\alpha
    - (\gamma_a)_{\gamma \beta} (\gamma^a \chi)^{\alpha}~, &\quad
T_{\tgamma \tbeta}{}^\talpha &=  
    2\,\chi_{(\tgamma} \delta_{\tbeta)}{}^\talpha
    - (\gamma_a)_{\tgamma \tbeta} (\gamma^a \chi)^{\talpha}~, \\
T_{\gamma b}{}^\alpha
    &= - \frac{1}{8} H_{bcd} \,(\gamma^{c d})_{\gamma}{}^\alpha~, &\quad
T_{\tgamma b}{}^\talpha
    &= \frac{1}{8} H_{bcd} \,(\gamma^{c d})_{\tgamma}{}^\talpha~, \\
T_{\tgamma b}{}^\alpha 
    &= -2i \,S^{\alpha \tbeta} \,(\gamma_b)_{\tbeta\tgamma}~, &\qquad
T_{\gamma b}{}^\talpha &= 2i \,S^{\talpha \beta} \,(\gamma_b)_{\beta\gamma}~.
\end{alignat}
\end{subequations}
The dilatini $\chi_\alpha$ and $\chi_\talpha$ are given by the spinor derivatives of the dilaton
\begin{align}
D_\alpha \varphi = \chi_\alpha~, \qquad
D_\talpha \varphi = \chi_\talpha~.
\end{align}
The non-vanishing components of the Kalb-Ramond field strength are
\begin{align}
H_{\gamma \beta a} 
    = -i (\gamma_a)_{\gamma \beta}, \qquad
H_{\tgamma \tbeta a} 
    = +i (\gamma_{a})_{\tgamma \tbeta}~, \qquad H_{abc}~.
\end{align}
The supercovariant Ramond-Ramond bispinor can be written as
\begin{align}
S^{\alpha \tbeta} 
    &= \frac{e^\varphi}{32i} \times
    \begin{cases}
    \sum_p
        \frac{1}{p!} \widehat \cF_{a_1 \cdots a_p} (\gamma^{a_1 \cdots a_p})^{\alpha \beta}\, \delta_\beta{}^\tbeta
        & \text{IIB/IIB$^*$ ($p$ odd)} \\[2ex]
    \sum_p
        \frac{1}{p!} \widehat \cF_{a_1 \cdots a_p} (\gamma^{a_1 \cdots a_p})^\alpha{}_\beta \, \delta^{\beta \tbeta}
        & \text{IIA/IIA$^*$ ($p$ even)}
    \end{cases}~.
\end{align}

\section{Gauged superspace \texorpdfstring{$\sigma$}{sigma}-models}
\label{A:GaugedSigmaModels}
In this appendix, we provide a concise extension of the work of Hull and Spence \cite{Hull:1989jk,Hull:1990ms} to superspace (see also \cite{Jack:1989ne} and \cite{Hull:2006qs}). In large part, this is merely a relabeling of indices and the addition of a grading, but we include it here for the reader's convenience.

\subsection{Target space supergeometry}
A superspace $\sigma$-model comes equipped with a graded symmetric rank-two tensor $G_{MN}$ and a super two-form $B_{MN}$. We presume there exist certain superisometries
$k_{\Iso1} = k_{\Iso1}{}^M \pa_M$, 
which leave $G_{MN}$ and $H = \rd B$ invariant.  The latter condition means that
\begin{align}
\Lie_{\Iso1} H = 0 \quad \implies \quad
k_{\Iso1} \lrcorner H = \rd v_{\Iso1}
\label{E:kintoH}
\end{align}
for some one-form $v_{\Iso1} = \rd Z^M v_{M \Iso1}$.
We use the convenient shorthand $\Lie_{\Iso1} \equiv \Lie_{k_{\Iso1}}$.
The Killing supervectors $k_{\Iso1}$ obey the algebra
$[k_{\Iso1}, k_{\Iso2}] = f_{\Iso1 \Iso2}{}^{\Iso3} k_{\Iso3}$.
The following conditions hold:
\begin{subequations}
\begin{alignat}{3}
k_{\Iso1} \lrcorner k_{\Iso2} \lrcorner H &= k_{\Iso1} \lrcorner \rd v_{\Iso2} 
&\quad &\implies &\quad
k_{\Iso1} \lrcorner \rd v_{\Iso2} &= 
    -k_{\Iso2} \lrcorner \rd v_{\Iso1} \,(-1)^{\grad{\Iso1} \grad{\Iso2}} ~, \\
\Lie_{\Iso1} k_{\Iso2} \lrcorner H &= f_{\Iso1 \Iso2}{}^{\Iso3} k_{\Iso3} \lrcorner H
&\quad &\implies &\quad
    \Lie_{\Iso1} \rd v_{\Iso2} &= f_{\Iso1 \Iso2}{}^{\Iso3} \rd v_{\Iso3}~, \\
\rd \Big(
k_{\Iso1} \lrcorner k_{\Iso2} \lrcorner H
\Big) &=  f_{\Iso1 \Iso2}{}^{\Iso3} \rd v_{\Iso3}
&\quad &\implies &\quad
k_{\Iso1} \lrcorner k_{\Iso2} \lrcorner H 
    &= - \rd\Lambda_{\Iso1 \Iso2} + f_{\Iso1 \Iso2}{}^{\Iso3} v_{\Iso3}~, \\
\rd \Big(
    k_{\Iso1} \lrcorner k_{\Iso2} \lrcorner k_{\Iso3} \lrcorner H
\Big) &= -3 f_{[\Iso1 \Iso2|}{}^{\Iso4} \rd \Lambda_{\Iso4 |\Iso3]}
&\quad &\implies &\quad
k_{\Iso1} \lrcorner k_{\Iso2} \lrcorner k_{\Iso3} \lrcorner H &= - c_{\Iso1 \Iso2 \Iso3}
    - 3 f_{[\Iso1\Iso2|}{}^{\Iso4} \Lambda_{{\Iso4} |{\Iso3}]}
\end{alignat}
\end{subequations}
where we introduce a locally defined, 
(graded) antisymmetric scalar function $\Lambda_{\Iso1 \Iso2}$
and the (graded) antisymmetric constant $c_{\Iso1 \Iso2 \Iso3}$.
As a consequence of the above equations, one can show
\begin{align}\label{E:dLambda1}
\Lie_{\Iso1} v_{\Iso2} - f_{\Iso1 \Iso2}{}^{\Iso3} v_{\Iso3}
    = \rd (k_{\Iso1} \lrcorner v_{\Iso2} - \Lambda_{\Iso1 \Iso2})
\end{align}
The closed one-form $w$ introduced in \cite{Hull:1989jk} corresponds to
$\rd (k_{\Iso1} \lrcorner v_{\Iso2} - \Lambda_{\Iso1 \Iso2})$ here.

There is some gauge redundancy in these quantities:
\begin{align}
\delta v_{\Iso1} = \rd \rho_{\Iso1}~, \qquad
\delta \Lambda_{\Iso1 \Iso2} = f_{\Iso1 \Iso2}{}^{\Iso3} \rho_{\Iso3} + c_{\Iso1 \Iso2}~, \qquad
\delta c_{\Iso1 \Iso2 \Iso3} = -3 f_{[\Iso1 \Iso2|}{}^{\Iso4} c_{\Iso4 |\Iso3]}~,
\label{E:rhoambiguity}
\end{align}
where $c_{\Iso1 \Iso2}$ is an antisymmetric constant and $\rho_{\Iso1}(Z)$ is a scalar function of the target space coordinates, with a residual ``gauge-for-gauge symmetry'' of
$\delta \rho_{\Iso1} = c_{\Iso1}$ and
$\delta c_{\Iso1 \Iso2} = -f_{\Iso1 \Iso2}{}^{\Iso3} c_{\Iso3}$.

One can also define the isometry on the $B$ field directly,
$\Lie_{\Iso1} B = \rd \Big(v_{\Iso1} + k_{\Iso1} \lrcorner B\Big) = -\rd \omega_{\Iso1}$.
Then in the context of generalized geometry, one can speak of the generalized vector
$\xi_{\Iso1} = k_{\Iso1} + \omega_{\Iso1}$
with Dorfman bracket
$[\xi_{\Iso1}, \xi_{\Iso2}]_D = f_{\Iso1 \Iso2}{}^{\Iso3} \xi_{\Iso3} +
\rd (\Lambda_{\Iso1 \Iso2} - k_{\Iso1} \lrcorner v_{\Iso2} )$
obeying a generalization of the non-abelian algebra of the $k_{\Iso1}$. The additional
one-form term above is a trivial transformation from the perspective of double field theory.

\subsection{Gauged \texorpdfstring{$\sigma$}{sigma}-model}
The ungauged $\sigma$-model is given as the sum of a kinetic and Wess-Zumino term,
\begin{align}
\cL = - \frac{1}{2} \rd Z^M \wedge \star \rd Z^N G_{NM}
    - \frac{1}{2} \rd Z^M \wedge \rd Z^N B_{NM}~.
\end{align}
It possesses a global symmetry $\delta Z^M = \lambda^{\Iso1} k_{\Iso1}{}^M$.
To gauge it, we introduce a worldsheet one-form $A^{\Iso1}$ that transforms as
\begin{align}
\delta_\lambda A^{\Iso1} = -\rd \lambda^{\Iso1} - A^{\Iso2} \lambda^{\Iso3} f_{\Iso3 \Iso2}{}^{\Iso1}
\end{align}
so that $D Z^M := \rd Z^M + A^{\Iso1} k_{\Iso1}{}^M$ transforms as
$\delta D Z^M = \lambda^{\Iso1} D Z^N \pa_N k_{\Iso1}{}^M$.
The kinetic term is then invariant by simply replacing $\rd Z^M \rightarrow D Z^M$.

The Wess-Zumino term is more involved. Let us simply give the answer:
\begin{align}
\cL_{\rm WZ} = 
    - B 
    - A^{\Iso1} \wedge v_{\Iso1} 
    - \frac{1}{2} A^{\Iso1} \wedge A^{\Iso2} \Lambda_{\Iso2 \Iso1}
    + F^{\Iso1} \chi_{\Iso1}~.
\end{align}
Pullbacks to the worldsheet are implicitly assumed in the above equations. In the final term, we have used the field strength
\begin{align}
F^{\Iso1} = \rd A^{\Iso1} - \frac{1}{2} A^{\Iso2} \wedge A^{\Iso3} f_{\Iso3 \Iso2}{}^{\Iso1}~, \qquad
\delta F^{\Iso1} = -F^{\Iso2} \lambda^{\Iso3} f_{\Iso3 \Iso2}{}^{\Iso1}
\end{align}
and included a Lagrange multiplier $\chi_{\Iso1}$ whose equation of motion
enforces that $A^{\Iso1}$ is pure gauge. Strictly speaking the gauged $\sigma$-model
lacks the Lagrange multiplier term, but we will include it since we are interested
in performing a duality transformation.

In order for the Wess-Zumino term to be invariant, we must impose two conditions.
First, the Lagrange multiplier field $\chi_{\Iso1}$ must transform as
\begin{align}
\delta \chi_{\Iso1} = 
    \lambda^{\Iso2} f_{\Iso2 \Iso1}{}^{\Iso3} \chi_{\Iso3}
    + \lambda^{\Iso2} (\Lambda_{\Iso2 \Iso1} - k_{\Iso2} \lrcorner v_{\Iso1})
\end{align}
With this condition, the Wess-Zumino term varies (up to a total derivative) into
\begin{align}
\delta \cL_{\rm WZ} = -\frac{1}{2} A^{\Iso1} \wedge A^{\Iso2} \,\lambda^{\Iso3} c_{\Iso3 \Iso2 \Iso1}
\end{align}
and so invariance actually requires this constant to vanish,
$c_{\Iso3 \Iso2 \Iso1}=0~$.
This is a crucial consistency condition for the ability to gauge the action.

The Lagrange multiplier field must transform under the residual symmetry \eqref{E:rhoambiguity} as $\delta \chi_{\Iso1} = - \rho_{\Iso1}(Z)$.
The residual constant $c_{\Iso1 \Iso2}$ shift is no longer a symmetry of the action: it instead leads to \emph{different gauged actions} whose $\Lambda_{\Iso1 \Iso2}$ factors differ by such a constant. Because $c_{\Iso1 \Iso2 \Iso3}$ vanishes, such shifts must obey the cocycle condition
$f_{[\Iso1 \Iso2|}{}^{\Iso4} c_{\Iso4 |\Iso3]} = 0$. 

In a standard gauging, the Lagrange multiplier is
absent and so one must be able to consistently fix $\chi_{\Iso1} = 0$, leading to
\begin{align}
\label{E:GaugingConstraint}
k_{\Iso2} \lrcorner v_{\Iso1} = \Lambda_{\Iso2 \Iso1} \quad \implies \quad 
k_{(\Iso2} \lrcorner v_{\Iso1)} = 0~.
\end{align}
This is a key condition discussed in \cite{Hull:1989jk}. It turns out that a consequence of
\eqref{E:GaugingConstraint} is that $c_{\Iso1 \Iso2 \Iso3}$ vanishes, so we can consider the former condition as fundamental. 
Once the condition \eqref{E:GaugingConstraint}
is imposed, the residual symmetry parameter $\rho_{\Iso1}$ in \eqref{E:rhoambiguity} is restricted to obey $\cL_{\Iso1} \rho_{\Iso2} = f_{\Iso1 \Iso2}{}^{\Iso3} \rho_{\Iso3}$.

In principle, the duality can proceed directly by integrating out the gauge fields $A^{\Iso1}$.
The resulting action admits the local $\lambda^{\Iso1}$ gauge symmetry, implying that ${\rm dim\,} G$ coordinates are unphysical and can be eliminated by a gauge-fixing. A simpler procedure is to go to adapted coordinates.

\subsection{Adapted coordinates}
If the isometries act freely, one can select out ${\rm dim\,} G$ coordinates so that
$k_{\Iso1} = k_{\Iso1}{}^{\cIso1} \pa_{\cIso1}$. In these adapted coordinates
$Z^M = (Z^\uM, Y^{\cIso1})$ where $Z^\uM$ are spectator coordinates.
We do not address the non-free case, but one can follow a very similar line of reasoning.

Let $g(Y)$ be a group element for the group $G$ we are gauging. The left and right-invariant one-forms are
$e^{\Iso1} t_{\Iso1} = g^{-1} \rd g$ and
$k^{\Iso1} t_{\Iso1} =  \rd g g^{-1}$
with the generators obeying \eqref{E:superLieAlgebra}.
The  Killing vectors $k_{\Iso1}$ obey
$k_{\Iso1} \lrcorner k^{\Iso2} = \delta_{\Iso1}{}^{\Iso2}$
and
$k_{\Iso1} \lrcorner e^{\Iso2} = (\Adj g^{-1})_{\Iso1}{}^{\Iso2}$.
We define
\begin{align}
\tilde A^{\Iso1} := D Z^{\cIso1} e_{\cIso1}{}^{\Iso1} 
    = \rd Z^{\cIso1} e_{\cIso1}{}^{\Iso1} 
    + A^{\Iso2} k_{\Iso2}{}^{\cIso1} e_{\cIso1}{}^{\Iso1}
    = (k^{\Iso2} + A^{\Iso2}) (\Adj g^{-1})_{\Iso2}{}^{\Iso1}~.
\end{align}
This is a gauge-invariant one-form, $\delta_\lambda \tilde A^{\Iso1} = 0$.
The kinetic term can be written
\begin{align}
\cL_{\rm kin} = -\Big(
    \rd Z^\uM \wedge \star \rd Z^\uN \,G_{\uN \uM}
    + 2 \tilde A^{\Iso1} \wedge \star \rd Z^\uN \,G_{\uN \Iso1 }
    + \tilde A^{\Iso1} \wedge \star \tilde A^{\Iso2} \,G_{\Iso2 \Iso1}
    \Big)
\end{align}
where we have flattened the $\cIso1$ indices on the metric with $e_{\Iso1}{}^{\cIso1}$.
Every piece above is separately gauge invariant. 
For the metric, the invariance condition reduces
to independence of $Y^{\cIso1}$.

The Wess-Zumino term is more involved. First, trade $A^{\Iso1}$ for $\tilde A^{\Iso1}$. The result is structurally identical and reads
\begin{align}
\cL_{\rm WZ} &= 
    -\tilde B 
    - \tilde A^{\Iso1} \wedge \tilde v_{\Iso1} 
    - \frac{1}{2} \tilde A^{\Iso1} \wedge \tilde A^{\Iso2} \tilde \Lambda_{\Iso2 \Iso1}
    + \tilde F^{\Iso1} \tilde \chi_{\Iso1}~,
\end{align}
where the tilded quantities are defined as
\begin{alignat}{2}
\tilde B &= B - k^{\Iso1} \wedge v_{\Iso1} 
    + \frac{1}{2} k^{\Iso1} \wedge k^{\Iso2} \Lambda_{\Iso2 \Iso1}~, &\quad
\tilde v_{\Iso1} &= (\Adj g)_{\Iso1}{}^{\Iso2} \Big(v_{\Iso2} - k^{\Iso3} \Lambda_{\Iso3 \Iso2}\Big)~, \eol
\tilde \Lambda_{\Iso2 \Iso1} &=
    (\Adj g)_{\Iso2}{}^{\Iso2'} (\Adj g)_{\Iso1}{}^{\Iso1'} \Lambda_{\Iso2' \Iso1'}
    \, (-)^{\grad{\Iso2'} (\grad{\Iso1} + \grad{\Iso1'})}~, &\quad
\tilde \chi_{\Iso1} &= (\Adj g)_{\Iso1}{}^{\Iso2} \chi_{\Iso2}~, \eol
\tilde F^{\Iso1} &= \rd \tilde A^{\Iso1} 
    - \frac{1}{2} \tilde A^{\Iso2} \wedge \tilde A^{\Iso3} f_{\Iso3 \Iso2}{}^{\Iso1}
    = F^{\Iso2} (\Adj g^{-1})_{\Iso2}{}^{\Iso1}~.
\end{alignat}
The tilded quantities $\tilde v_{\Iso1}$ and $\tilde\Lambda_{\Iso2 \Iso1}$ obey
the useful relations:
\begin{align}
\rd \tilde v_{\Iso1} 
    &= 
    e_{\Iso1} \lrcorner H 
    + e^{\Iso2} \wedge (e_{\Iso2} \lrcorner e_{\Iso1} \lrcorner H)
    - \frac{1}{2} e^{\Iso2} \wedge e^{\Iso3} \,
        (e_{\Iso3} \lrcorner e_{\Iso2} \lrcorner e_{\Iso1} \lrcorner H)~, \\
\rd \tilde \Lambda_{\Iso2 \Iso1} - f_{\Iso2 \Iso1}{}^{\Iso3} \tilde v_{\Iso3}
    &= - e_{\Iso2} \lrcorner e_{\Iso1} \lrcorner H 
    + e^{\Iso3} \,(e_{\Iso3} \lrcorner e_{\Iso2} \lrcorner e_{\Iso1} \lrcorner H)~.
\end{align}
The right-hand sides of both these expressions are annihilated by $e_{\Iso1} \lrcorner$, so they are independent of $\rd Y^{\cIso1}$.
The field strength $\tilde F^{\Iso1}$ can be expanded out to rewrite the Wess-Zumino term as
\begin{align}
\cL_{\rm WZ} &= 
    - \tilde B 
    - \tilde A^{\Iso1} \wedge \Big(\tilde v_{\Iso1} + \rd \tilde \chi_{\Iso1}\Big)
    - \frac{1}{2} \tilde A^{\Iso1} \wedge \tilde A^{\Iso2} \Big(
        \tilde \Lambda_{\Iso2 \Iso1} + f_{\Iso2 \Iso1}{}^{\Iso3} \tilde \chi_{\Iso3}
        \Big)
\end{align}
In this form, it's very easy to show gauge invariance of the second and third terms using
\begin{align}
\label{delta.tchi}
\delta_\lambda \tilde \chi_{\Iso1} = -\lambda^{\Iso2} k_{\Iso2} \lrcorner \tilde v_{\Iso1}~, \quad
\delta_\lambda \tilde v_{\Iso1} = \rd \Big( \lambda^{\Iso2} k_{\Iso2} \lrcorner \tilde v_{\Iso1} \Big)~, \quad
\delta_\lambda \tilde \Lambda_{\Iso2 \Iso1}
    = f_{\Iso2 \Iso1}{}^{\Iso4} \lambda^{\Iso3} k_{\Iso3} \lrcorner \tilde v_{\Iso4}~.
\end{align}

To understand the meaning of $\tilde B$, it helps to rewrite it as
\begin{align}
\tilde B = B - e^{\Iso1} \wedge \tilde v_{\Iso1} 
    - \frac{1}{2} e^{\Iso1} \wedge e^{\Iso2} \tilde \Lambda_{\Iso2 \Iso1}~.
\end{align}
In this form, we can show that
\begin{align}
\tilde H 
    &= H
    - e^{\Iso1} \wedge e_{\Iso1} \lrcorner H 
    - \frac{1}{2} e^{\Iso1} \wedge e^{\Iso2} \wedge (e_{\Iso2} \lrcorner e_{\Iso1} \lrcorner H)
    + \frac{1}{6} e^{\Iso1} \wedge e^{\Iso2} \wedge e^{\Iso3}
        \,(e_{\Iso3} \lrcorner e_{\Iso2} \lrcorner e_{\Iso1} \lrcorner H ) \eol
    &= \frac{1}{3!} \rd Z^\uM \wedge \rd Z^\uN \wedge \rd Z^{\uP} \,H_{\uP \uN \uM}
\end{align}
This means that $\tilde H$ is independent of both $Y^{\cIso1}$ and $\rd Y^{\cIso1}$.
Up to a $B$-field transformation, the same condition can be imposed on $\tilde B$,
at least locally.
This means that we can expand
\begin{align}
B &= \frac{1}{2} \rd Z^{\uM} \wedge \rd Z^{\uN} \,B_{\uN \uM}
    + e^{\Iso1} \wedge \rd Z^{\uM} \,B_{\uM \Iso1}
    + \frac{1}{2} e^{\Iso1} \wedge e^{\Iso2} \,B_{\Iso2 \Iso1}~, \\
\tilde v_{\Iso1} &= \rd Z^{\uM} \tilde v_{\uM \Iso1} + e^{\Iso2} \tilde v_{\Iso2 \Iso1}~, \\
B_{\uM \Iso1} &= \tilde v_{\uM \Iso1}~, \qquad
B_{\Iso2 \Iso1} = 2 \,\tilde v_{\Iso2 \Iso1} + \tilde \Lambda_{\Iso2 \Iso1}~.
\end{align}

If we want the $\lambda$ gauge symmetry to be completely eliminated at this stage, 
so that e.g. $\tilde \chi_{\Iso1}$ is invariant, we should choose
$\tilde v_{\Iso2 \Iso1}= 0$, which is the consistency condition
\eqref{E:GaugingConstraint} discussed earlier. A consequence of this condition is that
\begin{align}
k_{\Iso1} \lrcorner B = -v_{\Iso1}~, \qquad \cL_{\Iso1} B = 0~.
\end{align}
This means that the Wess-Zumino term  can finally be written as
\begin{align}
\cL_{\rm WZ}
    &= - \frac{1}{2} D Z^M \wedge D Z^N B_{NM}
    + \tilde F^{\Iso1} \tilde \chi_{\Iso1} \eol
    &= -\frac{1}{2} \rd Z^\uM \wedge \rd Z^\uN \,B_{\uN \uM}
    - \tilde A^{\Iso1} \wedge \rd Z^\uM \,B_{\uM \Iso1}
    - \frac{1}{2} \tilde A^{\Iso1} \wedge \tilde A^{\Iso2} \,B_{\Iso2 \Iso1}
    + \tilde F^{\Iso1} \tilde \chi_{\Iso1}
\end{align}
in terms of the original $B$-field. The components of $B$ in the second line above, are each independent of the coordinate $Y^{\cIso1}$. Relabeling $\tilde \chi_{\Iso1}$ as $\nu_{\Iso1}$,
we recover the recipe for gauging reviewed in section \eqref{S:SNATD.worldsheet}.

\section{Flux tensors for \texorpdfstring{$\eta$}{eta} and \texorpdfstring{$\lambda$}{lambda} deformations}
\label{A:FluxTensors}
We summarize the structure constants $F_{\hcA \hcB \hcC}$ relevant for the $\eta$ and $\lambda$ deformations below by their dimension. Both cases can be given in terms of coefficients $c_1$ and $c_2$ (which are proportional to $a_i$ or $b_i$) as well as a function $\Gamma$.

\begin{alignat*}{4}
&\textbf{Dimension 0}
&\qquad \qquad
F_{\alpha \beta \rc} &= \sqrt{2}\, f_{\alpha \beta c} 
&\,
F_{\balpha \bbeta \bar{\rc}} &= -\sqrt{2}\, f_{\alpha \beta c}
\\
&&
F_{\iso1 \alpha}{}^{\beta} &= f_{\iso1 \alpha}{}^{\beta} 
&\,F_{\iso1 \balpha}{}^{\bbeta} &= f_{\iso1 \balpha}{}^{\bbeta} \\
&&
F_{\iso1 \ra \rb} &= f_{\iso1 a b} 
&\,
F_{\iso1 \rba \rbb} &= -f_{\iso1 a b}\\
&&
F_{\iso1 \iso2}{}^{\iso3} &= f_{\iso1 \iso2}{}^{\iso3} 
\\[2ex]
&\textbf{Dimension 1}
&\qquad \qquad
F_{\alpha \bar \rb}{}^\bgamma &= \frac{1}{\sqrt 2} c_1 c_2\, f_{\alpha b}{}^{\bgamma}
&\,
F_{\balpha \rb}{}^\gamma &= \frac{1}{\sqrt 2} c_1 c_2\, f_{\balpha b}{}^{\gamma}\\
&&
F_{\alpha \bbeta}{}^{\iso1} &= c_1 c_2\, f_{\alpha \bbeta}{}^{\iso1}  
\\[2ex]
&\textbf{Dimension 2}
&\qquad \qquad
F_{\ra \rb}{}^{\iso1} &= \frac{1}{2} (c_1)^4 \,\Gamma\, f_{a b}{}^{\iso1} 
&\,
F_{\ol{\ra \rb}}{}^{\iso1} &= \frac{1}{2} (c_2)^4 \, \Gamma\, f_{a b}{}^{\iso1} 
\\
&&
F_{\ra \bar\rb}{}^{\iso1} &= \frac{1}{2} (c_1 c_2)^2\,f_{a b}{}^{\iso1} \\
&&
F_{\alpha}{}^{\beta {\iso1}} &= \frac{1}{2} (c_1)^4\, \Gamma \, f_\alpha{}^{\beta {\iso1}}
&\,
F_{\balpha}{}^{\bbeta {\iso1}} &= -\frac{1}{2} (c_2)^4\, \Gamma \, f_\balpha{}^{\bbeta {\iso1}}
\\
&&
F_\ra{}^{\beta \gamma} &= \frac{1}{\sqrt 2} (c_1)^4 \,\Gamma\, f_{a}{}^{\beta \gamma} 
&\,
F_{\bar\ra}{}^{\beta \gamma} &= \frac{1}{2\sqrt 2} (c_1 c_2)^2 \, f_{a}{}^{\beta \gamma} 
\\
&&
F_{\bar\ra}{}^{\bbeta \bgamma} &= -\frac{1}{\sqrt 2} (c_2)^4 \,\Gamma \, f_{a}{}^{\bbeta \bgamma}   
&\,
F_{\ra}{}^{\bbeta \bgamma} &= -\frac{1}{2\sqrt 2} (c_1 c_2)^2 \, f_{a}{}^{\bbeta \bgamma}  \\[2ex]
&\textbf{Dimension 3}
&\qquad \qquad
F^{\alpha \bbeta {\iso1}} &= -\frac{1}{4} (c_1 c_2)^3\,f^{\alpha \bbeta {\iso1}}
\\[2ex]
&\textbf{Dimension 4}
&\qquad \qquad
F^{\iso1 \iso2 \iso3} &= - \frac{1}{4} (c_1 c_2)^4\, (1-\Gamma^2) \, f^{\iso1 \iso2 \iso3}
\end{alignat*}

The coefficients $c_i$ appear in the fluxes only quadratically. In terms of the generators $T_\cA$ given in sections \ref{S:SUGRA.eta} and \ref{S:SUGRA.lambda}, the coefficients become
\begin{align}
c_i c_j = a_i a_j \,(1+\eta^2) = b_i b_j \,\lambda^{-1}
\end{align}
Specifying the coefficients quadratically circumvents introducing a square root. One can check that the two expressions for $c_i c_j$ go into each other under the analytic continuation
\eqref{E:eta.lambda.connection}. The function $\Gamma$ is given in the two cases by
\begin{align}
\Gamma = \frac{1-6\eta^2 + \eta^4}{(1+\eta^2)^2}
    = \frac{1+\lambda^4}{2\lambda^2}~.
\end{align}
For the $\eta$ deformation, $|\Gamma| \leq 1$ for all values of $\eta$ and vanishes at
$|\eta| = \sqrt{2} \pm 1$. For the $\lambda$ deformation, $\Gamma\geq1$ and saturates the lower bound at $\lambda=1$.
The highest dimension structure constant $F^{\iso1 \iso2 \iso3}$ involves
\begin{align}
1-\Gamma^2 = 
\left(
\frac{4 \eta \,(1-\eta^2)}{(1+\eta^2)^2}
\right)^2
    = - \left(
    \frac{1-\lambda^4}{2\lambda^2}
    \right)^2~.
\end{align}
For reference, we also give some of the relations above in terms of $\varkappa = \frac{2\eta}{1-\eta^2}$:
\begin{align}
-i \,\varkappa = \frac{1-\lambda^2}{1+\lambda^2}~, \qquad
\Gamma = \frac{1-\varkappa^2}{1+\varkappa^2}~, \qquad
1 - \Gamma^2 = \frac{4 \varkappa^2}{(1+\varkappa^2)^2}~.
\end{align}
Upon truncation to the bosonic sector, it is $\varkappa$ and $\lambda^2$ that play the role of the parameters for the conventional $\eta$ and $\lambda$ deformations for a group $G$.

After the redefinitions to go the supergravity frame, the derivatives $\widehat D_\hcA$ in
\eqref{E:nabla.eta} and \eqref{E:nabla.lambda} have flux tensors with $\cF_{\hcA \hcB \hcC}$ formally given by the $F_{\hcA \hcB \hcC}$ above, but with the replacements
\begin{align}
c_i c_j =
\begin{cases}
\hat a_i \hat a_j \times \dfrac{(1+\eta^2)}{(1-\eta^2)} & \text{$\eta$-deformation} \\
\hat b_i \hat b_j \times \lambda^{-1} & \text{$\lambda$-deformation}
\end{cases}~,
\end{align}
where $\hat a_i$ and $\hat b_i$ denote the phases of those quantities. In section \ref{S:SUGRA.eta}, we chose $\hat a_1 = \hat a_2 = 1$.

\bibliography{literature.bib}
\bibliographystyle{utphys_mod_v4}

\end{document}